\documentclass[useAMS,usenatbib]{mnras}
\usepackage{mncite}
\usepackage{graphicx}
\usepackage{hyperref}
\usepackage{amssymb}
\usepackage{aas_macros}
\usepackage{subfig}
\usepackage{ulem}
\usepackage{lscape}
\usepackage{multirow}
\usepackage{color}

\title[The effect of RM models on $v\sin I_{\rm s}$ estimates]{Rossiter--McLaughlin models and their effect on estimates of stellar rotation, illustrated using six WASP systems\thanks{based on observations (under proposal 090.C-0540) made using the HARPS high resolution {\'e}chelle spectrograph mounted on the ESO 3.6\,m  at the ESO La Silla observatory, and completed by photometry obtained the Swiss 1.2m {\it Euler} Telescope, also at La Silla.}}
\author[D. J. A. Brown et al.]
{
D. J. A. Brown$^{1,2}$\thanks{E-mail: d.j.a.brown@warwick.ac.uk},
A. H. M. J. Triaud$^{3,4}$,
A. P. Doyle$^{1}$,
M. Gillon$^{5}$,
M. Lendl$^{6,7}$,
\newauthor
D. R. Anderson$^{8}$,
A. Collier Cameron$^{9}$,
G. H{\'e}brard$^{10,11}$
C. Hellier$^{8}$,
C. Lovis$^{7}$,
\newauthor
P. F. L. Maxted$^{8}$,
F. Pepe$^{7}$,
D. Pollacco$^{1}$,
D. Queloz$^{12,7}$,
B. Smalley$^{8}$
\\
$^{1}$ Department of Physics, University of Warwick, Gibbet Hill Road, Coventry CV4 7AL, UK.\\
$^{2}$ Astrophysics Research Centre, School of Mathematics \&\ Physics, Queen's University, University Road, Belfast BT7 1NN, UK.\\
$^{3}$ Centre for Planetary Sciences, University of Toronto at Scarborough, 1265 Military Trail, Toronto, ON M1C 1A4, Canada\\
$^{4}$ Department of Astronomy \& Astrophysics, University of Toronto, Toronto, ON M5S 3H4, Canada\\
$^{5}$ Institut d'Astrophysique et de G{\'e}ophysique, Universit{\'e} de Li{\`e}ge, All{\'e}e du 6 Ao{\^u}t 17, 4000 Li{\'e}ge 1, Belgium\\
$^{6}$ Austrian Academy of Science, Space Research Institute, Schmiedlstra\ss e 6, A-8042 Graz, Austria \\
$^{7}$ Observatoire Astronomique de l'Universit{\'e} de Gen{\`e}ve, Chemin des Maillettes 51, CH-1290 Sauverny, Switzerland \\
$^{8}$ Astrophysics Group, Keele University, Staffordshire ST5 5BG, UK.\\
$^{9}$ SUPA, School of Physics and Astronomy, University of St Andrews, North Haugh, St Andrews, Fife KY16 9SS, UK.\\
$^{10}$ Institut d'Astrophysique de Paris, UMR7095 CNRS, Universit{\'e} Pierre \& Marie Curie, 98bis boulevard Arago, F-75014 Paris, France \\
$^{11}$ Observatoire de Haute Provence, CNRS/OAMP, F-04870 St Michel l'Observatoire, France. \\
$^{12}$ Cavendish Laboratory, J J Thomson Avenue, Cambridge CB3 0HE, UK\\
}

\begin{document}

\date{Accepted 0000 December 00. Received 0000 December 00; in original form 0000 October 00}

\pagerange{\pageref{firstpage}--\pageref{lastpage}} \pubyear{2014}

\maketitle

\label{firstpage}

\begin{abstract}
We present new measurements of the projected spin--orbit angle $\lambda$ for six WASP hot Jupiters, four of which are new to the literature (WASP-61, -62, -76, and -78), and two of which are new analyses of previously measured systems using new data (WASP-71, and -79). We use three different models based on two different techniques: radial velocity measurements of the Rossiter--McLaughlin effect, and Doppler tomography. Our comparison of the different models reveals that they produce projected stellar rotation velocities ($v \sin I_{\rm s}$) measurements often in disagreement with each other and with estimates obtained from spectral line broadening. The Bou{\'e} model for the Rossiter--McLaughlin effect consistently underestimates the value of $v\sin I_{\rm s}$ compared to the Hirano model. Although $v \sin I_s$ differed, the effect on $\lambda$ was small for our sample, with all three methods producing values in agreement with each other. Using Doppler tomography, we find that WASP-61\,b ($\lambda=4^\circ.0^{+17.1}_{-18.4}$), WASP-71\,b ($\lambda=-1^\circ.9^{+7.1}_{-7.5}$), and WASP-78\,b ($\lambda=-6^\circ.4\pm5.9$) are aligned. WASP-62\,b ($\lambda=19^\circ.4^{+5.1}_{-4.9}$) is found to be slightly misaligned, while WASP-79\,b ($\lambda=-95^\circ.2^{+0.9}_{-1.0}$) is confirmed to be strongly misaligned and has a retrograde orbit. We explore a range of possibilities for the orbit of WASP-76\,b, finding that the orbit is likely to be strongly misaligned in the positive $\lambda$ direction.
\end{abstract}


\section{Introduction}
\label{sec:intro}
All eight planets of the Solar system orbit in approximately the same plane, the ecliptic, which is inclined to the solar equatorial plane by only $7.155\pm0^\circ.002$ \citep{2005ApJ...621L.153B}. The orbital axes for the Solar system planets therefore exhibit near spin-orbit alignment with the Sun's rotation axis (the origin of the slight divergence from true alignment is unknown). There is no guarantee, however, that this holds true for extrasolar planets, as it is known that from binary stars that spin-orbit angles can take a wide variety of values \citep[e.g.][]{1982ApSS..81..357H,1994AJ....107..306H,2007AA...474..565A,2009Natur.461..373A,2014Natur.511..567J,2014ApJ...785...83A}. Compared to such systems, measurement of the alignment angle (`obliquity') for an extrasolar planet is more difficult owing to the greater radius and luminosity ratios. This is compounded by the face that the host star of a close-in exoplanet generally rotates more slowly than does the primary star in a stellar binary of the same orbital period. We are also generally limited to measuring the alignment angle as projected on to the plane of the sky, generally referred to as $\lambda$. Measurement of the true obliquity ($\psi$) requires knowledge of the inclination of the stellar rotation axis to the line of sight, $I_{\rm s}$, an angle that is currently very difficult to measure directly. It is possible to infer a value for $I_{\rm s}$ using knowledge of the projected stellar rotation speed, $v\sin I_{\rm s}$, the stellar radius, $R_{\rm s}$, and the stellar rotation period, $P_{\rm rot}$ \citep[e.g.][]{2014AA...570A..54L} , but the last of these can in turn be tricky to determine \citep[e.g.][]{2014AA...568A..81L}.

HD\,209458\,b was the first extrasolar planet for which $\lambda$ was measured \citep{2000AA...359L..13Q}. Since that work the number of systems for which the projected spin-orbit alignment angle has been measured or inferred has been increasing at a steady rate, and is now closing in on 100. By far the majority of these are transiting, hot Jupiter extrextrasolara-solar planets, and for most of these the value of $\lambda$ has been modelled using the Holt--Rossiter--McLaughlin (RM) effect \citep{1893AA.....12..646H,1910PAllO...1..123S,1916PAllO...3...23S,1924ApJ....60...15R,1924ApJ....60...22M}, the spectroscopic signature that is produced during a transit by the occultation of the red- and blue-shifted stellar hemispheres. Other, complementary methods such as Doppler tomography (DT, \citealt{2010MNRAS.403..151C}), consideration of the gravity darkening effect \citep[e.g.][]{2011ApJS..197...10B}, modelling of photometric star-spot signatures \citep[e.g.][]{2011ApJ...740L..10N,2011ApJ...733..127S,2015MNRAS.450.1760T}, measurement of the chromospheric RM effect in the Ca \textsc{II} H \& K lines \citep{2012AA...539A.150C}, and analysis of photometric variability distributions \citep{2015ApJ...801....3M} have also contributed to the tally.

While the vast majority of the measurements have been made via the RM effect, the models that have been used to model this effect have changed over time, becoming more complex and incorporating more detailed physics. The first models in widespread use were those of \citet{2005ApJ...622.1118O, 2009ApJ...690....1O} and \citet{2006ApJ...650..408G}, but these were superseded by the more detailed models of \citet{2011ApJ...742...69H} and \citet{2013AA...550A..53B}, which take different approaches to the problem. A recent addition to the stable of RM models is that of \citet{2015MNRAS.454.4379B}. This assortment of models means, combined with the variety of instruments with which RV measurements are made, might be introducing biases into the parameters that we measure, particularly $v\sin I_{\rm s}$ and $\lambda$. These have yet to be fully explored.

In this work, we present analysis of the spin-orbit alignment in six hot Jupiter systems, found by the WASP consortium \citep{2006PASP..118.1407P}, with the aim of shedding new light on the problems discussed above. We observed WASP-61 \citep{2012MNRAS.426..739H}; WASP-62  \citep{2012MNRAS.426..739H}; WASP-71 \citep{2013AA...552A.120S} WASP-76 \citep{2013arXiv1310.5607W}; WASP-78 \citep{2012AA...547A..61S}, and WASP-79 \citep{2012AA...547A..61S}. 

These six systems were observed with HARPS under programme ID 090.C-0540 (PI Triaud). Our earlier RM observation campaigns selected systems across a wide range of parameters and aimed to increase the number of spin-orbit measurements with as few preconceptions as possible. Here, we instead selected six particular objects. At the time, \citet{2010ApJ...719..602S}, \citet{2010ApJ...718L.145W} and \citet{2011AA...534L...6T} had noticed intriguing relations between some stellar parameters and the projected spin-orbit angle. Our selection of planets, orbiting stars with $T_{\rm eff}$ around 6250 K, was meant to verify these.

\section{Methods}
\label{sec:methods}
As in \citet{2012MNRAS.423.1503B,2012ApJ...760..139B}, we analyse the complete set of available data for each system: WASP photometry; follow-up photometric transit data from previous studies; follow-up spectroscopic data from previous studies; newly acquired photometric transit data, and newly acquired in-transit spectroscopic measurements of the RM effect using HARPS. New data are described in the appropriates subsections of Section\,\ref{sec:results}, and are available both in the appendix and online as supplementary information. Our modelling has been extensively described in previous papers from the SuperWASP collaboration \citep[e.g.]{2007MNRAS.380.1230C, 2008MNRAS.385.1576P, 2012ApJ...760..139B}, but we summarize the process here for new readers.

Our analysis is carried out using a Markov Chain Monte Carlo (MCMC) algorithm using the Metropolis--Hastings decision maker \citep{1953JChPh..21.1087M,Hastings1970}. Our jump parameters are listed in Table\,\ref{tab:jump}, and have been formulated to minimize correlations and maximize mutual orthogonality between parameters. We use $\sqrt{e}\sin(\omega)$, $\sqrt{e}\cos(\omega)$, $\sqrt{v\sin I}\sin(\lambda)$, and $\sqrt{v\sin I}\cos(\lambda)$ to impose uniform priors on $e$ and $v\sin I_{\rm s}$ and avoid bias towards higher values \citep{2006ApJ...642..505F, 2011ApJ...726L..19A}. Several of our parameters (namely impact parameter, $T_{\rm eff}$, [Fe/H], and the Bou{\'e} model parameters when appropriate) are controlled by Gaussian priors by default (see Table\,\ref{tab:priors}). Others may be controlled by a prior if desired, see Section\,\ref{sec:priors}.

At each MCMC step, we calculate models of the photometric transit (following \citealt{2002ApJ...580L.171M}), the Keplerian RV curve, and the RM effect (see following section). Photometric data is linearly decorrelated to remove systematic trends. Limb-darkening is accounted for by a four-component, non-linear model, with wavelength appropriate coefficients derived at each MCMC step by interpolation through the tables of \citet{2000AA...363.1081C, 2004AA...428.1001C}. The RV Keplerian curve, and thus the orbital elements, is primarily constrained by the existing spectroscopic data, as our new, in-transit spectroscopy covers only a small portion of the orbital phase. Quality of fit for these models is determined by calculating $\chi^2$.

Other parameters are derived at each MCMC step using standard methods. Stellar mass, for example, is calculated using the $T_{\rm eff}--M_{\rm s}$ calibration of \citet{2010AARv..18...67T}, with updated parameters from \citet{southworth2011}. Stellar radius is calculated from $R_{\rm s}/a$ (derived directly from the transit model) and the orbital period, via Kepler's third law.

We use a burn-in phase with a minimum of 500 steps, judging the chain to be converged (and thus burn-in complete) when $\chi^2$ for that step is greater than the median $\chi^2$ of all previous values from the burn-in chain \citep{2008ApJ...673..526K}. This is followed by a phase of 100 accepted steps, which are used to re-scale the error bars on the primary jump parameters, and a production run of $10^4$ accepted steps. Five separate chains are run, and the results concatenated to produce the final chain of length $5\times10^4$ steps. The reported parameters are the median values from this final chain, with the $1\sigma$ uncertainties taken to be the values that enclose $68.3$\,percent of the distribution.

We test for convergence of our chains using the statistics of (\citealt{geweke1992}, to check inter-chain convergence) and (\citealt{gelman1992}, to check intra-chain convergence). We also carry out additional visual checks using trace plots, autocorrelation plots, and probability distribution plots (both one- and two-dimensional). If an individual chain is found to be unconverged then we run a replacement chain, recalculating the reported parameters and convergence statistics. This process is repeated as necessary until the convergence tests indicate a fully-converged final chain.

We have used the UTC time standard and Barycentric Julian Dates in our analysis. Our results are based on the equatorial solar and jovian radii, and masses, taken from Allen's Astrophysical Quantities.

\begin{table*}
	\caption{Details of the jump parameters that we use for our MCMC analysis. These parameters have been selected to maximize mutual orthogonality, and minimize correlations. For detail of the priors, see Table\,\ref{tab:priors}. Some composite jump parameters are indirectly controlled by priors: $e\sin w$ and $e\cos w$ are controlled by the prior on orbital eccentricity, $e$, while $v\sin l$ and $v\cos l$ are controlled by the prior on $v\sin I_{\rm s}$.}
	\label{tab:jump}
	\begin{tabular}{llll}
		\hline
		Parameter					& Units						& Symbol				& Prior?	\\ [2pt]
		\hline \\
		Epoch					& ${\rm BJD}_{\rm TDB}-2450000$	& $t_0$				& No		\\
		Orbital period				& days						& $P_{\rm orb}$		& No		\\
		Transit width				& days						& $W$				& No		\\
		Transit depth				& --							& $d$				& No		\\
		Impact parameter			& Stellar radii					& $b$				& Yes	\\
		Effective temperature		& K							& $T_{\rm eff}$			& Yes	\\
		`Metallicity'				& dex						& $[{\rm Fe}/{\rm H}]$	& Yes	\\
		RV semi-amplitude			& km s$^{-1}$					& $K$				& No		\\
		$\sqrt{e}\sin(\omega)$		& --							& $e\sin w$			& indirectly; Yes/No \\
		$\sqrt{e}\cos(\omega)$		& --							& $e\cos w$			& indirectly; Yes/No \\
		Long-term RV trend			& --							& $\dot{\gamma}$		& Yes/No	\\
		$\sqrt{v\sin I}\sin(\lambda)$	& (km s$^{-1}$)$^{-1/2}$			& $v\sin l$				& indirectly; Yes/No\\
		$\sqrt{v\sin I}\cos(\lambda)$	& (km s$^{-1}$)$^{-1/2}$			& $v\cos l$			& indirectly; Yes/No \\
		Barycentric RV for CCFs		& km s$^{-1}$					& $\gamma_{\rm RM}$	& Yes/No 	\\
		FWHM for CCFs			& km s$^{-1}$					& FWHM$_{\rm RM}$	& No		\\
		Barycentric RV for RM data	& km s$^{-1}$					& $\gamma_{\rm boue}$	& Yes	\\
		Bou{\'e} model Gaussian width	& km s$^{-1}$					& $\sigma_{\rm boue}$	& Yes	\\
		\hline
	\end{tabular}
\end{table*}

\begin{table*}
	\caption{Details of the Bayesian priors that we apply during our MCMC analysis, and the values that were applied during the \textit{final} analysis of each system. Priors marked $^\dagger$ are only applied during tomographic analyses, and are taken from the headers of the relevant CCF FITS files. Priors marked $^\ddagger$ are only applied during analyses using the Bou{\'e} model for the RM effect, and are estimated from the CCF FWHM in the FITS file headers.}
	\label{tab:priors}
	\begin{tabular}{llllll}
		\hline
		\multirow{2}{*}{System}	& \multicolumn{5}{l}{Parameter} \\ [2pt]
							& $T_{\rm eff}$		& $[{\rm Fe}/{\rm H}]$	& $b$			& $\gamma_{\rm RM}^\dagger$	& $\sigma_{\rm boue^\ddagger}$	\\ [2pt]
		\hline \\			
		WASP-61				& $6250\pm150$	& $-0.10\pm0.11$		& $0.09\pm0.08$	& $18.970\pm0.002$		& $15.1\pm0.5$	\\
		WASP-62				& $6230\pm80$	& $0.04\pm0.06$		& $0.29\pm0.11$	& $14.970\pm0.005$		& $12.8\pm0.5$	\\
		WASP-71				& $6050\pm100$	& $0.14\pm0.08$		& $0.39\pm0.14$	& $7.799\pm0.003$		& $13.7\pm0.5$	\\
		WASP-76				& $6250\pm100$	& $0.19\pm0.10$		& $0.14\pm0.10$	& $-1.102\pm0.001$		& $8.5\pm0.5$		\\
		WASP-78				& $6100\pm150$	& $-0.35\pm0.14$		& $0.42\pm0.11$	& $0.456\pm0.002$		& $10.9\pm0.5$	\\
		WASP-79				& $6600\pm100$	& $0.03\pm0.10$		& $0.71\pm0.03$	& $4.9875\pm0.0004$	& $25.0\pm0.5$	\\
		\hline \\
	\end{tabular}
\end{table*}
 
\subsection{Modelling spin-orbit alignment}
\label{sec:RMmodels}
Our first model for the RM effect is that of \citet{2011ApJ...742...69H}. This has become the de facto standard thanks to its rigourous approach to the fitting procedure, which cross-correlates an in-transit spectrum with a template, and maximizes the cross-correlation function (CCF). This method requires prior knowledge of several broadening coefficients, specifically the macroturbulence, $v_{\rm mac}$, and the Lorentzian ($\gamma_H$) and Gaussian ($\beta_H$) spectral line dispersions. For this work we assumed $\gamma_H=0.9$\,km\,s$^{-1}$ in line with \citeauthor{2011ApJ...742...69H}, and also assumed that the coefficient of differential rotation, $\alpha_{\rm rot}=0$\footnote{Whilst several of the systems under consideration are rapidly rotating, without knowledge of the inclination of their stellar rotation axes it is difficult to place a value on $\alpha_{\rm rot}$.}. $\beta_H$ is calculated individually for each RV data set, and depends on the instrument used to collect the data as it is a function of the spectral resolution.

\citet{2013AA...550A..53B} pointed out that the \citeauthor{2011ApJ...742...69H} model is poorly optimized for instruments which use a CCF based approach to their data reduction. For iodine cell spectrographs (e.g. HIRES at the Keck telescope), the \citet{2011ApJ...742...69H} model works well, but for the HARPS data that we obtained for our sample the \citet{2013AA...550A..53B} model (as available via the AROME library\footnote{\url{http://www.astro.up.pt/resources/arome/}}) should be more appropriate. The model defines line profiles for the CCFs produced by the integrated stellar surface out-of-transit, the uncovered stellar surface during transit, and the occulted stellar surface during transit, and assumes them to be even functions. The correction needed to account for the RM effect is calculated through partial differentiation, linearization, and maximization of the likelihood function defined by fitting a Gaussian to the CCF of the uncovered stellar surface. This approach has been tested using simulated data, but has yet to be widely applied to real observations. In this paper, we will therefore compare its results to those from the two other models. To do so, we require values for the width of the Gaussian that is fit to the \textit{out-of-transit}, integrated surface CCF ($\sigma_0$), and for the width of the spectral lines expected if the star were not rotating ($\beta_0$). The latter we set equal to the instrumental profile appropriate to each datum, whilst we use the former as an additional jump parameter for our MCMC algorithm, using the average results given by the HARPS quick reduction pipeline as our initial estimate and applying a prior using that value.

The DT approach was developed by \citet{2010MNRAS.403..151C} for analysis of hot, rapidly rotating host stars that the RM technique is unable to deal with. It has since been applied to exoplanet hosts with a range of parameters \citep{2010MNRAS.407..507C, 2012ApJ...760..139B, 2012AA...543L...5G, 2015AA...579A..55B}. The alignment of the system is analysed through a comparison of the in-transit instrumental line profile with a model of the average out-of-transit stellar line profile. This latter model is created by the convolution of a limb-darkened stellar rotation profile, a Gaussian representing the local intrinsic line profile, and a term corresponding to the effect on the line profile of the `shadow' created as the planet transits its host star. This `bump' in the profile is time-variable, and moves through the stellar line profile as the planet moves from transit ingress to transit egress. Its width tells us the width, $\sigma$, of the local line profile, and is a free parameter. Since this width is measured independently, we can disentangle the turbulent velocity distribution of the local profile from the rotational broadening, measuring both $v\sin I_{\rm s}$ and $v_{\rm mac}$ directly. This gives DT an advantage over spectral analysis, as although it is possible to determine the turbulent velocity using the latter method it requires spectra with very high signal-to-noise ratio (SNR). For work such as ours it is usually necessary, therefore, to assume a value for $v_{\rm mac}$.

The path of the bump is dictated by $b$ and $\lambda$, and as the planet moves from transit ingress to transit egress its shadow covers regions of the stellar surface with different velocities. This leads to a relation between $b$, $\lambda$, and $v\sin I{\rm s}$, which must fit the observed stellar line profile when the local profile and rotational profile are convolved. We thus have two equations for two unknowns ($v\sin I_{\rm s}$ and $\sigma$, as both $b$ and $\lambda$ can be determined from the bump's trajectory), which are therefore well determined. Since $\lambda$ and $v\sin I_{\rm s}$ are independently determined using this method, it has the advantage of being able to break degeneracies that can arise between these two parameters in low impact parameter systems \citep[][e.g.]{2012ApJ...760..139B}. We note, however, that this breaks down in systems with very slow rotation, i.e. where the uncertainty on $v\sin I_{\rm s}$ is comparable to the rotation velocity.

Another advantage that is often observed with DT is the improved precision on measurements of $\lambda$ that it provides, as seen by \citet{2015AA...579A..55B} for the case of the rapidly rotating KOI-12 system. This method also has potential as a confirmation method for planetary candidates, as seen with the case of recent case of HATS-14\,b \citep{2015AJ....150..197H}, or conversely as a false positive identifier for difficult to confirm systems.

For all of these models we separate our RV measurements by instrument, and further treat spectroscopic data taken on nights featuring planetary transits as separate data sets. Our Keplerian RV model considers these separated sets of data to be independent. To account for stellar RV noise, an additional $1$\,m\,s$^{-1}$ is added in quadrature to the out-of-transit data; this is below the level of precision of the spectrographs used for this work.

\subsection{Exploring system architectures}
\label{sec:priors}
As in our previous work, we explore the possible solutions for each system using a combination of parameter constraints and initial conditions. We have four independent constraints that can be applied.
\begin{enumerate}
	\item Apply a Gaussian prior on $v\sin I_{\rm s}$. This indirectly controls the jump parameters $v\sin l$ and $v\cos l$.
	\item Force the planet's orbit to be circular, $e=0$. This indirectly controls the jump parameters $e\sin w$ and $e\cos w$.
	\item Force the barycentric system RV to be constant with time, $\dot{\gamma}=0$, neglecting long-term trends that are indicative of third bodies.
	\item Force the stellar radius, $R_{\rm s}$, to follow a main sequence relationship with $M_{\rm s}$, \textit{or} use the result from spectral analysis as a prior on $R_{\rm s}$.
\end{enumerate}
We consider all 16 possible combinations of these four constraints, analysing each case independently as described above. We discuss these analyses in the following sections. Once all combinations have been examined, we identify the most suitable combination by selecting that which provides the minimal value of the reduced chi-squared statistic, $\chi^2_{\rm red}$. This combination is then reported as the \textit{final solution} for each system.

\section{Results}
\label{sec:results}
\subsection{WASP-61}
\label{sec:W61}
WASP-61\,b orbits a solar metallicity, moderately rotating F7 star, and was initially identified using WASP-South. Follow-up observations using TRAPPIST \citep{2011Msngr.145....2J}, EulerCam (see \citealt{2012AA...544A..72L} for details of the instrument and data reduction procedure), and CORALIE \citep{2000fepc.conf..548Q} showed that the signal was planetary in origin \citep{2012MNRAS.426..739H}. The planet has a circular orbit with a period of $3.9$\,d, and has a relatively high density of $1.1$\,$\rho_{\rm Jup}$. 
 
\begin{figure}
	\centering
	\includegraphics[width=0.48\textwidth]{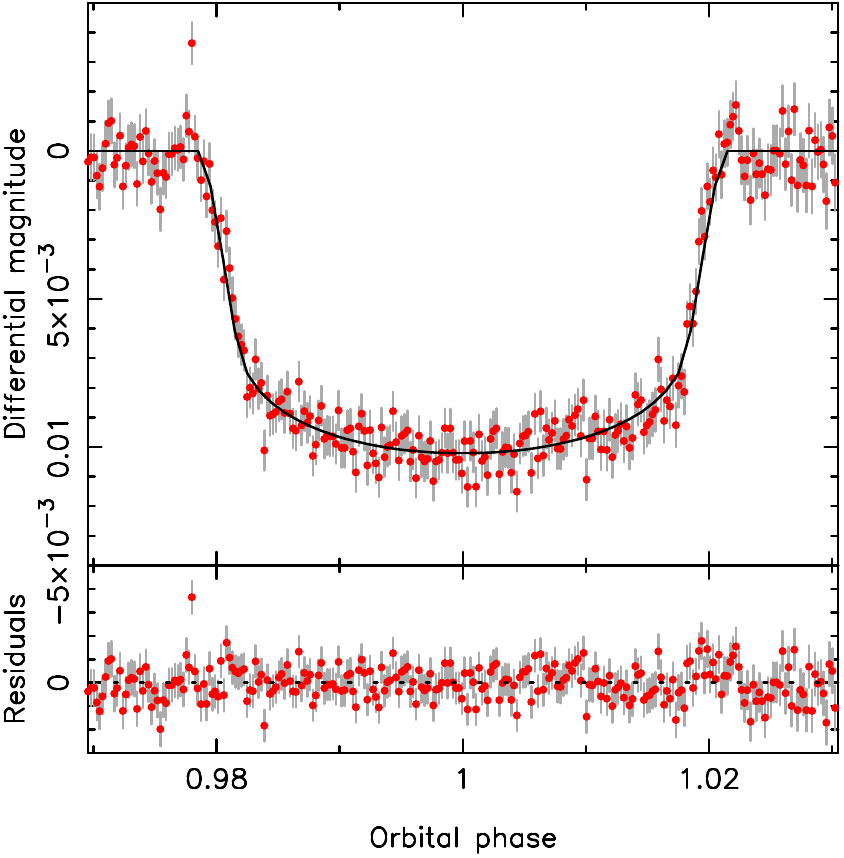}
	\caption{Upper panel: newly acquired EulerCam observations of the transit of WASP-61\,b, with the best-fitting model from MCMC analysis overlaid. No evidence of stellar activity is present. Lower panel: residuals of the data to the best-fitting model.}
	\label{fig:W61_phot}
\end{figure}

We observed the transit on the night of 2012 December 22 using the HARPS high-precision {\'e}chelle spectrograph \citep{2003Msngr.114...20M} mounted on the 3.6-m ESO telescope at La Silla. Fortuitously, we were able to simultaneously observe the same transit photometrically using EulerCam (white light; Fig.\,\ref{fig:W61_phot}). We use these new data in conjunction with all of the data presented in the discovery paper (including the original SuperWASP observations) to model the system using our chosen methods.

\subsubsection{Hirano model}
\label{sec:w61_hirano}
Trial runs with different combinations of input constraints revealed that the impact parameter of the system is low, $\sim0.1$. As expected, a degeneracy between $v\sin I_{\rm s}$ and $\lambda$ was observed to be present, with the distinct crescent shaped posterior probability distribution covering a wide range of angles and extending out to unphysical values of $v\sin I_{\rm s}$. Our $v\sin I_{\rm s} = 10.29\pm0.36$\,km\,s$^{-1}$ prior restricted the values of the two parameters as expected, but we felt that it was better to allow both to vary normally given our aim of comparing the various RM models. Tests with circular and eccentric orbital solutions showed no evidence for an eccentric orbit, with the F-test of \citet{1971AJ.....76..544L} returning a less than $5$\,percent significance for eccentricity. This is also expected, as our new near- and in-transit RV measurements do not help to constrain the orbital eccentricity. Relaxing the stellar radius constraint led to insignificant variations in stellar density (which is computed directly from the photometric light curve, and therefore is distinct from the mass and radius calculations), but for some combinations of input constraints the value of the stellar mass varied by $\approx1\sigma$. Tests for long-term trends in the barycentric velocity of the system returned results with strongly varying values of both positive and negative $\dot{\gamma}$, so we set this parameter to zero for our final runs.

\begin{figure}
	\centering
	\includegraphics[width=0.48\textwidth]{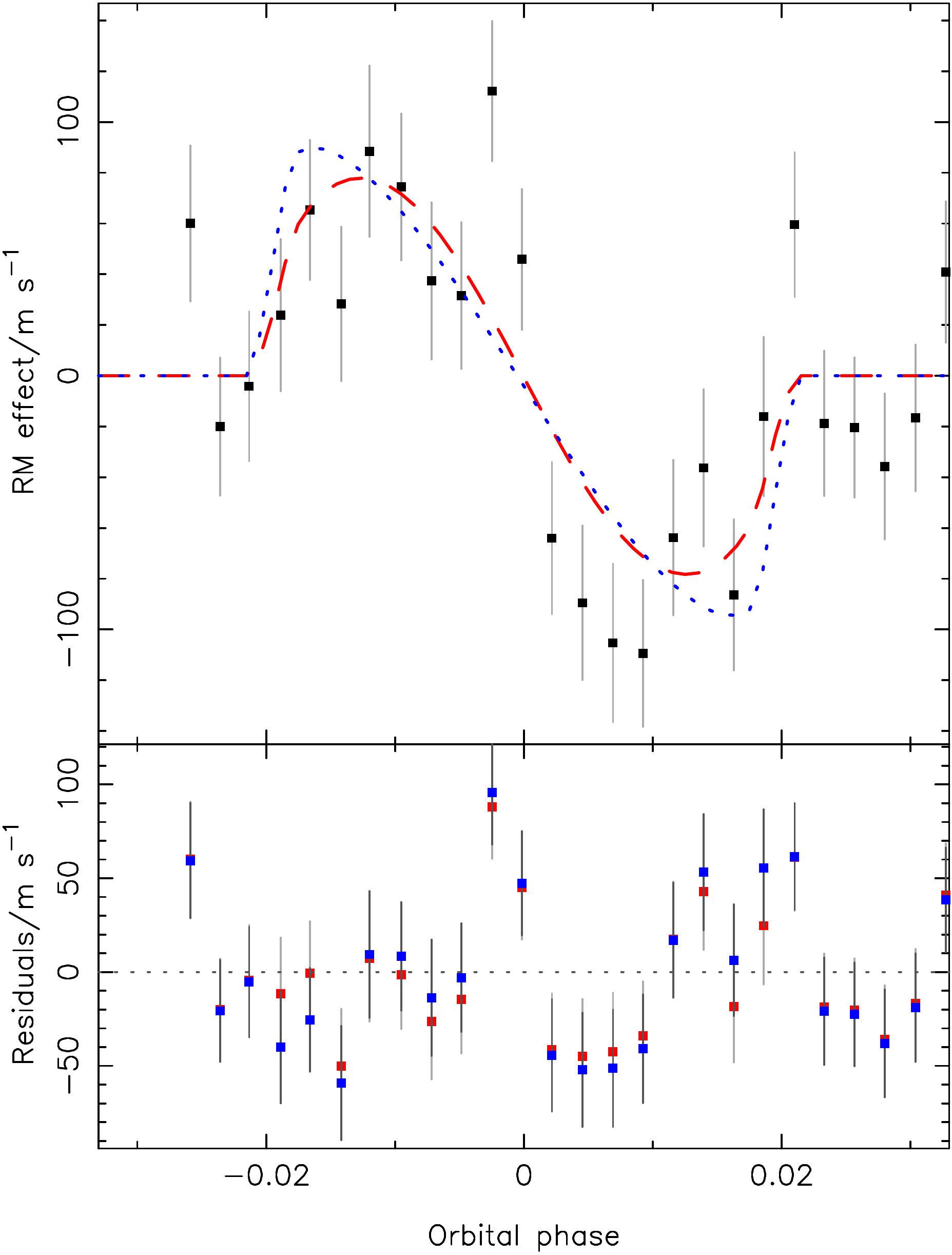}
	\caption{Upper panel: a close up of the RM anomaly in the RV curve of WASP-61, with the contribution from the Keplerian orbit subtracted to better display the form of the anomaly. The red, dashed line denotes the fit produced by the Hirano model, while the blue, dotted line denotes the fit produced by the Bou{\'e} model. The two models clearly produce different fits to the spectroscopic data. Lower panel: the residuals for the two model fits. Red data with light grey error bars represent residuals for the Hirano model, while blue data with dark grey errors bars represent those for the Bou{\'e} model fit.}
	\label{fig:W61_RV}
\end{figure}

The selected solution therefore does not apply a prior on $v\sin I_{\rm s}$, assumes a circular orbit, neglects the possibility of a long-term trend in RV, and neglects the stellar radius constraint. The fit to our data returns $\chi^2_{\rm red}=1.2$. This particular combination of applied constraints returns an projected spin-orbit alignment angle of $\lambda=1^\circ.3^{+18.8}_{-17.3}$, an impact parameter of $b=0.11^{+0.09}_{-0.07}$, and a projected rotation velocity of $v\sin I_{\rm s}=11.8^{+1.5}_{-1.4}$\,km\,s$^{-1}$, which is in agreement with the spectroscopic value of $v\sin I_{\rm s}=10.29\pm0.36$\,km\,s$^{-1}$ determined from the HARPS spectra using $v_{\rm mac}=5.04$\,km\,s$^{-1}$, itself derived using the calibration of \citet{2014MNRAS.444.3592D}. The RM fit produced by the best-fitting parameters is shown in Fig.\,\ref{fig:W61_RV}.

\subsubsection{Bou{\'e} model}
\label{sec:w61_boue}
Similarly to the Hirano model tests, we found no evidence for an eccentric orbit (as expected), no reason to apply a constraint on the stellar radius, and no long-term trend in $\gamma$. The interaction with the prior on $v\sin I_{\rm s}$ was more interesting; with no prior the same degeneracy between $v\sin I_{\rm s}$ and $\lambda$ was observed, but while applying the prior restricted the range of rotational velocities explored as expected, it led to a bimodal distribution in $\lambda$. Examination of the posterior probability distribution for the no-prior case revealed that this was caused by the Bou{\'e} model under-predicting $v\sin I_{\rm s}$ compared to the spectroscopic value and the Hirano model, such that the prior from spectral analysis restricted the MCMC algorithm to values within the `tails' of the crescent distribution. Checking posterior distributions for other parameters reveals that the MCMC chain is well converged, and our statistical convergence tests confirm this.

\begin{figure}
	\centering
	\includegraphics[width=0.48\textwidth]{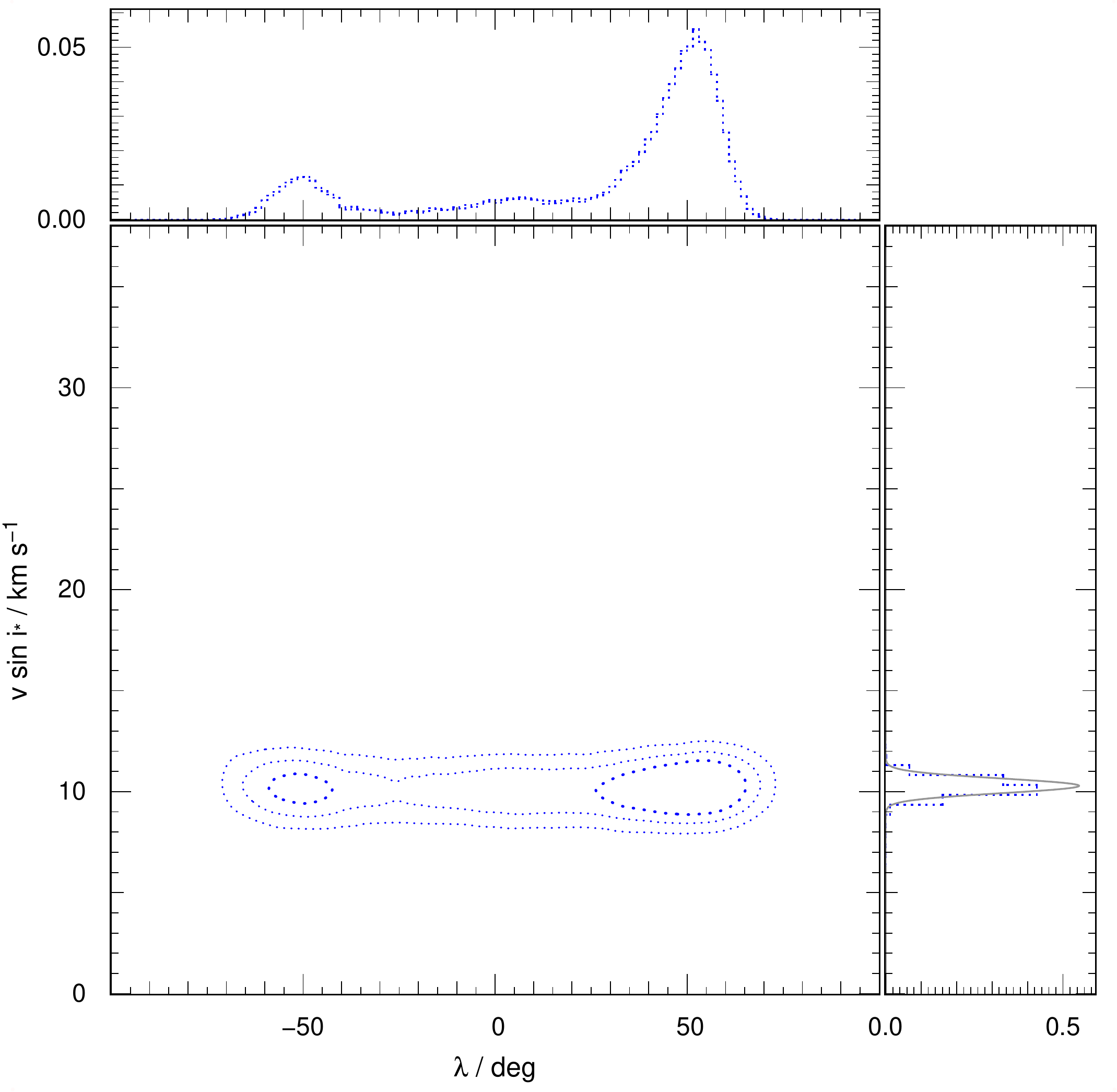}
	\caption{The posterior probability distribution in $i_{\rm orb}--\lambda$ parameter space for analysis of WASP-61 using the Bou{\'e} model while applying a prior on $v\sin I_{\rm s}$ using the value derived from spectral analysis. The contours mark the $1\sigma$, $2\sigma$, and $3\sigma$ confidence regions. Also displayed are the marginalized, one-dimensional distributions for the two parameters, with the additional, solid grey distribution in $v\sin I_{\rm s}$ representing the Bayesian prior. This distribution highlights the degeneracy that arises between solutions with ($i_{\rm orb}$,$\lambda$) and ($i_{\rm orb}-{\rm\pi}$,$-\lambda$)}
	\label{fig:W61_vsiprior}
\end{figure}

This highlights another degeneracy in the RM modelling problem, in addition to that between $v\sin I_{\rm s}$ and $\lambda$, where orbital configurations with ($i_{\rm orb}$,$\lambda$) and ($i_{\rm orb}-{\rm\pi}$,$-\lambda$) produce the same ingress and egress velocities and the same chord length \citep{2005ApJ...622.1118O, 2009ApJ...696.1230F}. By extension therefore these configurations are indistinguishable when considering the two-dimensional problem, and the degeneracy can only be broken by considering the true alignment angle, $\psi$. This is particularly pernicious in the case of orbits with $i_{\rm orb}\approx90^\circ$. Like \citeauthor{2009ApJ...696.1230F}, we limit the inclination to the range $0^\circ\leq i_{\rm orb}\leq90^\circ$, which leads to the distribution shown in Fig.\,\ref{fig:W61_vsiprior}, with solutions close to $\pm40^\circ$.

Ultimately, we adopt the same set of input constraints as for the Hirano model to enable strict comparison between the two models, and acquire final results of $v\sin I_{\rm s} = 8.9^{+3.2}_{-1.7}$\,km\,s$^{-1}$ (consistent with, but lower than the Hirano value as expected from our tests), $\lambda=13^\circ.9^{+35.7}_{-39.6}$, and $b=0.09^{+0.10}_{-0.06}$. The resulting RM fit is shown in blue in Fig.\,\ref{fig:W61_RV}, which clearly indicates that the two models are fitting the same RM effect in different ways. The Bou{\'e} fit exhibits steeper ingress and egress gradients, with sharper peaks at larger $|{\rm velocity}|$ than the Hirano model, which visually seems to provide a better fit to the RV data although neither model fits the second half of the anomaly particularly well. Comparing their reduced $\chi^2$ values though reveals that the Bou{\'e} model gives a slightly poorer fit at $\chi^2_{\rm red}=1.4$, compared to the Hirano model's $\chi^2_{\rm red}=1.2$.

\subsubsection{Doppler tomography}
\label{sec:w61_tomography}
\begin{figure}
	\includegraphics[width=0.48\textwidth]{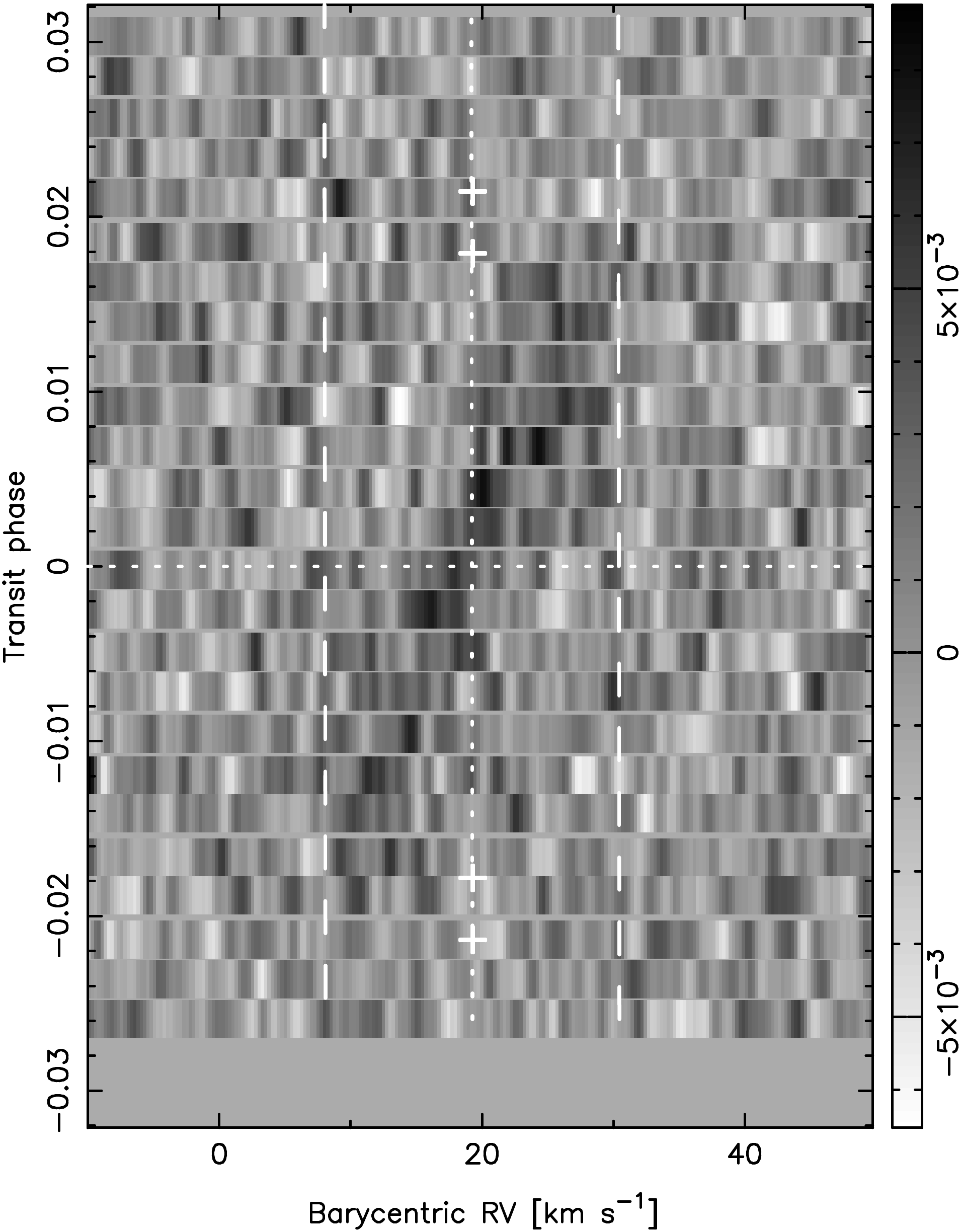}
	\caption{Time series map of the WASP-61 CCFs with the model stellar spectrum subtracted. The signature of the planet is just about visible moving from the lower left corner to the upper right corner, indicating a prograde orbit across both stellar hemispheres. The symmetry about the central line is indicative of an aligned system. Time (phase) increases vertically along the $y$-axis, with the horizontal dotted line marking the mid-transit time (phase). The vertical dotted line denotes the barycentric velocity $\gamma$, whilst the vertical dashed lines indicate $\pm v\sin I$ from this, effectively marking the position of the stellar limbs. The crosses mark the four contact points for the planetary transit. The background has been set to grey to aid clarity.}
	\label{fig:W61_DT}
\end{figure}

We applied the same set of constraints for our  DT analysis as for the other two methods: no prior on $v\sin I_{\rm s}$; $\dot{\gamma}=0$; $e=0$, and no constraint on the stellar radius. The stellar parameters returned were entirely consistent with those from both the Hirano and Bou{\'e} models. With results of $v\sin I_{\rm s}= 11.1\pm0.7$\,km\,s$^{-1}$, $\lambda=4^\circ.0^{+17.1}_{-18.4}$, and $b=0.10^{+0.10}_{-0.06}$, we find no discrepancy between DT analysis and the two other techniques for modelling the RM anomaly. The left-hand panel of Fig.\,\ref{fig:W61_DT} shows the time series of the CCFs, with the prograde signature of the planet barely visible. No sign of stellar activity is visible in the CCF residual map. 

\begin{figure}
	\centering
	\includegraphics[width=0.48\textwidth]{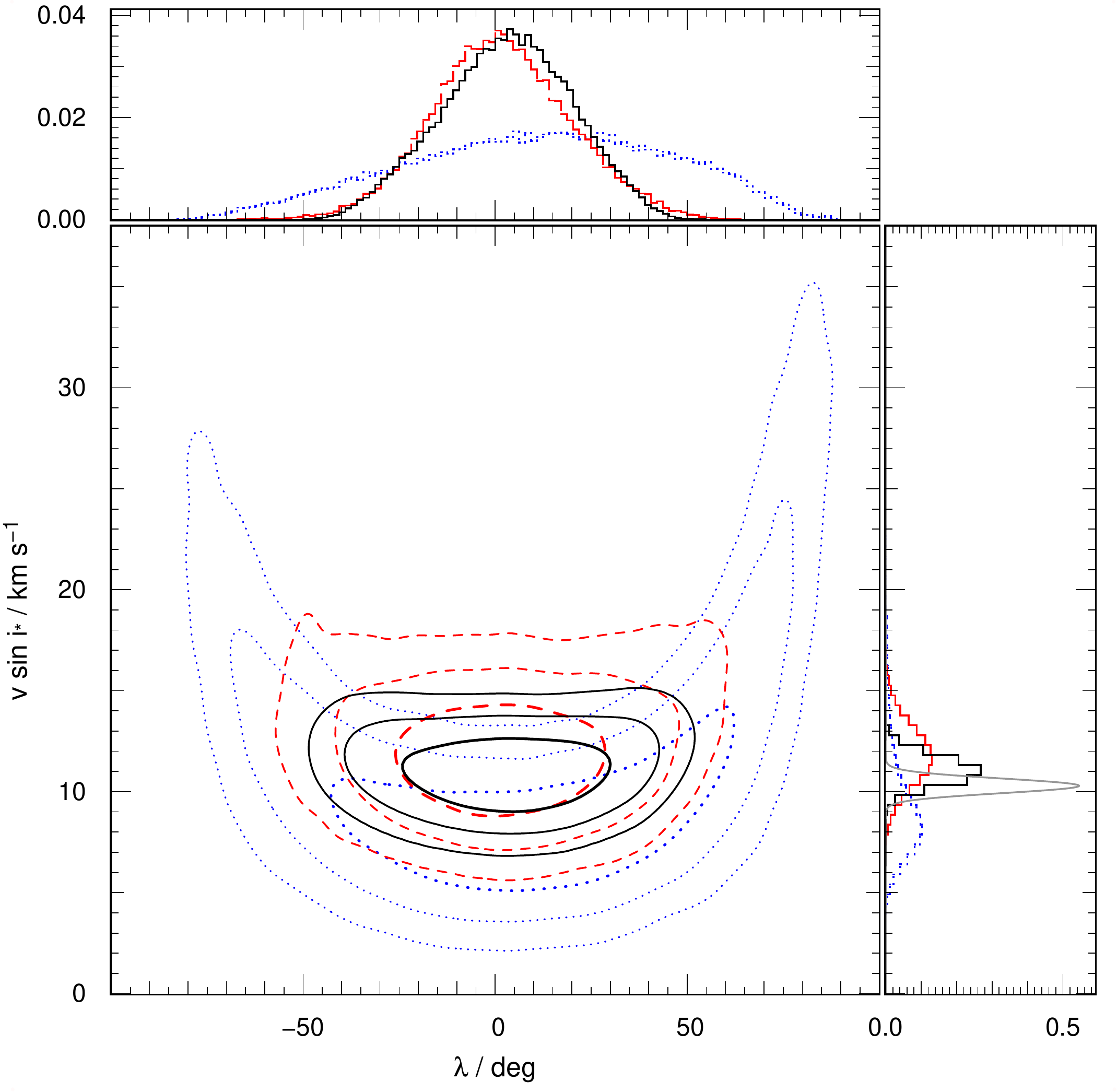}
	\caption{The posterior probability distribution in $v\sin I-\lambda$ parameter space for the Hirano model (red, dashed contours), Bou{\'e} model (blue, dotted contours), and DT (black, solid contours) analyses of WASP-61. The contours mark the $1\sigma$, $2\sigma$, and $3\sigma$ confidence regions. Also displayed are the marginalized, one-dimensional distributions for the two parameters; the different models are distinguished as for the main panel, and the additional, solid grey distribution in $v\sin I_{\rm s}$ represents the result from spectral analysis. The $\lambda$ 1D distribution shows that although tomography improves the precision in the alignment by a factor of $2$, it provides no improvement over the Hirano model. The improvement for this system comes in $v\sin I_{\rm s}$, where the uncertainty reduces by approximately $50$\,percent when tomography is used. Note the crescent shape of the Bou{\'e} distribution, even at the $1\sigma$ level, whereas the distributions for the other models show less structure.}
	\label{fig:W61_vsinI-lambda}
\end{figure}

Fig.\,\ref{fig:W61_vsinI-lambda} shows the posterior probability distributions for all three analysis methods. Interestingly, in this case tomography seems to provide little improvement in the uncertainties on the alignment angle over the Hirano or Bou{\'e} models. Instead, the improvement comes in the precision of the $v\sin I_{\rm s}$ measurement, with the uncertainty in the stellar rotation velocity reducing by approximately $50$\,percent compared to the RM modelling value. This improvement arises due to the different ways in which the different methods treat the spectroscopic data. The Hirano and Bou{\'e} models are analytic approximations of the behaviour of a Gaussian fit to the composite line profile. This is a valid approach when only the RV data are available, particularly when considering HARPS data as it mimics the calculations performed by the HARPS pipeline. But with the full CCF available, the tomographic method is able to treat the various components of the composite profile explicitly, and can use information from both the time-varying and time-invariant parts of the CCF directly.

\subsection{WASP-62}
\label{sec:W62}
Like WASP-61\,b, WASP-62\,b was discovered through a combination of WASP-South, EulerCam, and Trappist photometry, in conjunction with spectroscopy from CORALIE \citep{2012MNRAS.426..739H}. The host star is again a solar metallicity, F7-type star, and the planet has a circular orbit of period $4.4$\,d. WASP-62\,b is rather inflated ($R_{\rm p}=1.39\pm0.06$\,$R_{\rm Jup}$) compared to its mass ($M_{\rm p}=0.57\pm0.04$\,$M_{\rm Jup}$), leading to a much lower density of $0.21$\,$\rho_{\rm Jup}$. Analysis of the HARPS spectra gives $v\sin I_{\rm s}=8.38\pm0.35$\,km\,s$^{-1}$, with $v_{\rm mac}= 4.66$\,km\,s$^{-1}$ from the calibration of \citet{2014MNRAS.444.3592D}; it is these values that we use for our prior on $v\sin I_{\rm s}$.

HARPS was used to observe the spectroscopic transit on the night of 2012 October 12. Additional RV measurements were made using the same instruments on 2012 October 15--17 to help constrain the full RV curve. We use the full set of available data to characterize the system, including the weather-affected EulerCam light curve; as  \citet{2012MNRAS.426..739H} note, the MCMC implementation that underlies our analysis accounts for this poorer quality data.

\begin{figure}
	\centering
	\includegraphics[width=0.48\textwidth]{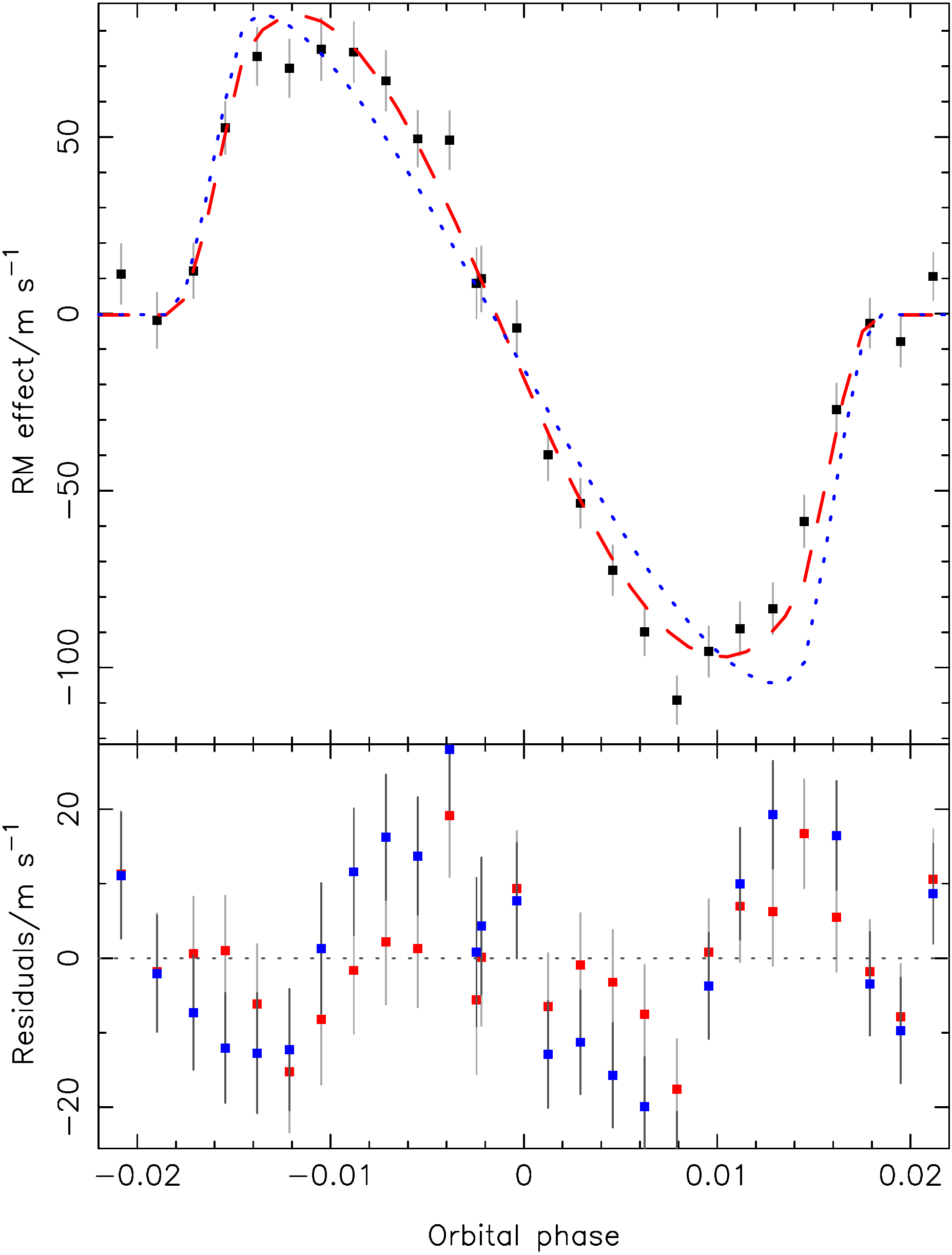}
	\caption{Upper panel: a close up of the RM anomaly in the RV curve of WASP-62, with the contribution from the Keplerian orbit subtracted to better display the form of the anomaly. Lower panel: the residuals for the two model fits. Legends for the two panels as for Fig.\,\ref{fig:W61_RV}.}
	\label{fig:W62_RV}
\end{figure}

\subsubsection{RM modelling}
\label{sec:w62_hirano}
Trial runs to test the effect of applying the four input constraints found that there was no long-term trend in barycentric velocity, and no evidence for an eccentric orbit, with either of the two RM models. Relaxation of the stellar radius constraint led to only minor changes in the reported stellar parameters, with stellar mass and radius being entirely consistent whether the constraint was enforced or not.

Unlike the WASP-61 system, the impact parameter was found to be $\sim0.2--0.3$ such that no degeneracy was expected between $v\sin I_{\rm s}$ and $\lambda$. This was found to be true for the Hirano model, but our examination of the posterior probability distribution produced using the Bou{\'e} model showed a long tail in $v\sin I_{\rm s}$ extending out to values that imply very rapid rotation of the host star. In general though, we again find that the Bou{\'e} model underpredicts $v\sin I_{\rm s}$ compared to the Hirano model - $v\sin I_{\rm s}=7.1^{+0.5}_{-0.4}$ as compared to $v\sin I_{\rm s}=10.5\pm0.4$\,km\,s$^{-1}$. The Bou{\'e} model therefore produces larger $1\sigma$ uncertainties in the value of $\lambda$ in order to compensate when trying to fit the RM effect. This effect can be seen in Fig.\,\ref{fig:W62_vsinI-lambda}, with the $1\sigma$ contours for these models being completely distinct. For both models, the alignment angle value remained pleasingly consistent across the different constraint combinations.

The full sets of results, which were produced from runs using no prior on $v\sin I_{\rm s}$, no constraint on $R_{\rm s}$, $e=0$, and $\dot{\gamma}=0$, can be found in Table\,\ref{tab:results}, and show that the larger impact parameter has enabled more stringent limits to be placed on the spin-orbit alignment angle than was the case for WASP-61. Fig.\,\ref{fig:W62_RV} shows that the two models again produce dissimilarly shaped best-fitting RM models; as with WASP-61, the Bou{\'e} model has steeper ingress and egress velocity gradients, and sharper peaks. However, the angles produced by the two models are entirely consistent, with the Hirano model finding $\lambda=19^\circ.1^{+6.4}_{-5.8}$ and the Bou{\'e} model $\lambda=18^\circ.9^{+11.5}_{-6.6}$.

\subsubsection{Doppler tomography}
\label{sec:w62_tomography}
\begin{figure}
	\includegraphics[width=0.48\textwidth]{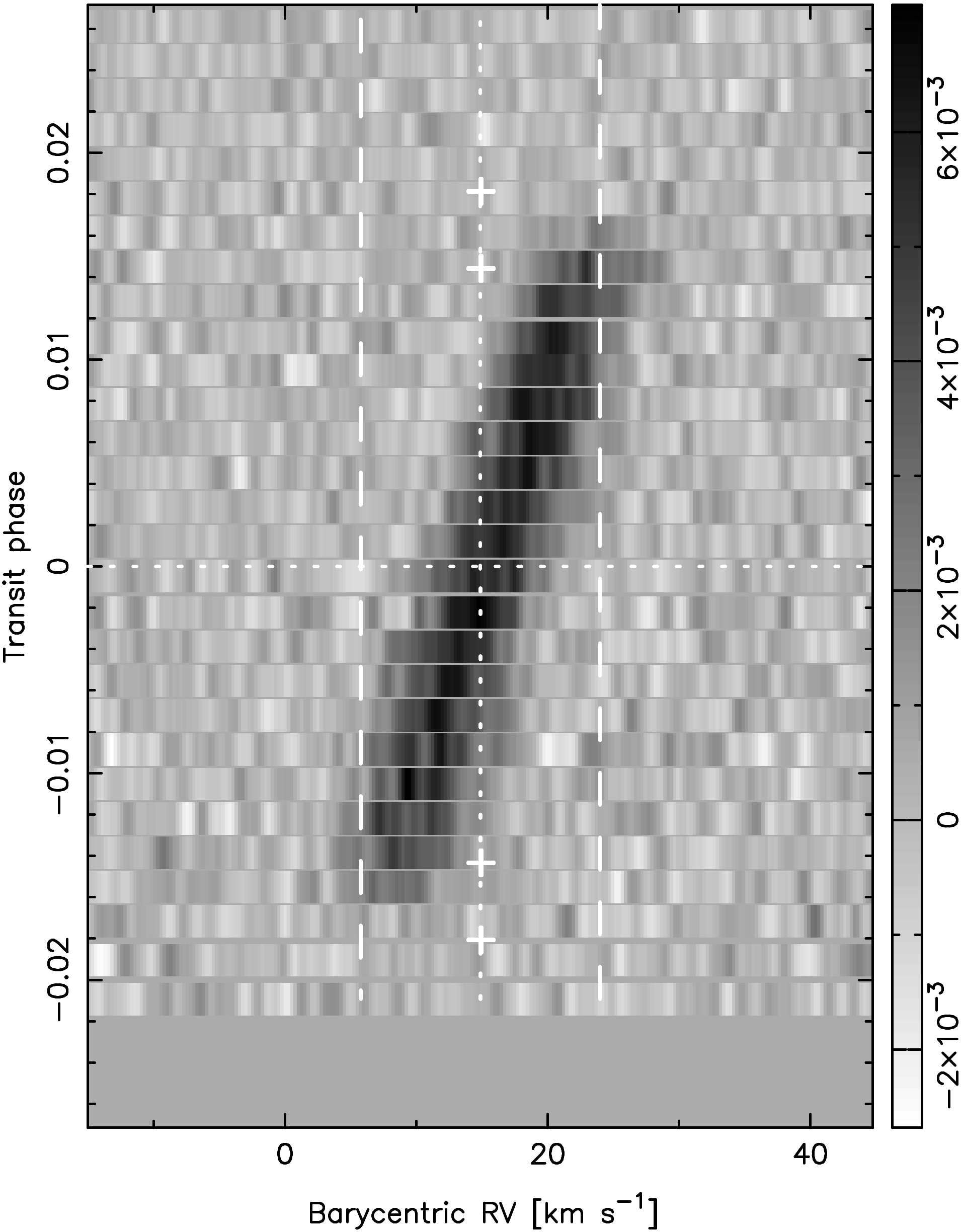}
	\caption{Time series map of the WASP-62 CCFs, following subtraction of a model stellar spectrum. The planetary signature, moving from lower left to upper right, is unambiguous. A prograde, symmetrical orbit is clearly implied, in agreement with the form of the RM effect. Legend as Fig.\,\ref{fig:W61_DT}.}
	\label{fig:W62_DT}
\end{figure}

Using the same set of input constraints as for our RM modelling, we again carried out DT analysis of the system, finding an alignment angle of $\lambda=19^\circ.4^{+5.1}_{-4.9}$. The planetary signature of WASP-62 is much stronger than that of WASP-61, and can be seen far more clearly in the CCF time series map (Fig.\,\ref{fig:W62_DT}). For this system, the tomographic method has improved the uncertainties on $\lambda$ by roughly a factor of $2$ compared to the Bou{\'e} model, but again provides little improvement over the precision afforded by the Hirano model. All three results are consistent with alignment according to the criterion of \citet{2010AA...524A..25T}, a conclusion which is supported by the trajectory of the planetary signal in Fig.\,\ref{fig:W62_DT}. However, this is based on an ad hoc criterion that reflects the typical uncertainty on $\lambda$ measurements at the time that it was formulated. Work since 2010 has improved the typical uncertainty, such that this criterion is no longer really applicable. We therefore classify WASP-62 as slightly misaligned; the resolution of the planet trajectory in our Doppler map is insufficient to distinguish this from a truly aligned orbit.

\begin{figure}
	\centering
	\includegraphics[width=0.48\textwidth]{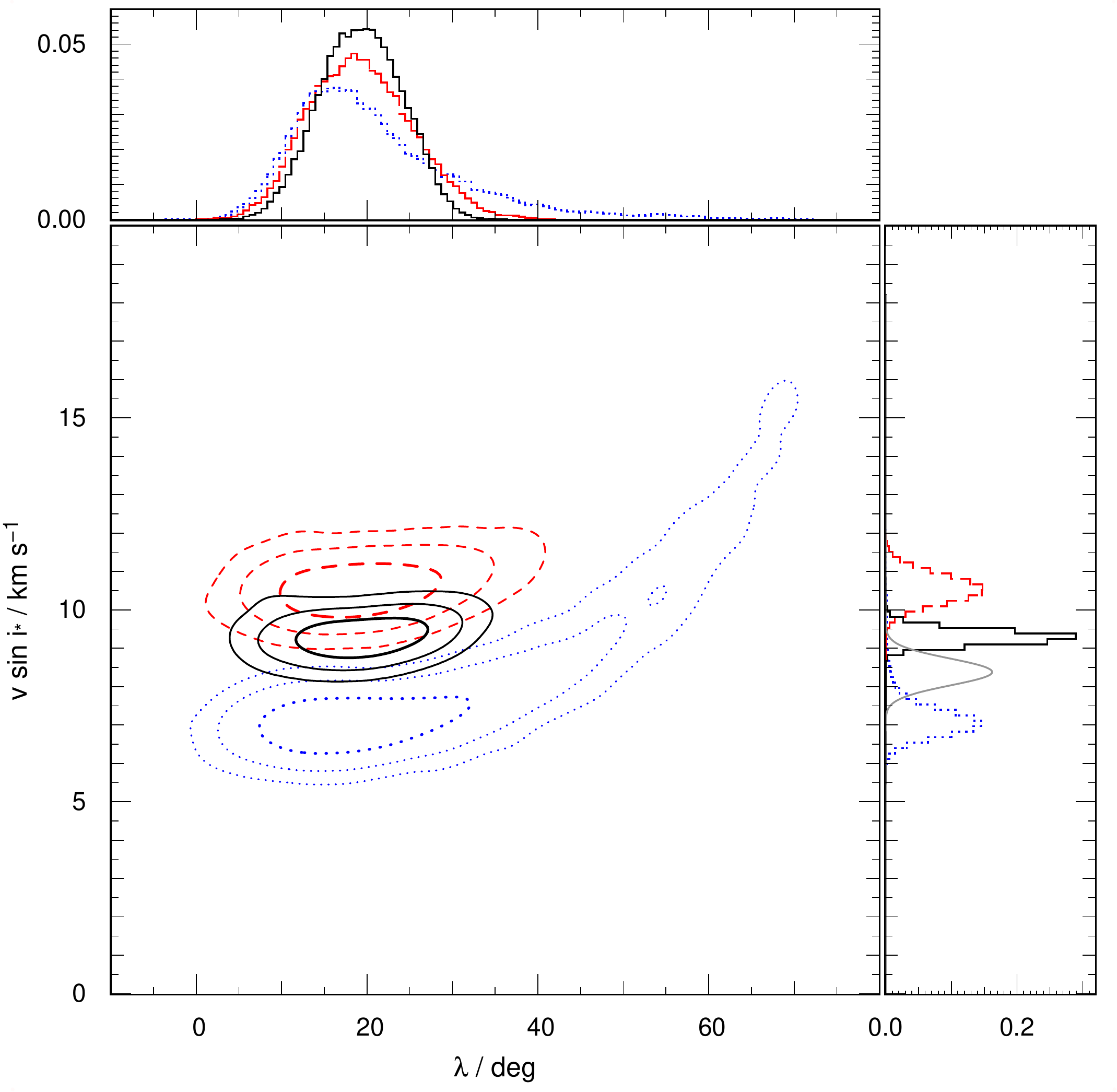}
	\caption{The posterior probability distributions in $v\sin I-\lambda$ parameter space for our analyses of WASP-62. Legend as for Fig.\,\ref{fig:W61_vsinI-lambda}. The three models give distinct $1\sigma$ solutions in $v\sin I_{\rm s}$ for this system, but provide similar precision in $\lambda$. Note that the Bou{\'e} distribution shows an extended tail in $\lambda$ compared to the other models.}
	\label{fig:W62_vsinI-lambda}
\end{figure}

As with WASP-61, it is the treatment of $v\sin I_{\rm s}$ by the three models that is interesting here. We have already noted that the Bou{\'e} model returns lower values than the Hirano model, but the DT result of $v\sin I_{\rm s}=9.3\pm0.2$ falls between the two whilst being consistent with neither thanks to the small error bars on all three estimates (see Fig.\,\ref{fig:W62_vsinI-lambda}). None of the $v\sin I_{\rm s}$ values that we find are consistent with the spectroscopic value of $8.38\pm0.35$\,km\,s$^{-1}$ derived from the HARPS spectra. We also note that the uncertainties in $\lambda$ using this method are smaller than for either the Hirano or Bou{\'e} models (see Table\,\ref{tab:results}).

\subsection{WASP-71}
\label{sec:W71}
\citet{2013AA...552A.120S} presented the discovery of WASP-71\,b using photometry from WASP-N, WASP-South and TRAPPIST, along with spectroscopy from CORALIE that included observations during transit made simultaneously with the TRAPPIST observations. The host star was found to be an evolved F8-type, and significantly larger and more massive than the Sun, whilst the planet was found to be inflated compared to the predictions of \citet[][e.g.]{2003ApJ...592..555B}, and to have a circular orbit with a period of $2.9$\,d. The spectroscopic transit observations made using CORALIE enabled \citeauthor{2013AA...552A.120S} to measure the projected spin-orbit alignment angle of the system. They found that the system was aligned, with $\lambda=20^\circ.1\pm9.7$, and rapidly rotating at $v\sin I_{\rm s}=9.4\pm0.5$\,km\,s$^{-1}$ (calculated assuming $v_{\rm mac}=3.3\pm0.3$\,km\,s$^{-1}$ following \citealt{2014MNRAS.444.3592D}).

We obtained additional spectroscopic data on the night of 2012 October 26, observing a complete transit, with further observations made on 2012 October 23 and 25. We combine these with the discovery photometry and spectroscopy to model the system. We do not, however, include the spectroscopic transit used by \citeauthor{2013AA...552A.120S} to measure $\lambda$, for two reasons. The first is that we wish to obtain an independent measurement of the spin-orbit alignment. The second is that, as noted by \citet{2013AA...550A..53B}, different instruments can produce different signals from the same measurement owing to their different analysis routines, and therefore RM data sets from different instruments should not be combined. Analysis of our new HARPS spectra gives $v\sin I=9.06\pm0.36$\,km\,s$^{-1}$ and $v_{\rm mac}=4.28$\,km\,s$^{-1}$.

\begin{figure}
	\centering
	\includegraphics[width=0.48\textwidth]{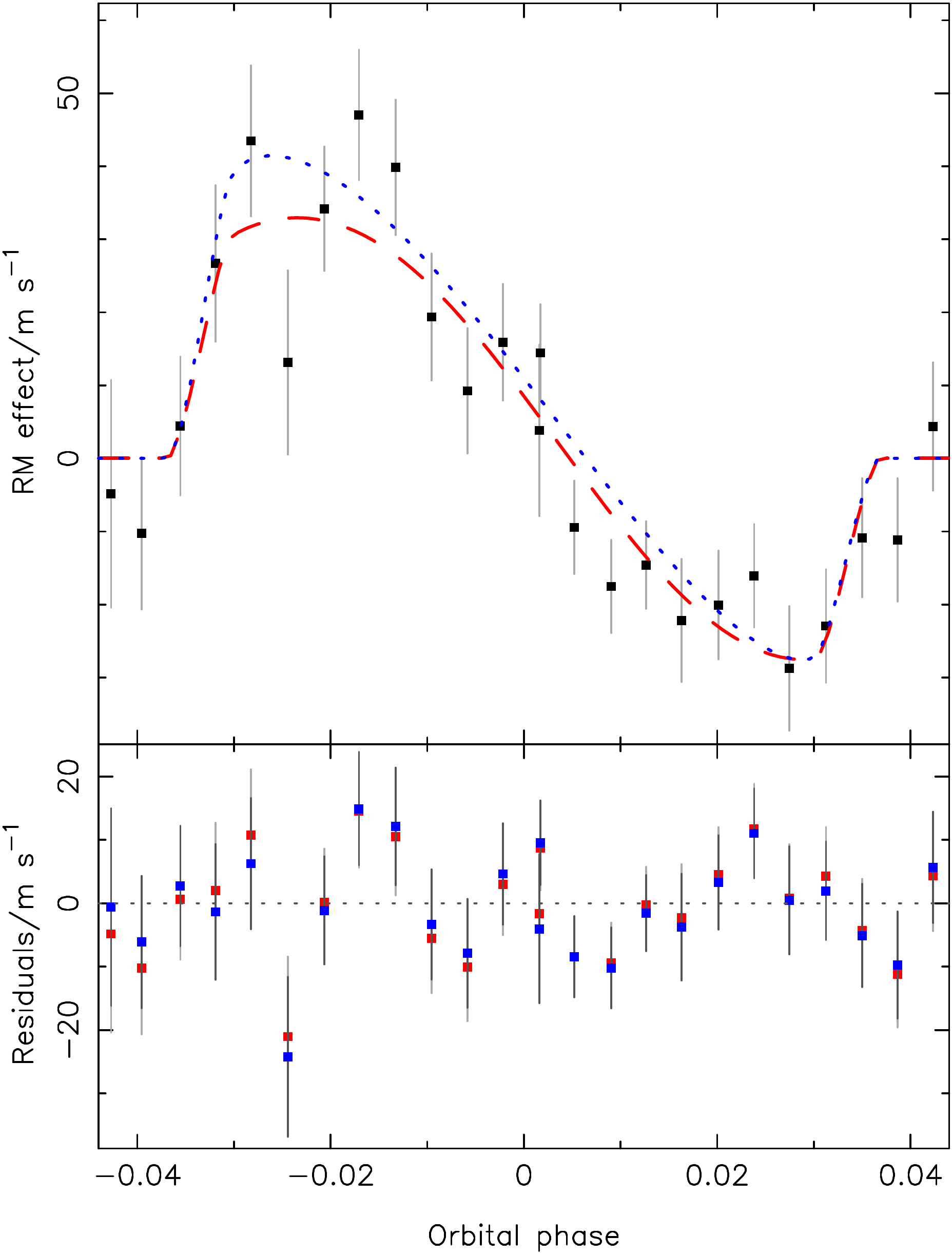}
	\caption{Upper panel: a close up of the RM anomaly in the RV curve of WASP-71, with the contribution from the Keplerian orbit subtracted to better display the form of the anomaly. Lower panel: the residuals for the two model fits. Legends for the two panels as for Fig.\,\ref{fig:W61_RV}.}
	\label{fig:W71_RV}
\end{figure}

\subsubsection{RM modelling}
\label{sec:w71_hirano}
Using the F-test of \citet{1971AJ.....76..544L} we found no indication of significant eccentricity in the system, in agreement with \citet{2013AA...552A.120S}, and therefore set $e=0$ in our final analysis. We also found no consistent evidence that there is a long-term trend in barycentric velocity, so set $\dot{\gamma}=0$.

The interaction between $v\sin I_{\rm s}$, $b$, and the stellar radius constraint is an interesting one for this system. Relaxing the constraint on $R_{\rm s}$ causes the stellar radius to decrease by $\sim25$\,percent, with the stellar mass increasing by approximately $6$\,percent. Relaxing the constraint also leads to a significant, approximately tenfold rise in the impact parameter from $\sim0.05$ to $\sim0.5$, with corresponding effect on the result for $v\sin I_{\rm s}$, which with the radius constraint active is almost unphysically large owing to the degeneracy that arises with both $\lambda$ and $R_{\rm s}$. \citet{2013AA...552A.120S} report an impact parameter of $0.39$, so we chose not impose the stellar radius constraint to allow the impact parameter to fit to what appears to be the more natural value. This does mean that we find a larger, less dense planet than the \citet{2013AA...552A.120S} result. We also note that applying the stellar radius constraint returns a stellar effective temperature which is $\sim200-300$\,K hotter than previous spectroscopic values, whilst neglecting the constraint gives a temperature more consistent with previous analyses. 

Once again, we find that the two different methods return similar results for the alignment angle and stellar parameters, but that the Bou{\'e} model gives a more slowly rotating star than is suggested by the Hirano model (see Table\,\ref{tab:results}). Fig.\,\ref{fig:W71_RV} shows the best-fitting models produced by both methods, with the Hirano model (the dashed, red line) having a shallower peak during the first half of the anomaly. There is substantial scatter in the RV measurements during this period however, and the Hirano model appears to better fit the second half of the anomaly, where there is less scatter in the radial velocities.

The final solutions that we report were taken from runs with no prior on $v\sin I_{\rm s}$, $e=0$, $\dot{\gamma}=0$, and no constraint applied to the stellar radius.

\subsubsection{Doppler tomography}
\label{sec:w71_tomography}
\begin{figure}
	\centering
	\includegraphics[width=0.48\textwidth]{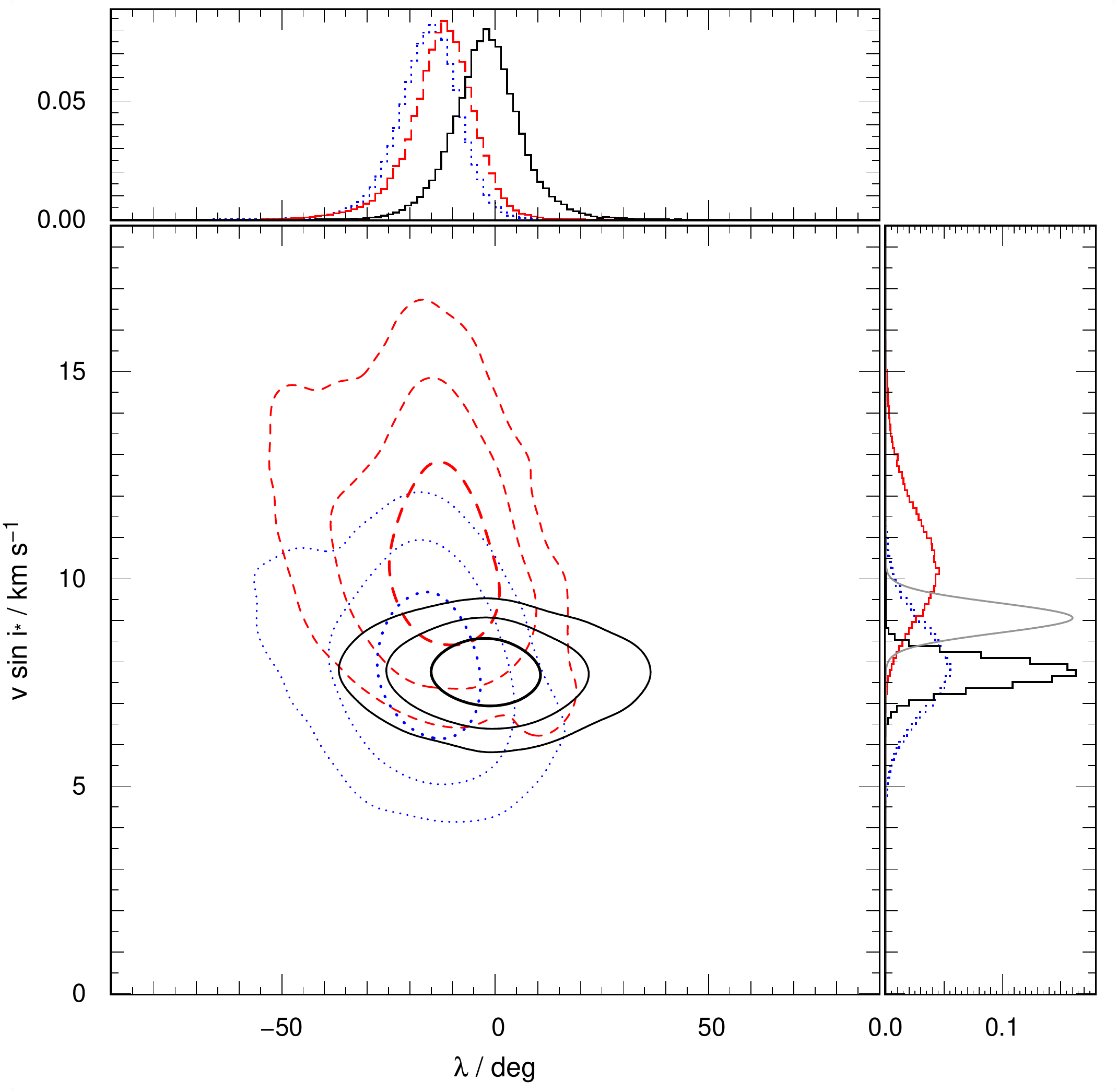}
	\caption{The posterior probability distributions in $v\sin I-\lambda$ parameter space for our analyses of WASP-71. Legend as for Fig.\,\ref{fig:W61_vsinI-lambda}. For this system DT gives a similar $v\sin I_{\rm s}$ result to the Bou{\'e} model, but returns an alignment angle that is shifted more towards $0$ than either the Hirano or Bou{\'e} models. Those models return distributions with very similar shapes, but shifted in $v\sin I_{\rm s}$.}
	\label{fig:W71_vsinI-lambda}
\end{figure}

\begin{figure}
	\includegraphics[width=0.48\textwidth]{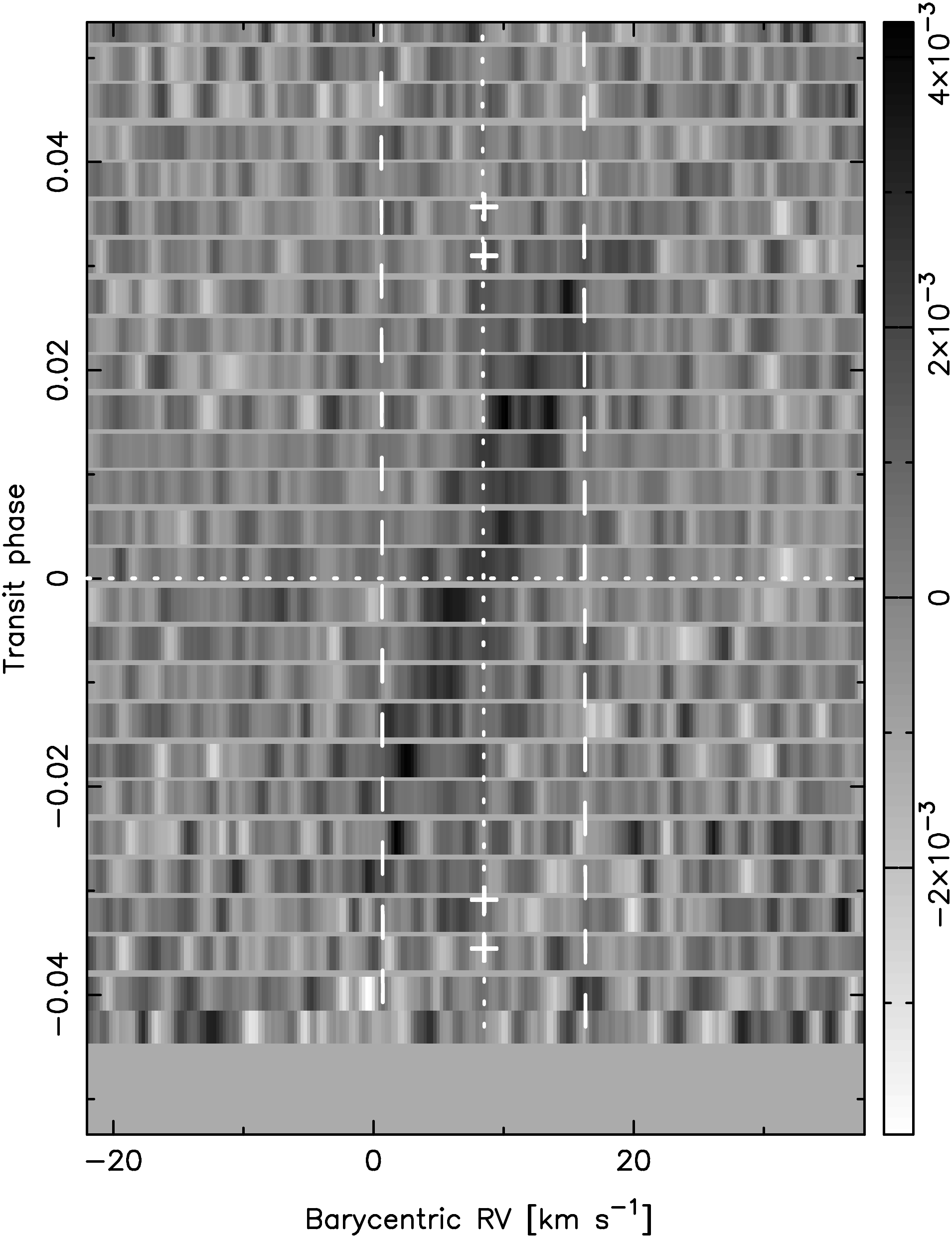}
	\caption{CCF time series, with model stellar spectrum subtracted, for WASP-71. The signature of the planet moves from lower left to top right across the plot, indicating a prograde orbit. The intersection of this signature with $\pm v\sin I_{\rm s}$ close to the phases of ingress and egress indicates a well-aligned orbit. Legend as Fig.\,\ref{fig:W61_DT}.}
	\label{fig:W71_DT}
\end{figure}

Whereas for the two previous systems the tomographic analysis supported the Hirano model with regards to the projected rotation velocity of the host star, for WASP-71 it is the Bou{\'e} model with which DT agrees (see Fig.\,\ref{fig:W71_vsinI-lambda}), although the value of $7.8\pm0.3$\,km\,s$^{-1}$ that we find is significantly lower than the spectroscopic value. The other parameter values that are found through DT are substantially different to either set of RM results. The impact parameter is lower, leading to a lower value of $\lambda$ that is consistent with $0^\circ$ (see Table\,\ref{tab:results}). Particularly interesting though is the difference in the physical stellar parameters found by this method, which imply a smaller star. As implied by our result of $\lambda=-1^\circ.9^{+7.1}_{-7.5}$, Fig.\,\ref{fig:W71_DT} shows that the system is well characterizedaligned, in agreement with the result from \citet{2013AA...552A.120S}, although we are not able to significantly improve on the precision that they report. 

\subsection{WASP-76}
\label{sec:W76}
WASP-76\,A \citep{2013arXiv1310.5607W} is another F7-type planet-hosting star, but is rotating significantly more slowly than either WASP-61 or WASP-62. The planet is substantially bloated, with a density of only $0.151\pm0.010$\,$\rho_{\rm Jup}$, and orbits its host every $1.8$\,days in a circular orbit. It was discovered and characterized using data from WASPSouth, TRAPPIST, EulerCam, SOPHIE \citep{2009AA...505..853B,2011SPIE.8151E..37P}, and CORALIE.

HARPS was used to observe the transit taking place on 2012 November 11, and to make additional measurements on 2012 November 12--14. We combined these measurements with the discovery paper's photometry for our analysis, excluding two spectra that were obtained at twilight. Spectral analysis of the new spectra returned $v_{\rm mac}=4.84$\,km\,s$^{-1}$ using the calibration of \citet{2014MNRAS.444.3592D}, leading to $v\sin I_{\rm s}=2.33\pm0.36$\,km\,s$^{-1}$. We use this for our prior on rotation velocity. The SNR of the spectra are relatively poor however, so we increased the lengths of our MCMC phases to $10000$ (minimum $5000$) for burn-in, and $20000$ for the production phase, leading to a final chain length of $10^5$ for the concatenated chain. We also approached the analysis of this system in a different manner to the other systems in our sample.

\subsubsection{Hirano model}
We began by testing the effects of constraints $2$, $3$, and $4$ (see Section\,\ref{sec:priors}). We found no evidence for a long-term trend in barycentric RV, so adopted the $\dot{\gamma}=0$ constraint for our final solution. We also set $e=0$ after finding no evidence for a significantly eccentric orbit. Despite the large number of photometric light curves available for the system, we found that applying the stellar radius constraint led to increases in both $M_{\rm s}$ and $R_{\rm s}$. We attribute this to a combination of poor photometric coverage of the transit ingress, and significant scatter in some of the light curves (see fig.\,1 of \citealt{2013arXiv1310.5607W}). Despite the varying stellar parameters there was no compelling reason to apply the constraint, and we therefore chose not to do so for the next phase of our analysis.

\paragraph{Exploring $\bf{v\sin I_{\rm s}}$}

Initial exploratory runs with $e=0$, $\dot{\gamma}=0$, no radius constraint, and no prior on $v\sin I_{\rm s}$ consistently found a small impact parameter of $b\sim0.1$, in agreement with the discovery paper value of $b=0.14^{+0.11}_{-0.09}$ but poorly constrained. These runs produced the expected degeneracy between $v\sin I_{\rm s}$ and $\lambda$, and the associated crescent-shaped posterior probability distributions (see Fig.\,\ref{fig:W76_vsinI-lambda}). The distribution shows a slight preference for positive $\lambda$, but the uncertainty on the value was large. Furthermore, our Geweke and Gelman--Rubin tests implied that the MCMC chains were poorly converged.

We thus applied constraint $1$, a prior on $v\sin I_{\rm s}$. The addition of this constraint leads to a bimodal distribution in $\lambda$, with both minima being tightly constrained (see Fig.\,\ref{fig:W76_vsinI-lambda}). Further investigation revealed that individual chains were split roughly $50:50$ between the positive and negative minima in $\chi^2$ space, dependent on the chain's exploration of parameter space during the burn-in phase; this naturally led to poor convergence when analysing the concatenated MCMC chain, but inspection of trace plots, autocorrelation data, running means, and statistics from the Geweke test showed that each \textit{individual} chain was well converged. We thus ran additional chains to collect five that favoured the positive minimum and five that favoured the negative minimum, and tested the convergence of the two minima. \textit{Both were found to be well converged.} We also carried out tests whereby the chain was started at $\pm90^\circ$; the results matched our expectations, with each chain remaining in the associated positive / negative minimum and being well converged.

As with some of our modelling of the WASP-61 (see Section\,\ref{sec:w61_boue}), this is an example of the degeneracy inherent in the RM problem, whereby solutions with ($i_{\rm orb}$, $\lambda$) are indistinguishable in terms of fitting the data from solutions with ($i_{\rm orb}-{\rm\pi}$, $-\lambda$).

Interestingly, chains that explored the positive minimum consistently returned a higher value of the impact parameter than those chains that explored the negative minimum. However, in the case of the positive minimum the median impact parameter of $0.09^{+0.04}_{-0.06}$ was consistent only with the lower end of the impact parameter given in the discovery paper, $b=0.14^{+0.11}_{-0.09}$, and in the case of the negative minimum the value of $0.02^{+0.02}_{-0.01}$ did not agree with discovery paper at all.

\paragraph{Impact parameter}

We therefore elected to explore the option of applying an additional constraint on the system, this time on the impact parameter using the value from the discovery paper as a prior. We tested the application of this prior to cases both with and without a prior on $v\sin I_{\rm s}$.

When we applied the prior on $b$ but \textit{not} the prior on $v\sin I_{\rm s}$, we found very similar results to those obtained in the corresponding case without the impact parameter prior, albeit with much improved convergence of our MCMC chains. The concatenated chain showed a preference for the ${\rm positive-}\lambda$ minimum, with sizeable uncertainty on the result; the major difference with that earlier example was the more tightly constrained impact parameter distribution. Testing chains with initial alignment of $\lambda_0=\pm90^\circ$ showed completely consistent results with the free $\lambda_0$ case, with both cases favouring the positive minimum, albeit with substantial uncertainty on $\lambda$.

When priors on both the impact parameter and $v\sin I_{\rm s}$ were both applied to the free $\lambda_0$ case, the MCMC chains were forced into the positive minimum. The results from the concatenated chain give an impact parameter in agreement with the discovery paper's value, and in addition provide a more precise determination of $\lambda$ than any of the other combinations of constraints. The $\lambda_0=\pm90^\circ$ gave solutions in the corresponding minima, but it is notable that the impact parameter for the negative case is significantly lower than the value expected from the discovery paper. This suggests that the positive $\lambda$ solution should be favoured.

Results from these analyses are shown in Table\,\ref{tab:W76models}.

\begin{figure}
	\centering
	\includegraphics[width=0.48\textwidth]{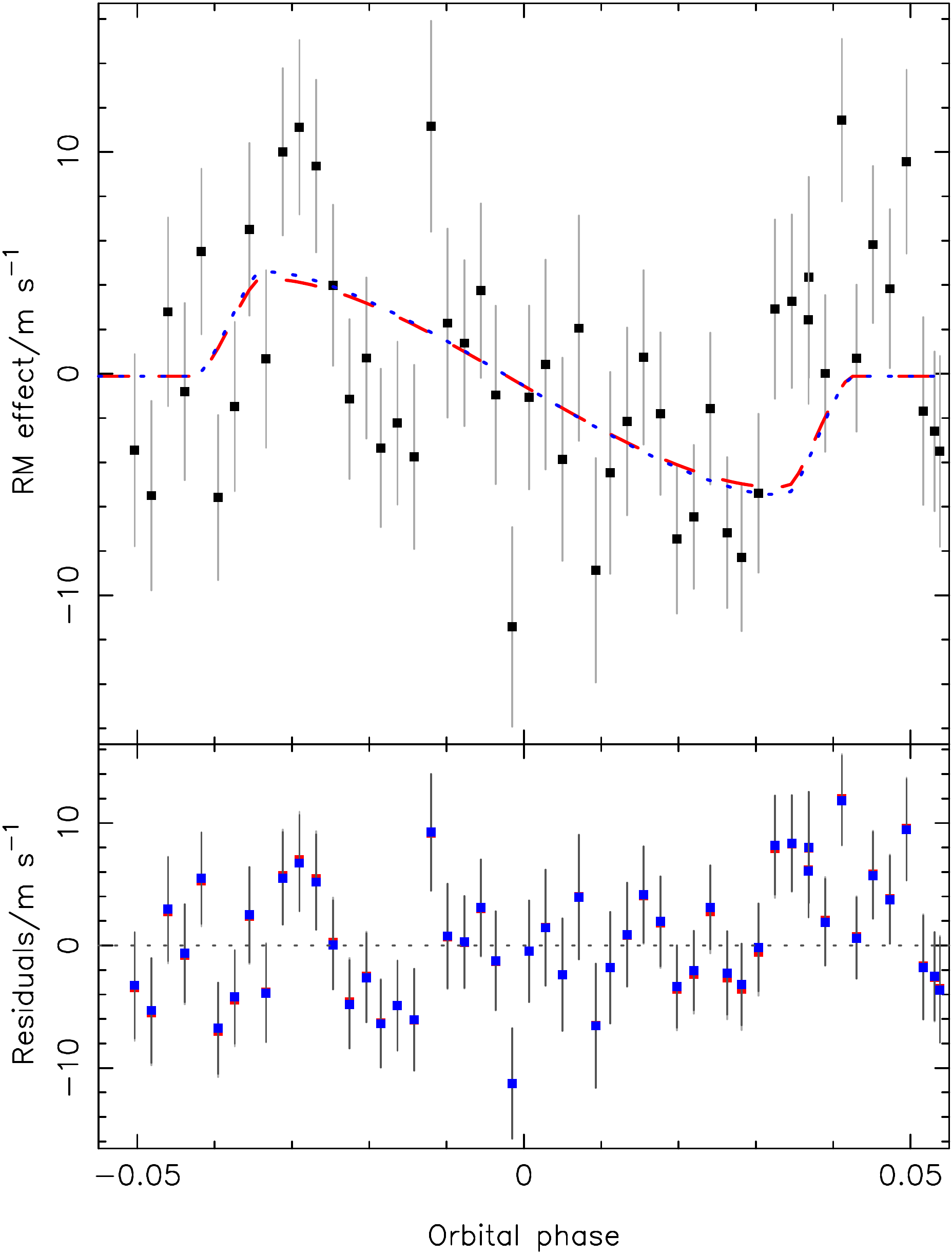}
	\caption{Upper panel: a close up of the RM anomaly in the RV curve of WASP-76, after correcting for a correlation between FWHM and RV residuals, with the contribution from the Keplerian orbit subtracted to better display the form of the anomaly. The two model fits are barely distinguishable. No prior on the impact parameter is applied. Lower panel: the residuals for the two model fits. Legends for the two panels as for Fig.\,\ref{fig:W61_RV}.}
	\label{fig:W76_RV}
\end{figure}

\subsubsection{Bou{\'e} model}
We investigate the system using the Bou{\'e} model, following the same methodology outline for the Hirano model. We again adopt a constraint of zero drift in the barycentric velocity owing to lack of evidence to the contrary. Applying the stellar radius constraint led to increases in $M_{\rm s}$, $R_{\rm s}$, $\rho_{\rm s}$, $M_{\rm p}$, $R_{\rm p}$, and $T_{\rm eff}$, but in several cases these parameters were unphysical. There was also no substantial improvement in fit when applying the radius constraint, and we therefore elected not to do so. We also adopted a circular solution as there was no evidence for a significantly eccentric orbit. In this we match our choice of constraints for the Hirano model, which is encouraging as it again shows that the two models are broadly consistent in their exploration of parameter space.

Initial tests without a prior on $v\sin I_{\rm s}$ also produced results consistent with those found using the Hirano model, including the crescent-shaped degeneracy between $v\sin I_{\rm s}$ and $\lambda$ with a preference for positive $\lambda$, though the stellar rotation velocity was found to be even slower at $0.4^{+0.5}_{-0.1}$\,km\,s$^{-1}$. When we applied the prior, we found that although convergence statistics were greatly improved, and the rotation velocity now agreed with our expectations from spectral analysis, the impact parameter was significantly lower than expected, and a bimodal distribution in $\lambda$ was obtained. When we forced $\lambda_0=\pm90^\circ$, the convergence statistics were again improved and the chains explored the expected minimum, but like the Hirano model tests the impact parameter remained lower than anticipated.

Applying a prior on the impact parameter, in the absence of the $v\sin I_{\rm s}$ prior, showed that the chains favoured the positive minimum, but with sizeable uncertainty on $\lambda$ and a slow rotation velocity. When we apply priors on both impact parameter \textit{and} $v\sin I_{\rm s}$, we found results consistent with the Hirano model equivalents.

\begin{table*}
	\caption{A summary of the results obtained during our investigation of the WASP-76 system. We explored different combinations of Gaussian priors on $v\sin I_{\rm s}$ and $b$, while allowing the stellar parameters to float freely and fixing $e=0$ and $\dot{\gamma}=0$. We also explored the effect of varying $\lambda_0$ between different local minima. We found that the application of a prior on $b$ leads to a positive solution for $\lambda$, irrespective of the value of $\lambda_0$. Behaviour in the presence of a prior on $v\sin I_{\rm s}$ varies with the choice of model, but does force the chains to limit themselves to a single minimum in $\lambda$ parameter space.}
	\label{tab:W76models}
	\begin{tabular}{lllrlll}
		\hline \\ [2pt]
		Model		& $v\sin I_{\rm s}$	& $b$	& $\lambda_0$				& $v\sin I_{\rm s}$				& $\lambda$				& $b$				\\ [2pt]
					& prior?			& prior?	& /$^\circ$					& /km\,s$^{-1}$				& /$^\circ$					& /$R_{\rm s}$			\\ [2pt]
		\hline \\ [2pt]
		Hirano		& No				& No		&						& $0.7^{+0.7}_{-0.2}$		& $37.6^{+31.4}_{-52.5}$		& $0.11^{+0.11}_{-0.08}$	\\ [2pt]
					& Yes			& No		& 						& $2.2\pm0.4$				& $73.4^{+6.2}_{-151.2}$ 		& $0.03^{+0.04}_{-0.02}$ \\ [2pt]
					& Yes			& No		& $\,90$					& $2.0\pm0.4$				& $76.5^{+3.8}_{-5.1}$		& $0.09^{+0.04}_{-0.06}$ \\ [2pt]
					& Yes			& No		& $-90$					& $2.2\pm0.4$				& $-76.8^{+4.5}_{-3.5}$		& $0.02^{+0.02}_{-0.01}$ \\ [2pt]
					
					& No				& Yes	&						& $0.7^{+0.5}_{-0.2}$		& $41.1^{+25.0}_{-50.1}$		& $0.130\pm0.003$	\\ [2pt]
					& No				& Yes	& $\,90$					& $0.7^{+0.5}_{-0.1}$		& $42.1^{+24.0}_{-49.9}$		& $0.13^{+0.01}_{-0.01}$ \\ [2pt]
					& No				& Yes	& $-90$					& $0.7^{+0.5}_{-0.2}$		& $41.8^{+24.7}_{-50.4}$		& $0.130^{+0.003}_{-0.004}$ \\ [2pt]
					
					& Yes			& Yes	&						& $1.9\pm0.3$				& $74.8^{+4.0}_{-5.2}$ 		& $0.130^{+0.001}_{-0.003}$ \\ [2pt]
					& Yes			& Yes	& $\,90$					& $1.9\pm0.3$				& $74.8^{+4.2}_{-5.5}$		& $0.130^{+0.001}_{-0.002}$ \\ [2pt]
					& Yes			& Yes	& $-90$					& $2.1\pm0.4$				& $-76.9^{+4.5}_{-3.3}$		& $0.02^{+0.02}_{-0.01}$ \\ [2pt]
		\hline \\
		Bou{\'e} 		& No				& No		& 						& $0.4^{+0.5}_{-0.1}$		& $34.7^{+36.8}_{-64.9}$		& $0.10^{+0.10}_{-0.08}$	\\ [2pt]
					& Yes			& No		&						& $2.2\pm0.4$				& $82.2^{+2.4}_{-164.8}$ 		& $0.02^{+0.02}_{-0.01}$ \\ [2pt]
					& Yes			& No		& $\,90$					& $2.2\pm0.4$				& $82.9^{+1.9}_{-2.6}$		& $0.03\pm0.02$		\\ [2pt]
					& Yes			& No		& $-90$					& $2.2\pm0.4$				& $-82.9^{+2.2}_{-1.7}$		& $0.01\pm0.01$		\\ [2pt]
		
					& No				& Yes	&						& $0.4^{+0.2}_{-0.1}$		& $37.1^{+27.7}_{-52.9}$		& $0.13^{+0.01}_{-0.01}$	\\ [2pt]
					& No				& Yes	& $\,90$					& $0.4^{+0.3}_{-0.1}$		& $34.6^{+29.3}_{-52.9}$		& $0.13^{+0.01}_{-0.01}$ \\ [2pt]
					& No				& Yes	& $-90$					& $0.4^{+0.3}_{-0.1}$		& $36.8^{+27.8}_{-51.9}$		& $0.13\pm0.004$ \\ [2pt]

					& Yes			& Yes	&						& $2.0^{+0.3}_{-0.4}$		& $70.0^{+7.0}_{-11.6}$ 		& $0.13\pm0.01$		 \\ [2pt]
					& Yes			& Yes	& $\,90$					& $2.0\pm0.3$				& $70.2^{+6.6}_{-11.7}$		& $0.13\pm0.003$		\\ [2pt]
					& Yes			& Yes	& $-90$					& $2.1\pm0.3$				& $69.6^{+6.8}_{-10.6}$		& $0.13\pm0.01$		\\ [2pt]
		\hline \\ 
		Tomography	& No				& No		&						& $1.2^{+0.9}_{-0.6}$ 		& $69.3^{+10.2}_{-26.7}$ 		& $0.11^{+0.09}_{-0.06}$ \\ [2pt]
					& Yes			& No		&						& $2.1\pm0.3$				& $-77.4^{+3.6}_{-2.9}$		& $0.01\pm0.01$		\\ [2pt]
					& Yes			& No		& $\,90$					& $2.2\pm0.3$				& $77.3^{+3.7}_{-2.9}$		& $0.01\pm0.01$		\\ [2pt]
					& Yes			& No		& $-90$					& $2.1\pm0.3$				& $-77.2^{+4.0}_{-3.0}$		& $0.01\pm0.01$		\\ [2pt]

					& No				& Yes	&						& $1.1^{+0.5}_{-0.4}$		& $64.6^{+10.0}_{-23.6}$		& $0.13^{+0.01}_{-0.01}$	\\ [2pt]
					& No				& Yes	& $\,90$					& $1.1^{+0.5}_{-0.4}$ 		& $64.1^{+10.0}_{-25.2}$		& $0.130\pm0.003$		\\ [2pt]
					& No				& Yes	& $-90$					& $1.1\pm0.5$	 			& $66.4^{+9.1}_{-21.1}$		& $0.13\pm0.01$		\\ [2pt]

					& Yes			& Yes	&						& $1.9\pm0.3$				& $76.5^{+3.4}_{-4.4}$		& $0.130\pm0.002$		\\ [2pt]
					& Yes			& Yes	& $\,90$					& $1.9\pm0.3$				& $77.7^{+3.1}_{-3.9}$		& $0.130\pm0.003$		\\ [2pt]
					& Yes			& Yes	& $-90$					& $1.9\pm0.3$				& $76.4^{+3.4}_{-4.4}$		& $0.129\pm0.003$		\\ [2pt]
		\hline \\ [2pt]
	\end{tabular}
\end{table*}

\subsubsection{Doppler tomography}
We adopted constraints of $e=0$ and $\dot{\gamma}=0$, but left the stellar mass and radius freely varying, in order to be consistent with our analyses using the Hirano and Bou{\'e} models.

\begin{figure*}
	\centering
	\subfloat{
		\centering
		\includegraphics[width=0.48\textwidth]{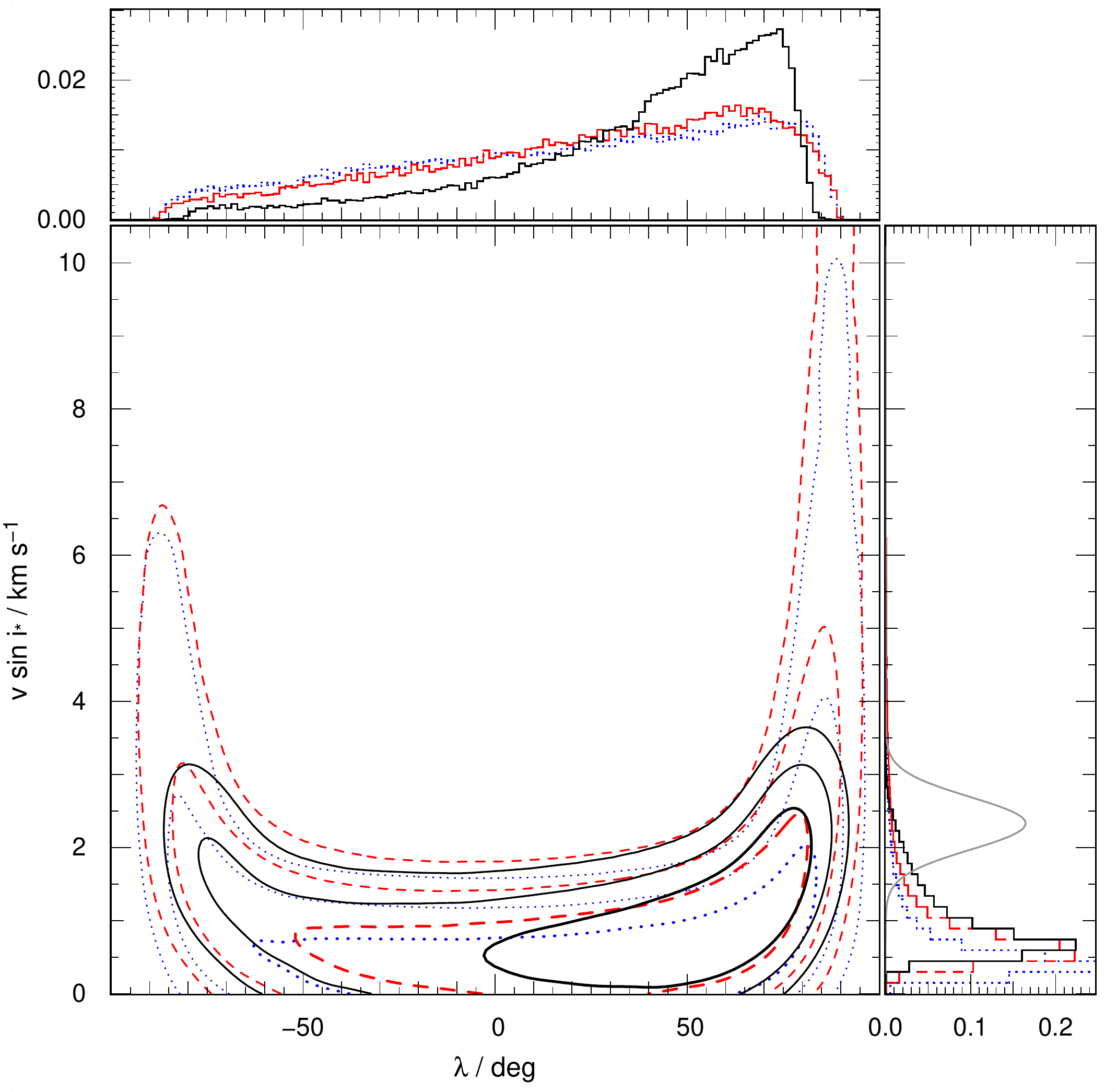}}
	\subfloat{
		\centering
		\includegraphics[width=0.48\textwidth]{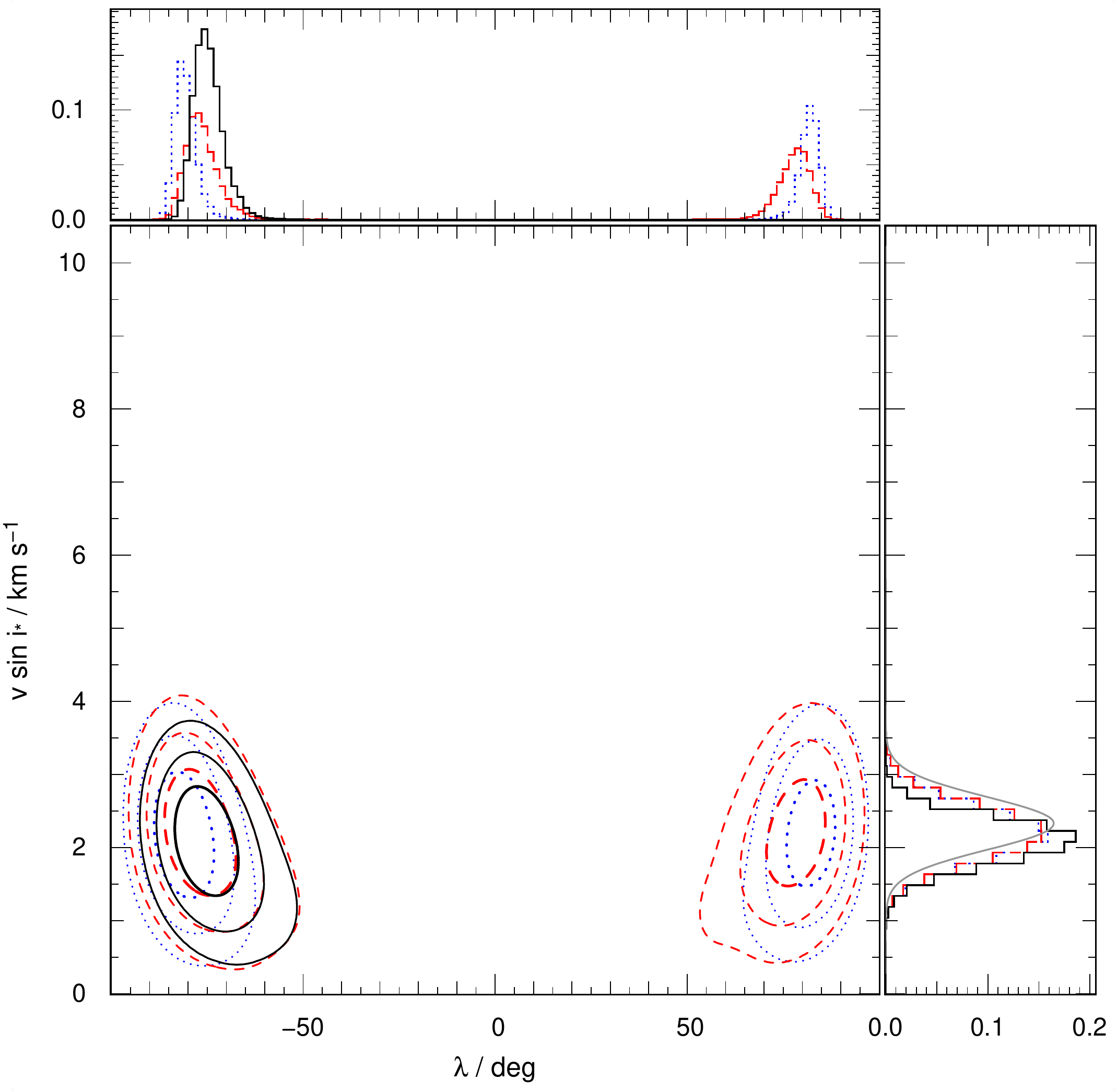}}
	\caption{The posterior probability distributions in $v\sin I-\lambda$ parameter space for our analyses of WASP-76. Legend as for Fig.\,\ref{fig:W61_vsinI-lambda}. No prior on impact parameter is applied. Left: results without a prior on $v\sin I_{\rm s}$. DT fails to break the degeneracy that arises between $v\sin I_{\rm s}$ and $\lambda$ as a result of the low impact parameter of the transit chord, although the length of the large $v\sin I_{\rm s}$, large $|\lambda|$ tails is strongly reduced. Right: results with a spectral analysis applied to $v\sin I_{\rm s}$. As with WASP-61, the degeneracy between solutions with ($i_{\rm orb}$, $\lambda$) and ($i_{\rm orb}-{\rm\pi}$, $-\lambda$) leads to a bimodal distribution when using the Hirano and Bou{\'e} models. Unlike that system however, the bimodality is on a chain-by-chain basis; each individual chain concatenated in our final MCMC chain is well converged on to either the positive or negative $\lambda$ solution. DT succeeds in breaking this degeneracy, with all chains produced using that model converging at the negative $\lambda$, retrograde solution. However, there is little reduction in the size of the distribution compared to the Bou{\'e} and Hirano models.}
	\label{fig:W76_vsinI-lambda}
\end{figure*}

As anticipated, using the DT method substantially reduced the degeneracy between $v\sin I_{\rm s}$ and $\lambda$, though it did not, for this system, remove it completely (see Fig.\,\ref{fig:W76_vsinI-lambda}). DT again favoured the positive minimum, though more strongly than the other two models, and again returned a more slowly rotating star than anticipated. Adding a prior on $v\sin I_{\rm s}$, however, produced different behaviour than shown previously. With DT, adding a prior on $v\sin I_{\rm s}$ forced the chains into the \textit{negative} minimum, with no bimodal distribution observed, though once again the impact parameter strongly disagreed with the value from the discovery paper and implied a central transit.

If we impose a prior on the impact parameter using the value from the discovery paper, then in the absence of a prior on $v\sin I_{\rm s}$ we again find that the chains favour the positive minimum in $\lambda$, irrespective of the value of $\lambda_0$. We also find a faster value of $v\sin I_{\rm s}$ than was returned by either the Hirano or Bou{\'e} models for the same combination of priors, though the values are consistent to $1\sigma$. If we add the prior on $v\sin I_{\rm s}$ then the results remain consistent, but the $1\sigma$ uncertainties are reduced in magnitude, particularly for $\lambda$ which also moves closer towards a value that implies a polar orbit.

\begin{figure}
		\includegraphics[width=0.48\textwidth]{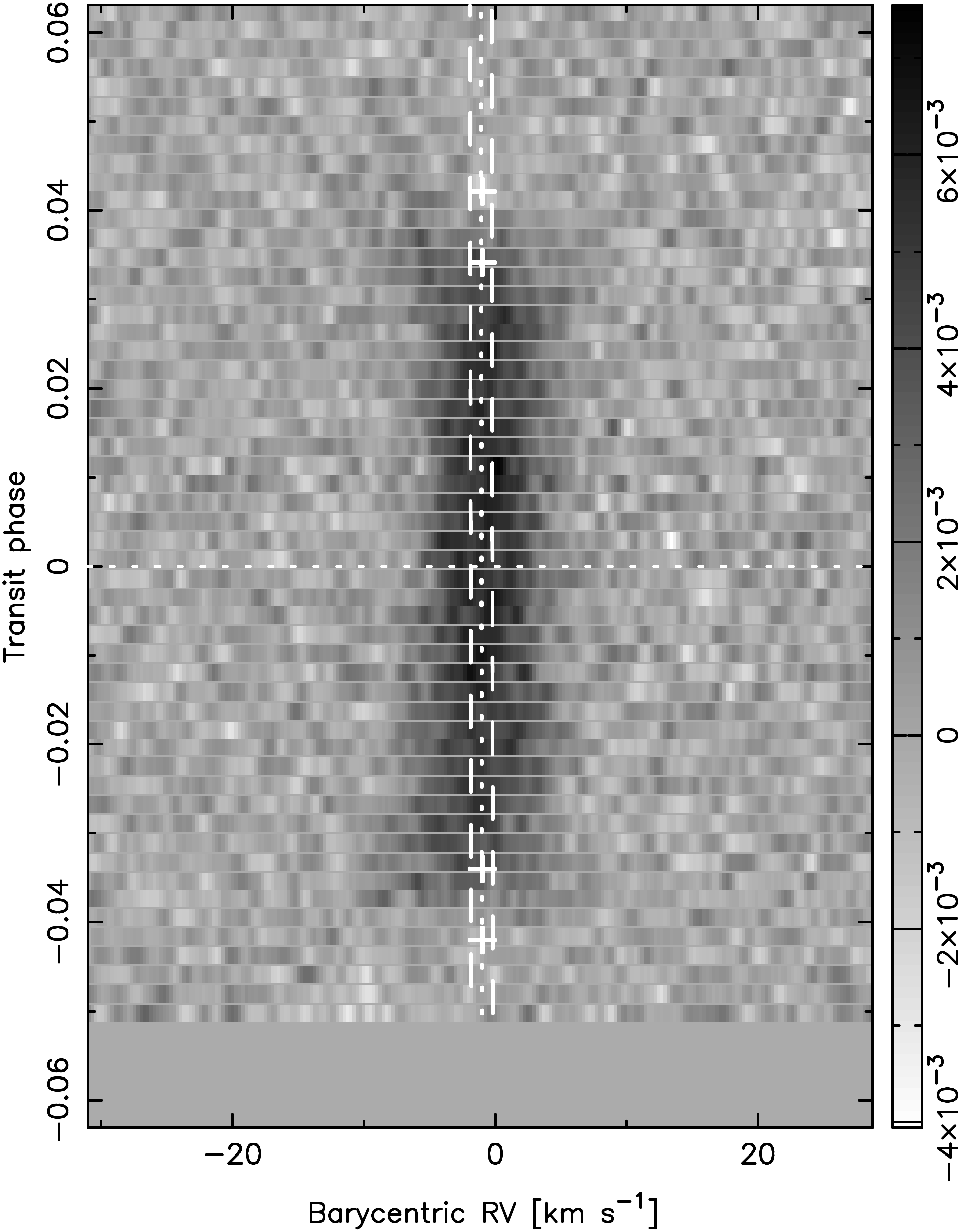}
	\caption{Map of WASP-76 time series CCFs with the model stellar spectrum subtracted, for the case with no application of a $v\sin I_{\rm s}$ prior from spectral analysis. The trajectory of the planet signature is difficult to determine owing to the slow rotation of the host star. The planetary signal appears to bleed outside the area of the plot denoting the stellar boundaries, perhaps indicating that this method is underestimating $v\sin I_{\rm s}$. Legend as Fig.\,\ref{fig:W61_DT}.}
	\label{fig:W76_DT}
\end{figure}	

\subsubsection{A possible polar orbit?}
\label{sec:W76_vsinI}
\begin{figure}
	\centering
	\includegraphics[width=0.48\textwidth]{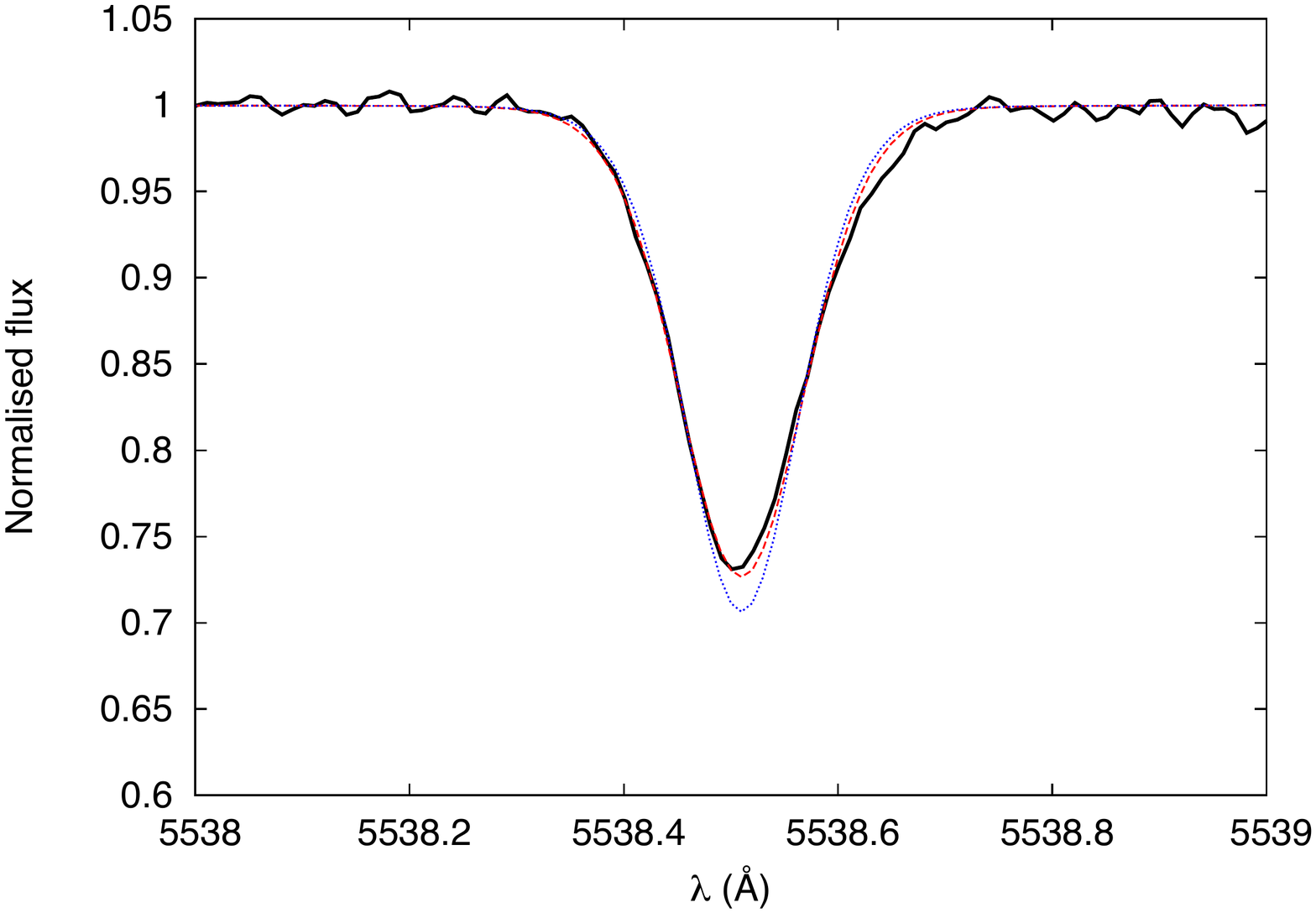}
	\caption{The line profile of the Fe\,\textsc{I} line at $5538.517$\,\AA of WASP-76\,A, as observed with HARPS (black, solid line). Overlaid are models for two different values of $v\sin I_{\rm s}$: $2.33$\,km\,s$^{-1}$, the spectroscopic value (red, dashed line), and $1.0$\,km\,s$^{-1}$, on the order of the results returned by the RM models (blue, dotted line). While the red model fits the observed line profile well, the blue model is both deeper and more narrow.This indicates that the host star is rotating more quickly than the RM models are able to account for if the planet is on a non-polar orbit.}
	\label{fig:W76rotation}
\end{figure}

Consecutive analyses of WASP-76 have gradually reduced our assessment of the stellar rotation velocity. Spectral analysis of the CORALIE data by \citet{2013arXiv1310.5607W} gave $v\sin I_{\rm s}=3.3\pm0.6$\,km\,s$^{-1}$, while analysis of our new HARPS spectra returned a value of $v\sin I_{\rm s}=2.33\pm0.36$\,km\,s$^{-1}$. In the absence of a prior on rotation, our modelling of the RM effect using any of the three methods gives a value significantly slower than this, at $\sim1$\,km\,s$^{-1}$. Yet inspection of both the CORALIE and HARPS spectra reveals visible rotation (see Fig.\,\ref{fig:W76rotation}), and the full width at half-maximum (FWHM) of the spectra are greater than for stars with similar ($B--V$) colour, such as WASP-20. This would suggest that the star is indeed oriented close to edge-on, rather than the pole-on orientation suggested by the slow rotation velocity returned by our MCMC chains.

We therefore consider the possibility that the orbit is oriented at close to $|\lambda|=90^\circ$, with a transit chord such that the path of the planet is almost parallel to the stellar rotation axis. This solution is consistent with the path of the planetary `bump' through the stellar line profile in Fig.\,\ref{fig:W76_DT}, which shows little movement in velocity space. To explore this possible system configuration, we carried out additional analyses both with and without a prior on $v\sin I_{\rm s}$, this time forcing the MCMC chain to adopt $\lambda=\pm90^{\circ}$ throughout. Note that these analyses used the constraints of $e=0$ and $\dot{\gamma}=0$, and applied no constraints on the stellar mass or radius, as before. We present these results in Table\,\ref{tab:W76polar} and Fig.\,\ref{fig:W76_RV}.

\begin{table*}
	\caption{A summary of the results obtained while investigating a potential polar orbit for WASP-76. We fixed $\lambda=\pm90^\circ$, and explored different combinations of Gaussian priors on $v\sin I_{\rm s}$ and $b$ while allowing the stellar parameters to float freely, and fixing $e=0$ and $\dot{\gamma}=0$. }
	\label{tab:W76polar}
	\begin{tabular}{lllrlll}
		\hline \\ [2pt]
		Model		& $v\sin I_{\rm s}$	& $b$	& $\lambda_0$				& $v\sin I_{\rm s}$				& $\lambda$				& $b$				\\ [2pt]
					& prior?			& prior?	& /$^\circ$					& /km\,s$^{-1}$				& /$^\circ$					& /$R_{\rm s}$			\\ [2pt]
		\hline \\ [2pt]
		Hirano		& Yes			& No		& $90$					& $2.2\pm0.4$				& $90$ (fixed)				& $0.06^{+0.04}_{-0.03}$ \\ [2pt]
					& 				& 		& $-90$					& $2.2\pm0.4$				& $-90$ (fixed)				& $0.01^{+0.02}_{-0.01}$ \\ [2pt]
					& 				& Yes	& $90$					& $2.0\pm0.4$				& $90$ (fixed)				& $0.13\pm0.02$ 		\\ [2pt]
					& 				& 		& $-90$					& $1.6^{+0.6}_{-0.4}$		& $-90$ (fixed)				& $0.13^{+0.003}_{-0.10}$ \\ [2pt]  
					& No				& No		& $90$					& $1.0^{+4.6}_{-0.9}$		& $90$ (fixed)				& $0.11^{+0.12}_{-0.09}$ \\ [2pt]
			 		&				& 		& $-90$					& $0.2^{+0.9}_{-0.2}$		& $-90$ (fixed)				& $0.05^{+0.11}_{-0.04}$ \\ [2pt]
			 		& 				& Yes	& $90$					& $0.8^{+0.7}_{-0.6}$		& $90$ (fixed)				& $0.13\pm0.02$ 		\\ [2pt]
			 		& 				& 		& $-90$					& $0.1^{+0.3}_{-0.1}$		& $-90$ (fixed)				& $0.13\pm0.02$ 		\\ [2pt]
		\hline \\ [2pt]
		Bou{\'e} 		& Yes			& No		& $90$					& $2.2\pm0.4$				& $90$ (fixed)				& $0.03\pm0.02$		\\ [2pt]
			 		& 				& 		& $-90$					& $2.2\pm0.4$				& $-90$ (fixed)				& $0.01\pm0.01$		\\ [2pt]
					& 				& Yes	& $90$					& $2.1^{+0.3}_{-0.4}$		& $90$ (fixed)				& $0.130\pm0.001$		\\ [2pt]
			 		& 				& 		& $-90$					& $1.8\pm0.3$				& $-90$ (fixed)				& $0.130\pm0.001$		\\ [2pt]
					& No				& No		& $90$					& $1.5^{+5.6}_{-1.3}$		& $90$ (fixed)				& $0.10^{+0.15}_{-0.08}$ \\ [2pt] 
					& 				& 		& $-90$					& $0.3^{+1.8}_{-0.3}$		& $-90$ (fixed)				& $0.07^{+0.10}_{-0.05}$ \\ [2pt] 
					& 				& Yes	& $90$					& $0.3^{+0.4}_{-0.3}$		& $90$ (fixed)				& $0.130\pm0.001$ 		\\ [2pt]
					& 				& 		& $-90$					& $0.1^{+0.2}_{-0.1}$		& $-90$ (fixed)				& $0.130\pm0.001$		\\ [2pt]
		\hline \\ [2pt]
		Tomography	& Yes			& No		& $90$					& $2.2\pm0.3$				& $90$ (fixed)				& $0.09\pm0.03$		\\ [2pt]
					& 				& 		& $-90$					& $2.3\pm0.4$				& $-90$ (fixed)				& $0.01\pm0.01$		\\ [2pt]
					& 				& Yes	& $90$					& $2.2\pm0.3$				& $90$ (fixed)				& $0.130\pm0.002$		\\ [2pt]
					& 				& 		& $-90$					& $1.8\pm0.3$				& $-90$ (fixed)				& $0.13\pm0.01$		\\ [2pt]
					& No				& No		& $90$					& $1.6\pm0.8$ 				& $90$ (fixed)				& $0.12^{+0.09}_{-0.05}$	\\ [2pt]
					& 				& 		& $-90$					& $0.1^{+1.0}_{-0.1}$ 		& $-90$ (fixed)				& $0.05^{+0.11}_{-0.04}$	\\ [2pt]
					& 				& Yes	& $90$					& $1.3^{+0.6}_{-0.7}$ 		& $90$ (fixed)				& $0.130\pm0.001$		\\ [2pt]
					& 				& 		& $-90$					& $0.01\pm0.01$	 		& $-90$ (fixed)				& $0.130\pm0.002$		\\ [2pt]
		\hline \\ [2pt]
	\end{tabular}
\end{table*}

We found that when applying the Bou{\'e} model, the MCMC chains took approximately twice as long to converge as when applying the Hirano model. The source of this difficulty with convergence is uncertain, but seems to be related to the ratio between $\sigma_{\rm Boue}$ and $v\sin I_{\rm s}$. Large steps in the latter that explore rapidly rotating solutions lead to unphysical values of this ratio, such that the step fails. This restricts the set of possible solutions to a more limited area of parameter space, such that a larger percentage of possible steps lead to poor solutions, and thus convergence of the chain proceeds more slowly.

\begin{figure}
	\centering
	\includegraphics[width=0.48\textwidth]{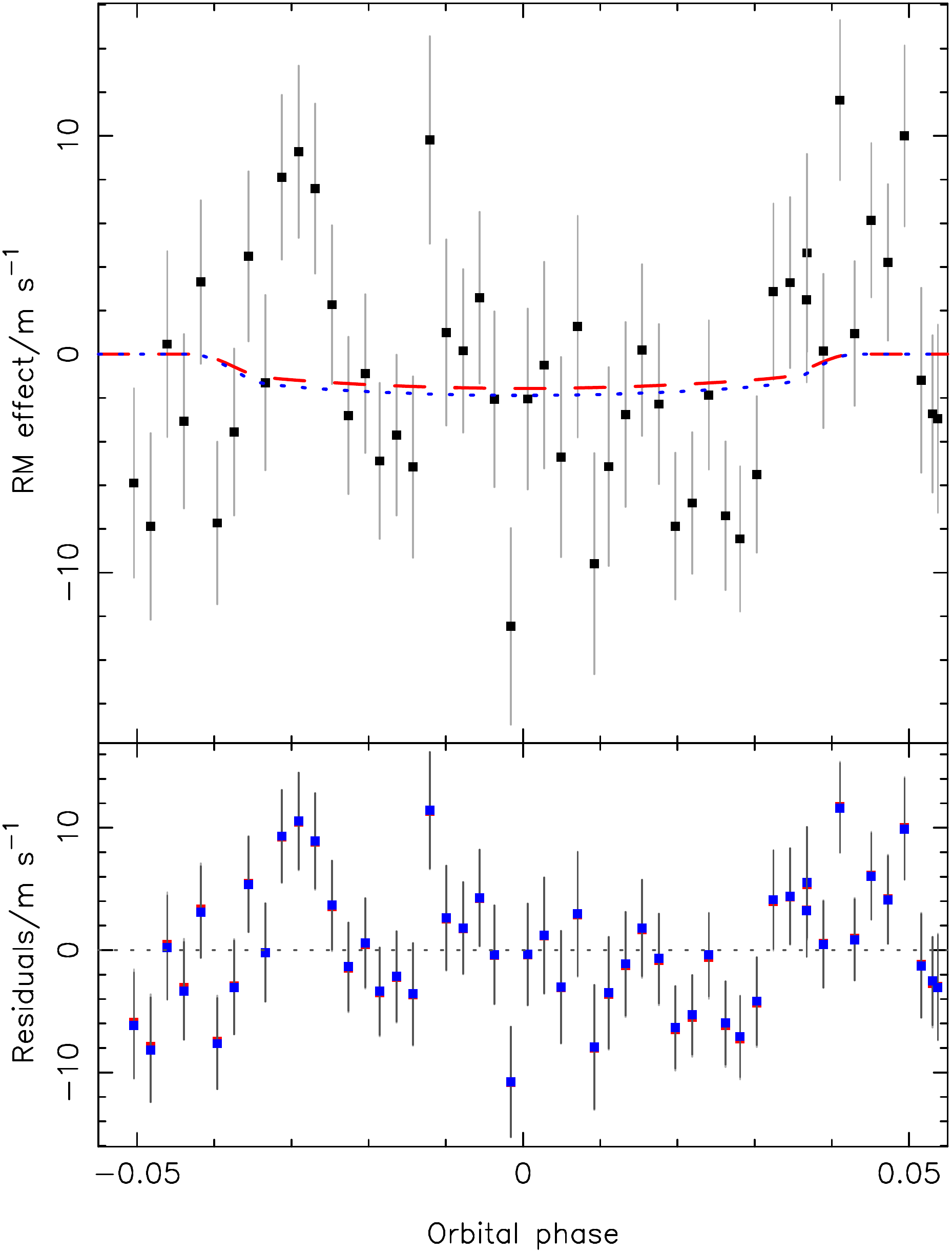}
	\caption{As for Fig.\,\ref{fig:W76_RV} for the RM effect of WASP-76 in the case where an orbit parallel to the stellar rotation axis, $\lambda=90^{\circ}$, was forced.}
	\label{fig:W76_lam90}
\end{figure}

The three analysis techniques generally produced consistent results. In the majority of cases, we found that the value for $b$ returned by the chains was in agreement with the discovery paper; the exceptions to this were the cases with $\lambda_0=-90^\circ$ and the $v\sin I_{\rm s}$ prior only. With the rotation prior \textit{inactive}, the only case to be consistently in agreement with predictions across all three techniques was the case with $b$ prior also inactive, and $lambda_0=90^\circ$; in the other cases, the stellar rotation was generally slower than the spectral analysis result (as noted in previous sections).

These results lend some small support to the hypothesis of a polar orbit, as we note that in the case of neither prior being applied the results were consistent with both spectral analysis and the discovery paper.

\subsubsection{A poorly constrained system?}
WASP-76 seems to represent a similar case to WASP-1 \citep{2011ApJ...738...50A}: a low impact parameter, combined with a poor SNR for the RM effect, leading to a weak detection. Here though, we find a three-way degeneracy between $\lambda$, $v\sin I_{\rm s}$, and $b$ that can only be broken through the application of appropriate Gaussian priors.

Our results show tentative support for a strongly misaligned orbit. Applying a prior on stellar rotation or on the impact parameter produces results that suggest strong misalignment, particularly when using the DT method. Forcing the system to adopt a polar orbit, or using a polar orbit as the initial condition, reveals that this is a plausible option for the system's configuration, though there is still substantial ambiguity in the precise orientation of the planet's orbit.

Although we have reported and discussed results for cases both with and without the various combinations of these two priors, in Section\,\ref{sec:discussion} we focus on the case \textit{with} a prior on the impact parameter, but \textit{without} a prior on $v\sin I_{\rm s}$. This combination maximizes the relevance of our cross-system comparison by allowing the different models to evaluate stellar rotation freely, while ensuring that the other system parameters are truly representative and derived from fully converged chains; for the WASP-76 system this necessitates the prior on $b$. We caution readers, however, that we cannot constrain the obliquity of the planet's orbit beyond the general statement that it is likely strongly misaligned in the positive $\lambda$, prograde direction.

\subsection{WASP-78}
\label{sec:W78}
WASP-78\,b \citep{2012AA...547A..61S} is large-radius, low-density hot Jupiter that orbits its host star every $2.2$\,d in a circular orbit. From the CORALIE RV data in the discovery paper, the orbit appears to be circular, and such a solution is adopted therein, but the authors state that additional RV data is required to pin-down the eccentricity. Spectral analysis shows that the host star is of spectral type F8, with $T_{\rm eff}=6100\pm150$\,K, $v_{\rm mac}=3.5\pm0.3$\,km\,s$^{-1}$, and $v\sin I_{\rm s}=7.2\pm0.8$\,km\,s$^{-1}$.

\begin{figure}
	\centering
	\includegraphics[width=0.48\textwidth]{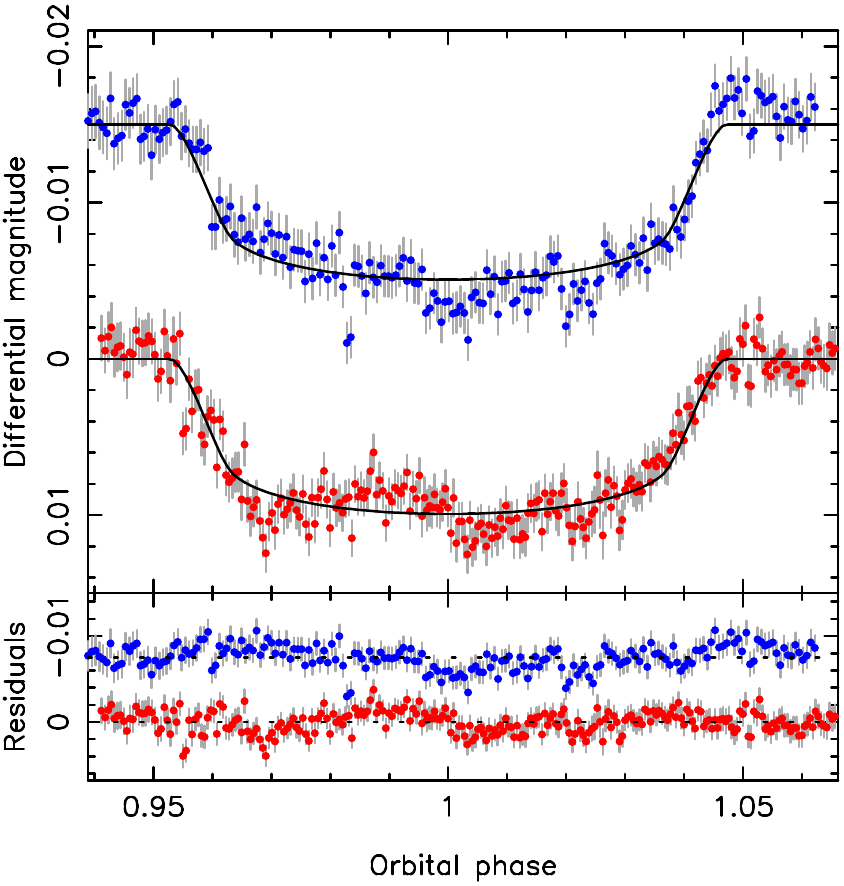}
	\caption{Upper panel: newly acquired EulerCam observations of two separate transits of WASP-78\,b, with best-fitting MCMC transit models. The two light curves have been offset by $0.015$\,mag for clarity. The lower, red data are the $R$-band light curve of the transit on 2012 November 26. The upper, blue data are the $I$-band light curve of the transit on 2012 November 2. The signature of stellar activity is detected in the $R$-band, with structured variation in the light curve. There is also some evidence for activity in the $R$-band light curve, but this is less conclusive. Lower panel: residuals of the fits of the data to the best-fitting transit models. The stellar activity signature from the $R$-band light curve is more clear in the residuals.}
	\label{fig:W78_phot}
\end{figure}

We obtained new photometry using EulerCam, in both the $I$ and $R$ bands, of transits on 2012 November 2 and 26, respectively, which we combine with the discovery photometry from WASPSouth and TRAPPIST. The $R$-band light curve shows clear modulation, almost certainly as a result of the presence of star-spots or related stellar activity features, as this modulation is not replicated in the $I$-band light curve (Fig.\,\ref{fig:W78_phot}). Our HARPS spectroscopic transit observations were made simultaneously with the $I$-band EulerCam observations, and used in conjunction with spectroscopy from CORALIE. Analysis of the HARPS spectra provides $v_{\rm mac}=4.85$\,km\,s$^{-1}$, and $v\sin I_{\rm s}=6.63\pm0.16$\,km\,s$^{-1}$.

\begin{figure}
	\centering
	\includegraphics[width=0.48\textwidth]{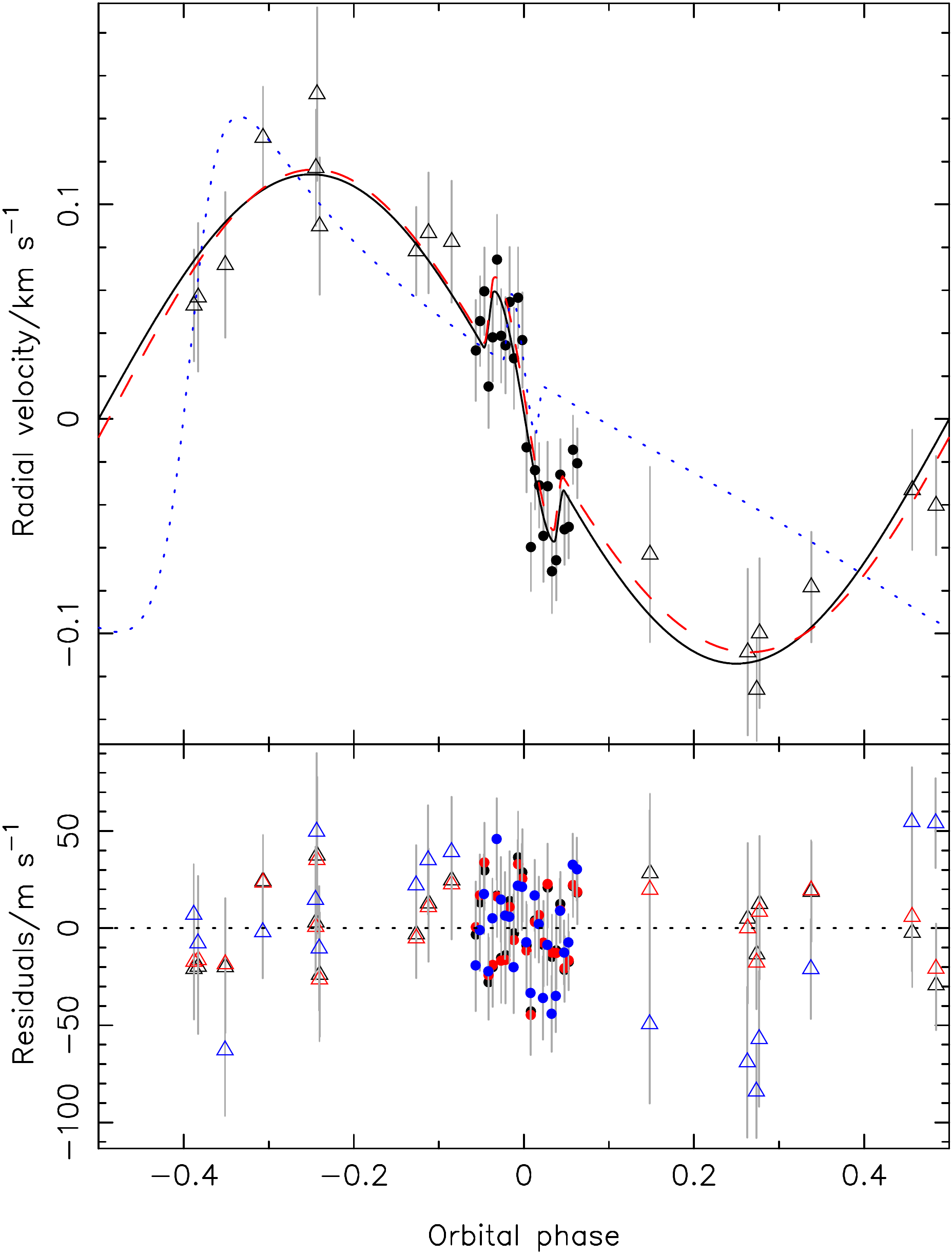}
	\caption{Upper panel: the RV curve of WASP-78, phase folded using the best-fitting ephemeris. HARPS data are denoted by solid circles, CORALIE data by open triangles. The best-fitting barycentric velocity for each set of data has been subtracted. Overplotted are models for a circular solution (solid, black line), a model with $e=0.05$ (dashed, red line), and a model with $e=0.6$ (dotted, blue line).  We adopt the circular solution in our final results, as the high eccentricity solution is clearly incorrect, and there is insufficient evidence for a small eccentricity. Lower panel: the residuals for the two model fits. Data are colour-coded according to the model being fit.}
	\label{fig:W78_RV}
\end{figure}

\begin{figure}
	\centering
	\includegraphics[width=0.48\textwidth]{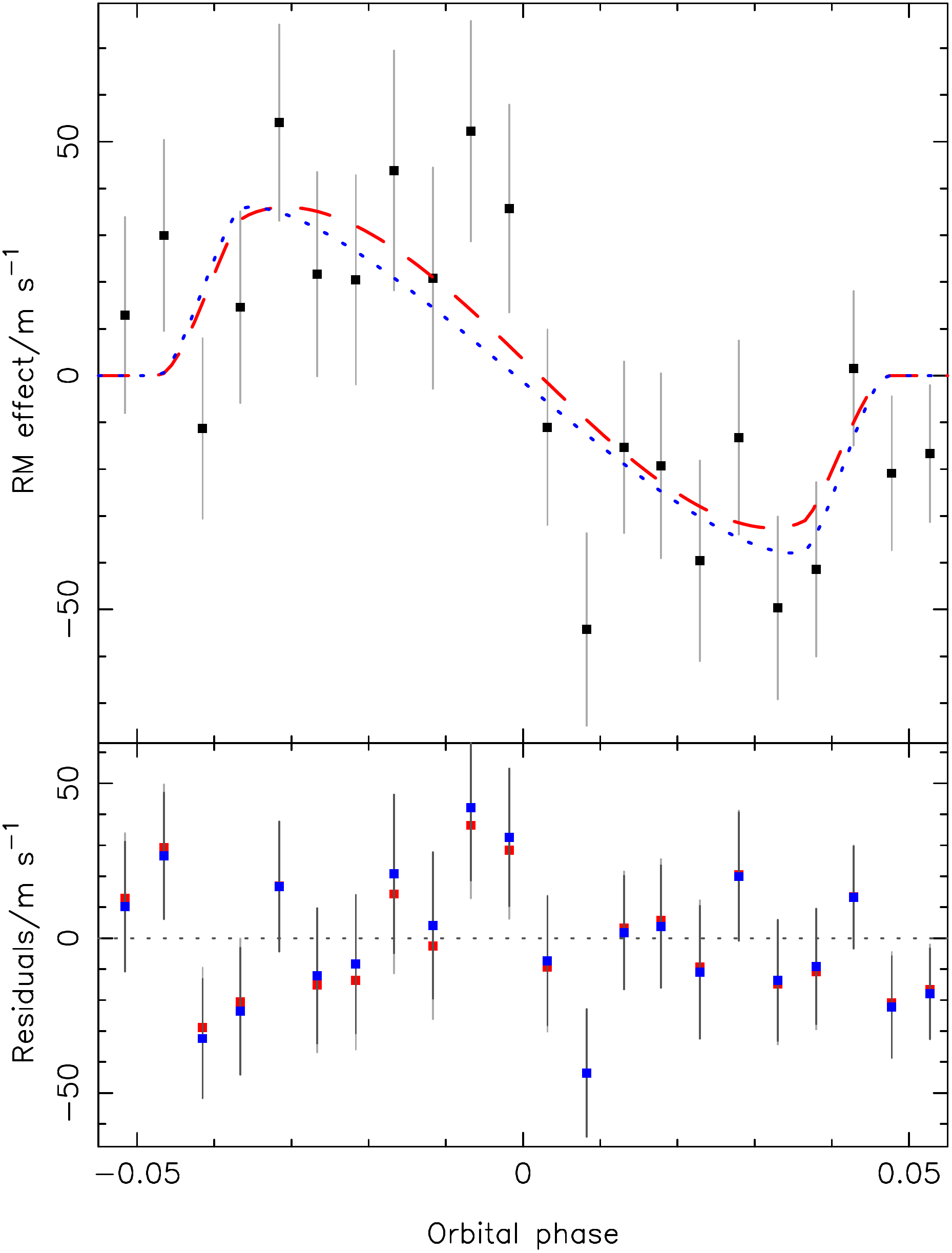}
	\caption{Upper panel: a close up of the RM anomaly in the RV curve of WASP-78, with the contribution from the Keplerian orbit subtracted to better display the form of the anomaly. Lower panel: the residuals for the two model fits. Legends for the two panels as for Fig.\,\ref{fig:W61_RV}.}
	\label{fig:W78_RM}
\end{figure}

\subsubsection{RM modelling}
\label{sec:w78_hirano}
As with the other systems in our sample, we find no evidence of a long-term barycentric velocity trend, and set $\dot{\gamma}=0$. We also find little difference in behaviour or results between the two models; the following discussion is applicable to both.

When allowing eccentricity to float we find that the MCMC algorithm returns two general sets of solutions, with the Hirano and Bou{\'e} models giving comparable estimates as expected. The imposition of the stellar radius constraint leads to $e\approx0.6$, but forces the stellar radius to be approximately half the value presented in \citet{2012AA...547A..61S}. Furthermore, examination of Fig.\,\ref{fig:W78_RV} shows that high eccentricity orbits are a very poor fit to the RV data. We do not consider these solutions plausible. When the stellar radius constraint is removed, the free eccentricity fit returns $e\approx0.05$. Testing these small eccentricity solutions using the method of \citet{1971AJ.....76..544L} reveals that they are not significant. Comparing eccentric and circular solutions \textit{with no other constraints applied}, we find that variations in physical parameters are within the $1\sigma$ uncertainties. We thus choose to force a circular orbit as {\rm it provides} the most plausible solution; the phase coverage of the HARPS data is insufficient to truly constrain any small eccentricity that might be present, and the pre-existing CORALIE RV data cannot provide a firm conclusion regarding circularity (or otherwise). We also note that hot Jupiters with $M_{\rm p}\approx1$\,M$_{\rm Jup}$ are generally observed to have zero (or small) eccentricity \citep{2012MNRAS.422.1988A}.

Application of the stellar radius constraint to a circular orbit increases the reported stellar mass, and also leads to an increased value of $T_{\rm eff}$ and a lower impact parameter. Our results in this scenario agree very well with those from \citet{2012AA...547A..61S}, as expected; the host star is both too massive and too large, and inconsistent with both spectral analysis of both the existing CORALIE spectra and our new HARPS data. Conversely, if the planet's orbit is allowed to be eccentric (against the evidence), then adding the stellar radius constraint decreases the mass of both bodies, and significantly decreases their radii.

We find a moderate impact parameter of $0.52\pm0.05$, with no correlation or degeneracy between $v\sin I_{\rm s}$ and $\lambda$, so do not apply a prior using the spectroscopic $v\sin I_{\rm s}$. Our chosen solution is therefore a circular orbit with no long-term velocity trend, and no application of the constraints on $v\sin I_{\rm s}$ or $R_{\rm s}$. The full set of results is shown in Table\,\ref{tab:results}, but the alignment angles found by the Hirano and Bou{\'e} models are consistent with each other, and with $\lambda=0$, implying a well-aligned orbit. We do however find that the Bou{\'e} model produces a slower rotation velocity, as we found for all of the previous systems, in this case inconsistent with the spectroscopic value. This cannot, however, be as a result of a poorly constrained impact parameters; the results in Table\,\ref{tab:results} show that the impact parameter returned by the Bou{\'e} model is in agreement with the results from other models, and has similar $1\sigma$ uncertainties.

The form of the best-fittingting models are very similar in Fig.\,\ref{fig:W78_RM}, as they were for WASP-76 (the other system with a low signal-to-noise anomaly), although in this case there are far fewer data during the transit.

Our selected solutions are taken from runs for which no prior was applied to $v\sin I_{\rm s}$, $e=0$, $\dot{\gamma}=0$, and no constraint was applied to the stellar radius.

\subsubsection{Doppler tomography}
\label{sec:w78_tomography}
\begin{figure*}
	\centering
	\subfloat{
		\centering
		\includegraphics[width=0.48\textwidth]{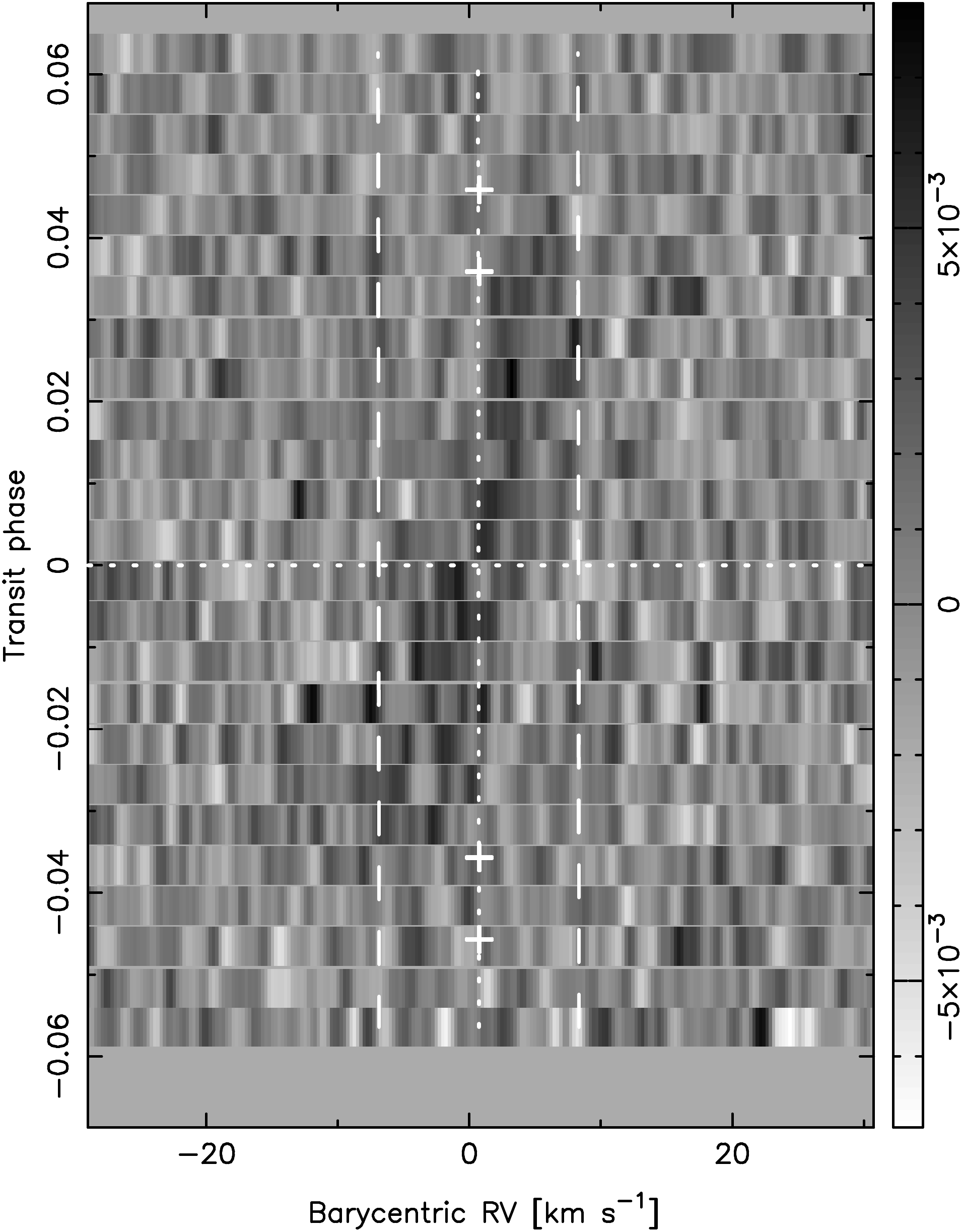}}
	\subfloat{
		\centering
		\includegraphics[width=0.48\textwidth]{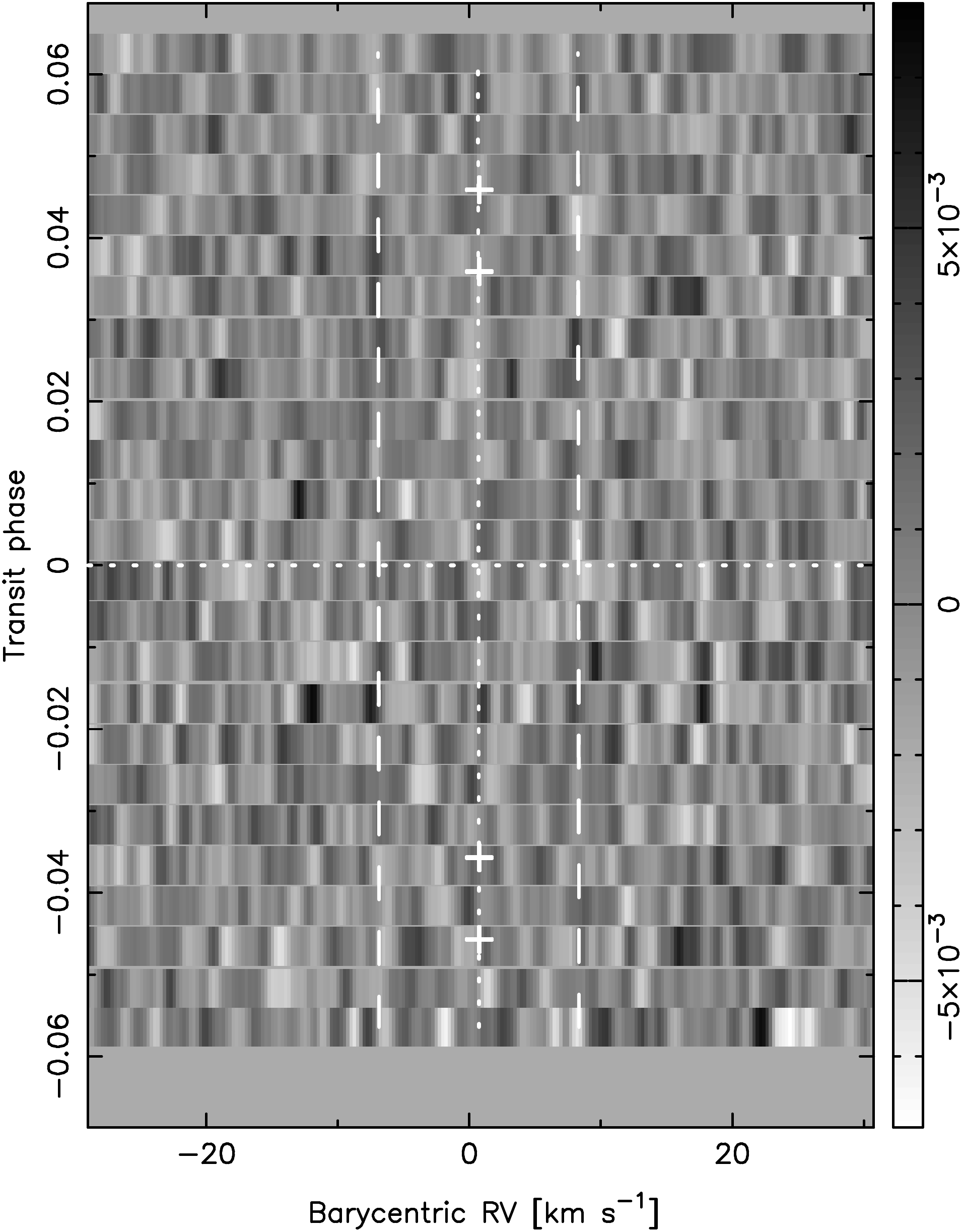}}
	\caption{Left: residual map of WASP-78 time series CCFs with the model stellar spectrum subtracted. The signature of the planet is relatively weak, reflecting the low signal to noise of the RM anomaly that can be seen in Fig.\,\ref{fig:W78_RM}. But it seems to moves from lower left to upper right, indicating a prograde orbit, and is apparently symmetrical, indicating alignment. Right: the best-fitting model for the time-variable planet feature has been subtracted, leaving the overall residual map. The lack of any strong features in this figure indicates a lack of large-scale stellar activity. Legend as Fig.\,\ref{fig:W61_DT}.}
	\label{fig:W78_DT}
\end{figure*}

\begin{figure}
	\centering
	\includegraphics[width=0.48\textwidth]{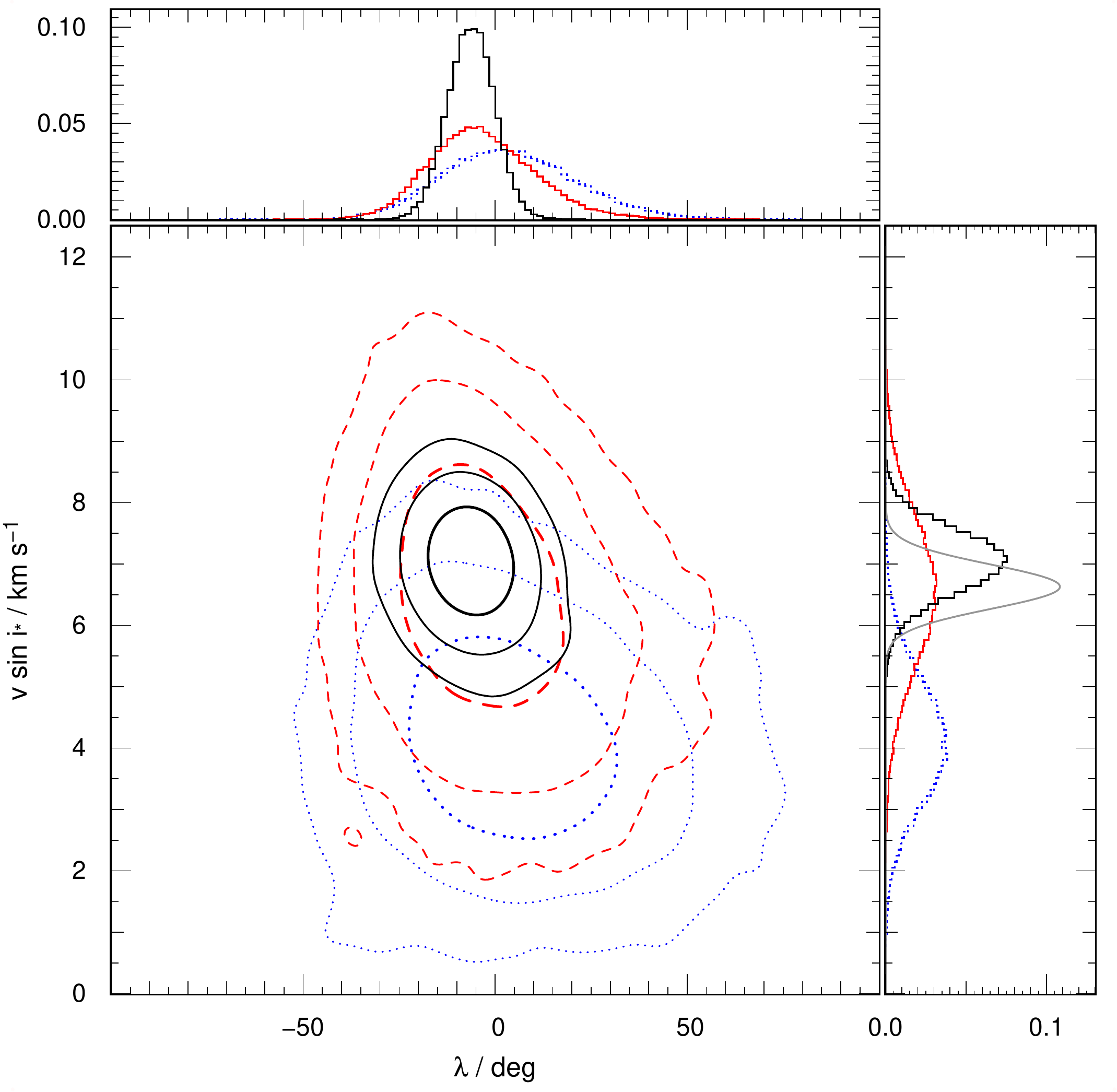}
	\caption{The posterior probability distributions in $v\sin I-\lambda$ parameter space for our analyses of WASP-78. Legend as for Fig.\,\ref{fig:W61_vsinI-lambda}. The Hirano and DT results agree well, but the latter gives a significant improvement in the precision of both quantities. The Bou{\'e} model also has a broader distribution in $v\sin I_{\rm s}$ than either DT or the Hirano model.}
	\label{fig:W78_vsinI-lambda}
\end{figure}

We apply the same set of constraints to our tomographic MCMC analysis (Fig.\,\ref{fig:W78_DT}) as for the RM modelling runs. We find that all parameters are in agreement with the results from the Hirano model and the case in which no RM modelling is carried out, and that all parameters except $v\sin I_{\rm s}$ are in agreement with the Bou{\'e} model results, though we note that the disagreement there originates with the Bou{\'e} model rather than DT. Unlike for some of the other systems studied herein, we find a roughly factor of $2$ improvement in the precision of our alignment angle measurement over the RM modelling methods when using tomographic analysis (see Fig.\,\ref{fig:W78_vsinI-lambda} and Table\,\ref{tab:results}).

Acquiring simultaneous photometry and spectroscopy provides us with a means to cross-check for stellar activity signatures, as any stellar activity which affects the light curve should be visible in the CCF time series map, Fig.\,\ref{fig:W78_DT}. Fig.\,\ref{fig:W78_phot} shows evidence of the presence of star spots in the $R$-band light curve, but the $I$-band light curve that was acquired simultaneously with our HARPS spectroscopy shows only hints of similar activity. Examination of the right-hand panel of Fig.\,\ref{fig:W78_DT} is inconclusive.

\subsection{WASP-79}
\label{sec:W79}
\citet{2012AA...547A..61S} found WASP-79\,b to have both a low density and a large radius, and to orbit its F5-type host star in a circular orbit with a period of $3.7$\,d. But they were unable to fully characterize the shape and duration of the transit signature owing to a lack of high-precision follow-up photometry, having access to only a single, partial transit light curve from TRAPPIST. This limited the accuracy of the physical and orbital parameters that they were able to obtain. \citet{2013ApJ...774L...9A} measured the spin-orbit alignment angle of the system using UCLES, modelling the RM effect during transit to obtain $\lambda=-106^{\circ\,\,+19}_{\,\,\,\,-13}$, indicating significant misalignment. 

\begin{figure}
	\centering
	\includegraphics[width=0.48\textwidth]{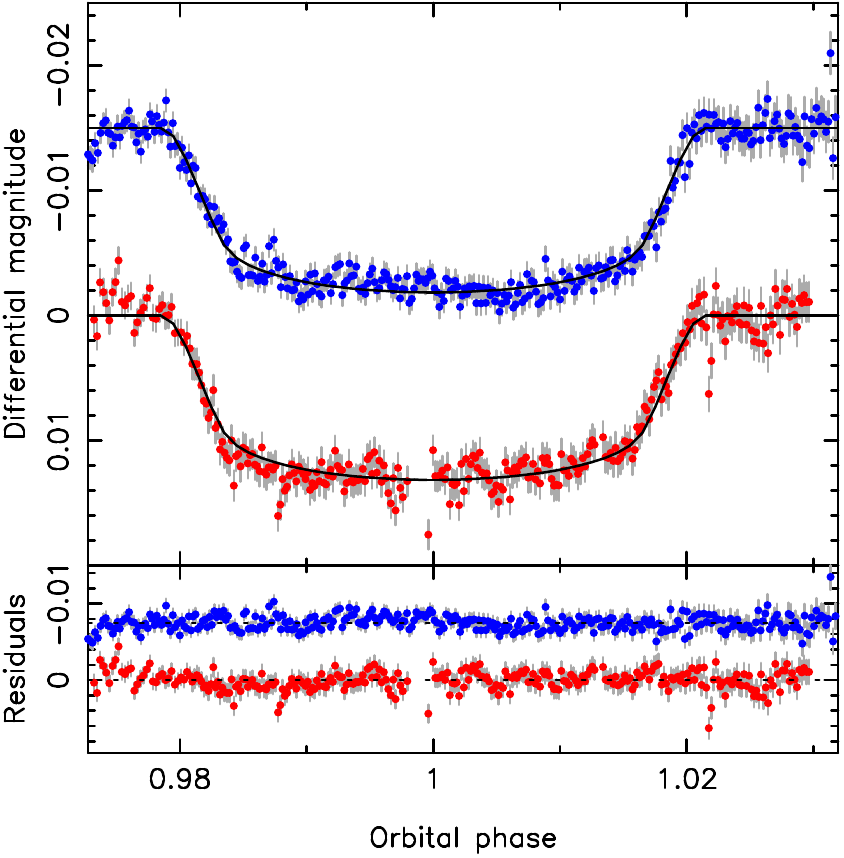}
	\caption{Upper panel: newly acquired EulerCam observations of two separate transits of WASP-79\,b, with best-fitting MCMC transit models. The two light curves have been offset by 0.015 mag for clarity. Lower, red data are for the $R$-band light curve of the transit on 2012 November 11. Upper, blue data are for the $R$-band light curve of the transit on 2012 December 4. Neither light curve shows any sign of stellar variability. Lower panel: residuals of the fits of the data to the best-fitting transit models.}
	\label{fig:W79_phot}
\end{figure}

We obtained $R$-band EulerCam observations of the transits on the nights of 2012 November 11 and December 4 (see Fig.\,\ref{fig:W79_phot}), as well as spectroscopic observations of the transit on 2012 November 13 using HARPS. We also used the data from WASPSouth, TRAPPIST, and CORALIE that was presented in \citeauthor{2012AA...547A..61S}. Spectral analysis of the HARPS spectra provides values of $v\sin I_{\rm s}=18.53\pm0.40$\,km\,s$^{-1}$ and $v_{\rm mac}=6.96$\,km\,s$^{_1}$ that we use as our prior on the rotation velocity. We note though that the value of $v_{\rm mac}$ is extrapolated owing to the star's effective temperature being outwith the range used for the calibration of \citet{2014MNRAS.444.3592D}. If $v_{\rm mac}$ is fixed to $0$, then we obtain $v\sin I_{\rm s}=19.91\pm1.14$\,km\,$s^{-1}$, which effectively places an upper limit on the stellar rotation velocity of $v\sin I_{\rm s}<21.05$\,km\,s$^{-1}$.

\begin{figure}
	\centering
	\includegraphics[width=0.48\textwidth]{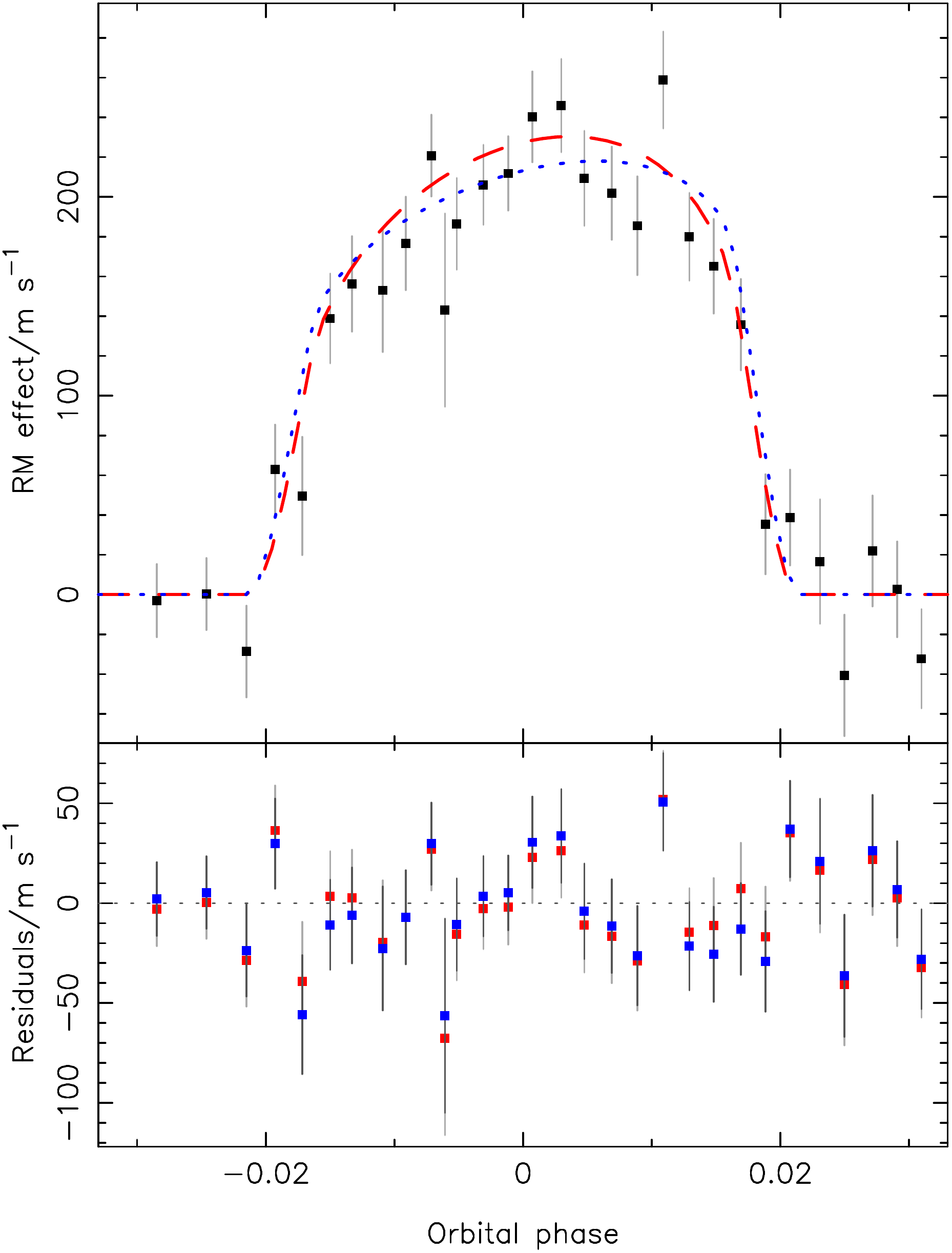}
	\caption{Upper panel: a close up of the RM anomaly in the RV curve of WASP-79, with the contribution from the Keplerian orbit subtracted to better display the form of the anomaly. Lower panel: the residuals for the two model fits. Legends for the two panels as for Fig.\,\ref{fig:W61_RV}.}
	\label{fig:W79_RV}
\end{figure}

\subsubsection{RM modelling}
\label{sec:w79_hirano}
We find no evidence for an eccentric orbit or for a long-term barycentric velocity trend, and find no difference between the stellar parameters obtained by MCMC when run with and without the stellar radius constraint. We also find no reason to apply a prior on $v\sin I_{\rm s}$; the impact parameter is sufficiently large for there to be no discrepancy between $v\sin I_{\rm s}$ and $\lambda$, although we do once more see a more slowly rotating star with the Bou{\'e} model than with the Hirano model. In this case the Bou{\'e} model is consistent with the spectroscopic rotation velocity, whilst the Hirano model seems to be over-predicting the speed of the stellar surface, even when we consider the upper limit set by the zero macroturbulence case. Of interest are the different forms of the [$v\sin I_{\rm s}$-$\lambda$] posterior probability distributions for the two models; that for the Hirano model shows a correlation between the two parameters and a more triangular shape, whilst the distribution for the Bou{\'e} model is more elliptical in shape.

Our selected solutions are taken from runs for which no prior was applied to $v\sin I_{\rm s}$, $e=0$, $\dot{\gamma}=0$, and no constraint was applied to the stellar radius. Both models return strongly misaligned results for the system at very high precision: $\lambda=-95.9^{\circ\,+2.8}_{\,\,\,-3.0}$ for the Hirano model, and $\lambda=-95.5^{\circ}\pm2.7$ for the Bou{\'e} model, in agreement with the result of \citet{2013ApJ...774L...9A}. The reason for such strong constraints are readily apparent from Fig.\,\ref{fig:W79_RV}. The RM anomaly is strongly asymmetrical and comprises only positive deviation from the out-of-transit velocity. This indicates that only one half of the star, the approaching hemisphere, is being traversed by the planet, and in combination with the angle that we find suggests a near-polar orbit for the planet.

\subsubsection{Doppler tomography}
\label{sec:w79_tomography}
\begin{figure}
	\centering
	\includegraphics[width=0.48\textwidth]{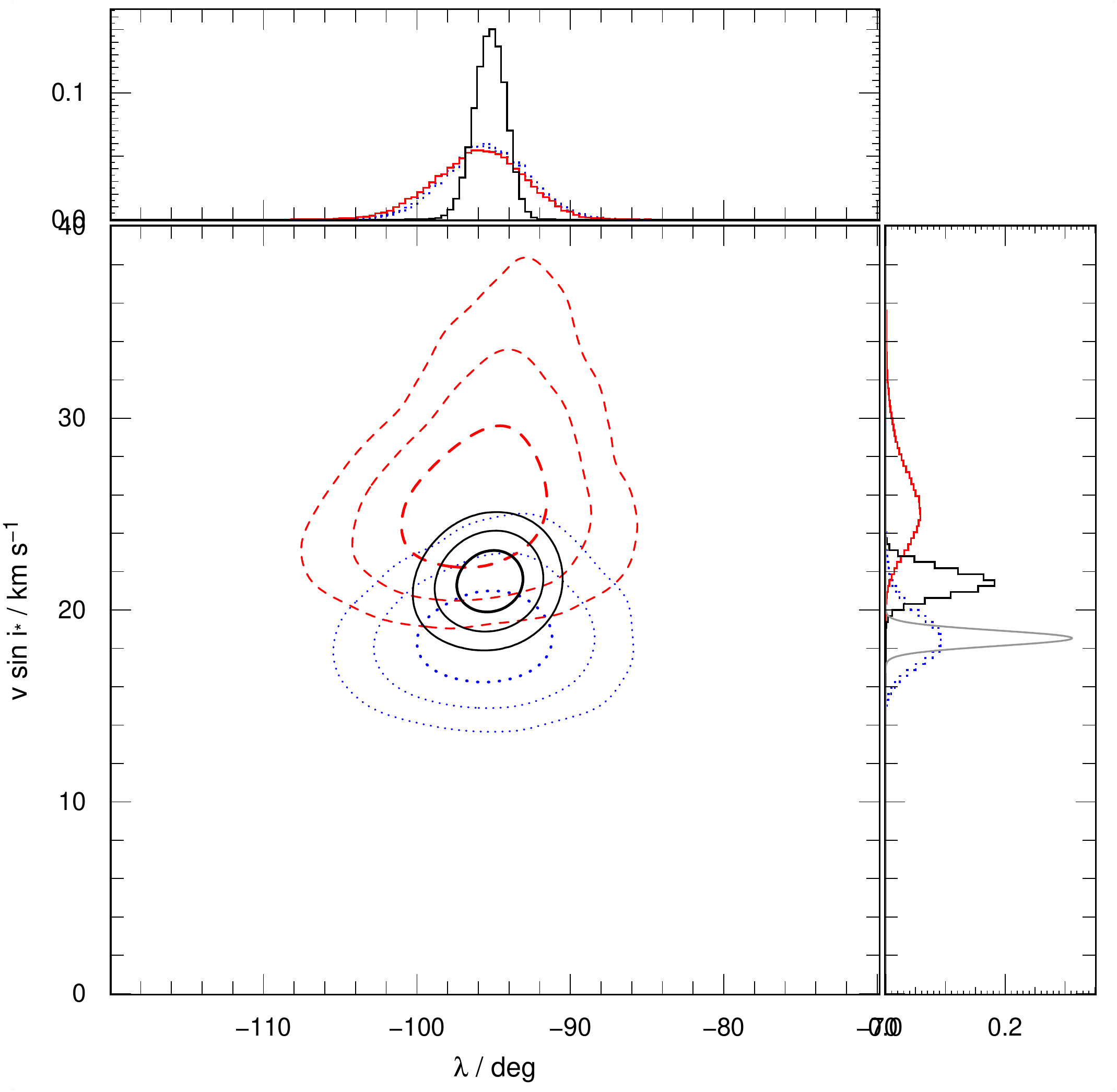}
	\caption{The posterior probability distributions in $v\sin I-\lambda$ parameter space for our analyses of WASP-79. Legend as for Fig.\,\ref{fig:W61_vsinI-lambda}. The DT result is more precise than either of the other two, for both $v\sin I_{\rm s}$ and $\lambda$, but falls between the two in this parameter space. The Hirano distribution displays a triangular shape, in contrast to the more ellipsoidal shapes shown by the distributions created by the other two models.}
	\label{fig:W79_vsinI-lambda}
\end{figure}

Our tomographic analysis is similarly well constrained, and returns an alignment angle of $\lambda=-95^\circ.2^{+0.9}_{-1.0}$, consistent with the results from both the Bou{\'e} and Hirano models. What is notable however is that the uncertainties have been reduced by a factor of $2$ over the already impressive results that we obtained with those models, and are an order of magnitude better than those obtained by \citet{2013ApJ...774L...9A}. $v\sin I_{\rm s}$ for our DT analysis falls between those of the two RM models, and is faster than the spectroscopic result, though it is in agreement with the upper limit set by the zero macroturbulence case (see Fig.\,\ref{fig:W79_vsinI-lambda} and Table\,\ref{tab:results}).

\begin{figure}
	\includegraphics[width=0.48\textwidth]{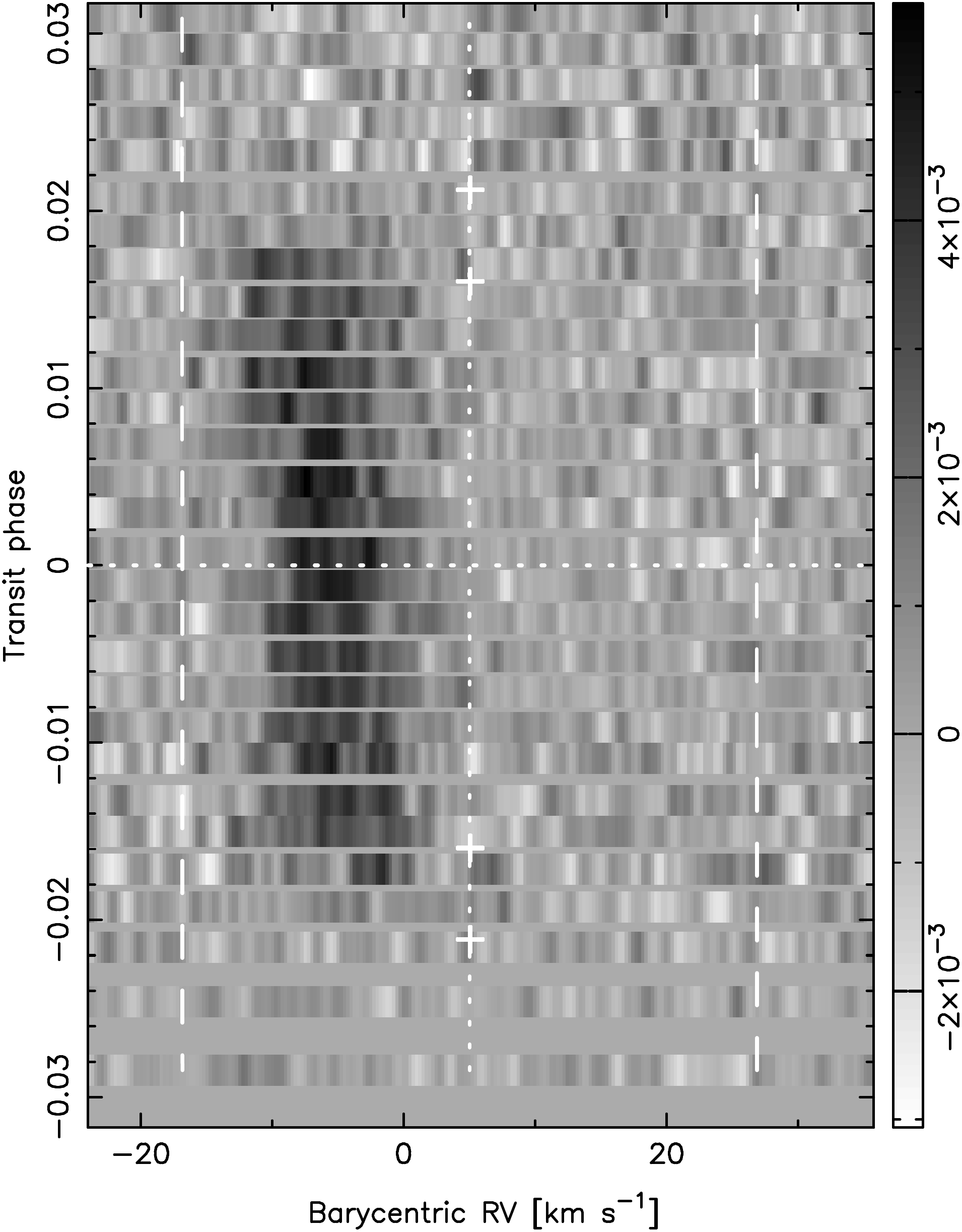}
	\caption{Residual map of WASP-79 time series CCFs with the model stellar spectrum subtracted. The signature of the planet is clear, and confined to one half of the spectral line, indicating a strongly misaligned orbit. The trajectory is slightly right to left, suggesting a mildly retrograde or polar orbit. Legend as Fig.\,\ref{fig:W61_DT}.}
	\label{fig:W79_DT}
\end{figure}

Fig.\,\ref{fig:W79_DT} again shows the strongly asymmetric signature of a misaligned polar orbit. The planetary trajectory is unlike that for any of the other five systems, being confined to one side of the stellar spectral line and moving from right to left (although any movement through the line is slight at best). This implies either a polar orbit, or a slightly retrograde one, as suggested by the value of $\lambda$ that we obtain.

\section{Discussion}
\label{sec:discussion}
We have obtained measurements of $v\sin I_{\rm s}$, $\lambda$, and $b$ using three different analyses of the RM effect. We show these results in Table\,\ref{tab:results}, along with other relevant parameters as determined from either spectral analysis of our new HARPS data ($T_{\rm eff}$), modelling of the available photometric and RV data ($M_{\rm s}$, $R_{\rm s}$), or isochronal fitting (${\rm Age}_{\rm Padova}$, see Section\,\ref{sec:lamage}). In the following discussion we will consider our results in the context of existing literature data on spin-orbit misalignment, treating the values obtained through the DT technique as definitive. We provide a more comprehensive list of parameters in Table\,\ref{tab:allresults} in the appendix.

\begin{table*}
	\scriptsize
	\caption{A summary of the results for the six systems studied herein, as returned by analyses incorporating the three different models that we have applied to our data. Effective temperatures are taken from spectral analysis of our HARPS observations. For the cases in which no RM effect is fitted, the value of $v\sin I_{\rm s}$ is fixed at the values assessed from the same spectral analysis. The spectroscopic $v\sin I_{\rm s}$ value for WASP-79 is calculated using an extrapolated macroturbulence value, as $T_{\rm eff}$ for the star lies outwith the range of the calibration used by \citet{2014MNRAS.444.3592D}. $\chi^2_{\rm RM, red}$ values are the reduced $\chi^2$ for the RM data set \textit{only}. For WASP-76, a prior on the impact parameter has been applied. We treat the tomography results, which shows smaller uncertainty in $\lambda$ than the other results, as definitive. WASP-61, -71, and -78 are aligned. WASP-62 is slightly misaligned, whereas -79 is strongly misaligned and in a retrograde orbit. Our results for WASP-76\,b suggest that it too may be on a strongly misaligned orbit, but detailed investigation of the system (see Section\,\ref{sec:W76}) shows that the alignment angle cannot be well constrained.}
	\label{tab:results}
	\begin{tabular}{llllllllll}
		\hline \\ [2pt]
		System	& Model		& $v\sin I_{\rm s}$\,/km\,s$^{-1}$	& $\lambda$\,/$^\circ$	& $b$\,/$R_{s}$		& $\chi^2_{\rm RM, red}$	& $M_{\rm s}$\,/${\rm M}_{\sun}$	& $R_{\rm s}$\,/${\rm R}_{\sun}$	& $T_{\rm eff, spec.}\,/K$ 	& ${\rm Age}_{\rm Padova}$ /Gyr\\ [2pt]
		\hline \\ [2pt]
		WASP-61	& Hirano		& $11.8^{+1.5}_{-1.4}$		& $1.3^{+18.8}_{-17.3}$	& $0.11^{+0.09}_{-0.07}$	& $1.4$	& $1.27\pm0.06$		& $1.39\pm0.03$		& $6250\pm150$		& $2.7^{+0.1}_{-0.6}$\\ 
				& Bou{\'e}		& $8.9^{+3.2}_{-1.7}$		& $13.9^{+35.7}_{-39.6}$	& $0.09^{+0.10}_{-0.06}$	& $1.2$	& $1.27\pm0.06$		& $1.39\pm0.03$		& $6250\pm150$		& $2.6\pm0.5$	\\ 
				& Tomography	& $11.1\pm0.7$				& $4.0^{+17.1}_{-18.4}$	& $0.10^{+0.10}_{-0.06}$	& -	& $1.27\pm0.06$		& $1.39\pm0.03$		& $6250\pm150$		& $2.7^{+0.1}_{-0.6}$\\ [2pt]
				& No RM fitting & $10.29\pm0.36$			& -				& $0.13^{+0.10}_{-0.08}$	& -	& $1.27\pm0.06$		& $1.40\pm0.03$		& $6250\pm150$		& $2.8^{+0.1}_{-0.8}$\\ 
		\hline \\ [2pt]
		WASP-62	& Hirano		& $10.5\pm0.4$			& $19.1^{+6.4}_{-5.8}$	& $0.25\pm0.07$		& $1.2$	& $1.28\pm0.04$		& $1.29\pm0.03$		& $6230\pm80$		& $1.1^{+0.6}_{-0.8}$\\ [2pt]
				& Bou{\'e}		& $7.1^{+0.5}_{-0.4}$		& $18.9^{+11.5}_{-6.6}$	& $0.28^{+0.09}_{-0.11}$	& $1.9$	& $1.27\pm0.04$		& $1.30^{+0.05}_{-0.04}$	& $6230\pm80$		& $0.8\pm0.5$	\\ [2pt]
				& Tomography	& $9.3\pm0.2$				& $19.4^{+5.1}_{-4.9}$	& $0.25^{+0.08}_{-0.06}$	& -	& $1.28\pm0.04$		& $1.29^{+0.04}_{-0.03}$	& $6230\pm80$		& $0.8\pm0.6$	\\ [2pt]
				& No RM fitting & $8.38\pm0.35$			& -				& $0.18\pm0.11$	& -		& $1.26\pm0.03$		& $1.27^{+0.04}_{-0.02}$ 	& $6230\pm80$		& $0.8^{+0.9}_{-0.4}$\\ [2pt]
		\hline \\ [2pt]
		WASP-71	& Hirano		& $10.5^{+1.5}_{-1.2}$		& $-12.0^{+6.5}_{-7.4}$	& $0.54^{+0.09}_{-0.14}$	& $1.0$	& $1.60\pm0.08$		& $2.47^{+0.22}_{-0.21}$	& $6050\pm100$		& $3.2^{+0.3}_{-1.2}$\\ [2pt]
				& Bou{\'e}		& $7.9\pm1.1$				& $-15.3^{+6.5}_{-7.3}$	& $0.57^{+0.07}_{-0.10}$	& $1.2$	& $1.62^{+0.08}_{-0.07}$	& $2.54\pm0.19$		& $6050\pm100$		& $3.4^{+0.1}_{-1.3}$\\ [2pt]
				& Tomography	& $7.8\pm0.3$				& $-1.9^{+7.1}_{-7.5}$	& $0.30^{+0.16}_{-0.19}$	& -	& $1.53^{+0.07}_{-0.06}$	& $2.17^{+0.18}_{-0.10}$	& $6050\pm100$		& $3.6^{+1.6}_{-1.0}$\\ [2pt]
				& No RM fitting & $9.06\pm0.36$			& -				& $0.55^{+0.08}_{-0.12}$	& -	& $1.61\pm0.08$		& $2.50\pm0.21$	& $6050\pm100$		& $3.4^{+5.6}_{-1.3}$\\ [2pt]
		\hline \\ [2pt]
		WASP-76 & Hirano 		& $0.7^{+0.5}_{-0.2}$		& $41.1^{+25.0}_{-50.1}$	& $0.130\pm0.003$		& $1.9$		& $1.46\pm0.06$		& $1.70\pm0.03$		& $6250\pm100$		& $$\\ [2pt]
				& Bou{\'e} 	& $0.4^{+0.2}_{-0.1}$		& $37.1^{+27.7}_{-52.9}$	& $0.13\pm0.01$		& $1.9$		& $1.45^{+0.07}_{-0.06}$	& $1.72\pm0.03$		& $6250\pm100$		& $$\\ [2pt]
				& Tomography & $1.1^{+0.5}_{-0.4}$ 		& $64.6^{+10.0}_{-23.6}$ & $0.13\pm0.01$		& -	& $1.46\pm0.06$ 		& $1.70\pm0.03$		& $6250\pm100$		& $$\\ [2pt]
				& No RM fitting & $2.33\pm0.36$			& -				& $0.12^{+0.12}_{-0.08}$	& -	& $1.46\pm0.07$		& $1.71\pm0.04$		& $6250\pm100$		& $2.7^{+0.1}_{-0.6}$\\ [2pt]
			\hline \\ [2pt]
		WASP-78	& Hirano		& $6.6\pm1.3$				& $-4.1^{+14.3}_{-12.5}$	& $0.52\pm0.05$		& $0.8$	& $1.40^{+0.09}_{-0.08}$	& $2.37^{+0.10}_{-0.09}$	& $6100\pm150$		& $3.5^{+8.2}_{-1.0}$\\ [2pt]
				& Bou{\'e}		& $4.1\pm1.1$				& $3.2^{+18.4}_{-16.5}$	& $0.52^{+0.04}_{-0.05}$	& $0.8$	& $1.39^{+0.09}_{-0.08}$	& $2.37^{+0.09}_{-0.10}$	& $6100\pm150$		& $3.5^{+1.6}_{-1.0}$\\ [2pt]
				& Tomography	& $7.1\pm0.5$				& $-6.4\pm5.9$			& $0.51\pm0.05$		& -	& $1.39^{+0.09}_{-0.08}$	& $2.35^{+0.10}_{-0.09}$	& $6100\pm150$		& $2.8^{+1.6}_{-0.3}$\\ [2pt]
				& No RM fitting & $6.63\pm0.16$			& -				& $0.58\pm0.03$		& -	& $1.42^{+0.09}_{-0.08}$ & $2.47\pm0.07$		& $6100\pm150$		& $3.4^{+1.5}_{-0.8}$\\ [2pt]
		\hline \\ [2pt]
		WASP-79	& Hirano		& $25.5^{+2.6}_{-2.0}$		& $-95.9^{+2.8}_{-3.0}$	& $0.50\pm0.03$		& $1.1$	& $1.39\pm0.06$		& $1.51\pm0.04$		& $6600\pm100$		& $1.4\pm0.3$	\\ [2pt]
				& Bou{\'e}		& $18.5^{+1.4}_{-1.3}$		& $-95.5\pm2.7$		& $0.51\pm0.03$		& $1.3$	& $1.39\pm0.06$		& $1.51\pm0.04$		& $6600\pm100$		& $1.7^{+0.1}_{-0.7}$\\ [2pt]
				& Tomography	& $21.5\pm0.7$			& $-95.2^{+0.9}_{-1.0}$	& $0.50\pm0.02$		& -	& $1.39\pm0.06$		& $1.51^{+0.04}_{-0.03}$	& $6600\pm100$		& $1.4\pm0.3$	\\ [2pt]
				& No RM fitting & $18.53\pm0.40$			& -				& $0.49\pm0.04$		& -	& $1.39\pm0.06$		& $1.49\pm0.05$		& $6600\pm100$		& $1.4\pm0.3$	\\ [2pt]
		\hline \\ [2pt]
	\end{tabular}
 \end{table*}

\subsection{Alignment angle as a function of temperature}
\label{sec:lamtemp}
\begin{figure}
	\centering
	\includegraphics[width=0.48\textwidth]{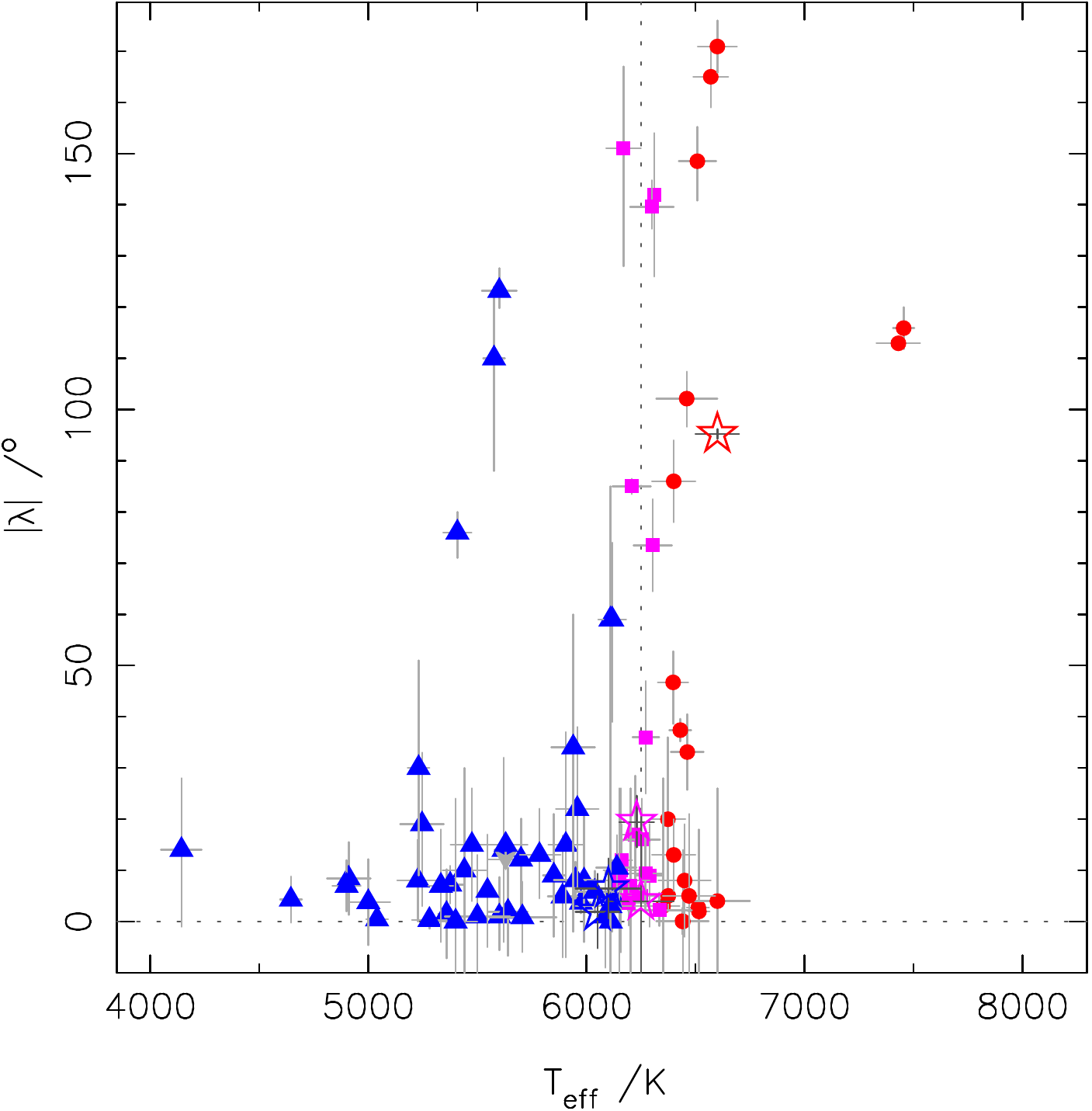}
	\caption{Absolute value of the spin-orbit alignment angle, $|\lambda|$, as a function of stellar effective temperature for hot Jupiter systems for which the angle has been measured; no distinction is made by method. Several systems are omitted from this plot (see the text for details). The horizontal dashed line marks $0^\circ$. The vertical dashed line denotes the dividing temperature of \citet{2010ApJ...718L.145W}, $T_{\rm crit}=6250$\,K. Blue symbols mark systems with $T_{\rm eff}<6150$\,K, red symbols systems with $T_{\rm eff}>6350$\,K, and magenta symbols those systems that occupy the region of uncertainty surrounding the position of the transition, $6150\leq T_{\rm eff}\leq6350$\,K. Open symbols represent the systems presented herein.}
	\label{fig:lamtemp}
\end{figure}

As noted in the introduction to this paper, one of the most commonly discussed trends in the exoplanetary alignment literature is that of $\lambda$ with $T_{\rm eff}$ -- cool stars ($T_{\rm eff}\lesssim6250$\,K) preferentially host aligned systems, hot stars preferentially host misaligned systems. Counter-trend examples, such as KOI-368 \citep{2014ApJ...786..131A} and HAT-P-18b \citep{2014AA...564L..13E}, do exist, but the general correlation has been confirmed \citep{2014ApJ...790L..31D}. The placement of our new results in $[\lambda-T_{\rm eff}]$ parameter space (Fig.\,\ref{fig:lamtemp}) is therefore an obvious place to start examining the influence of our new results. 

In Fig.\,\ref{fig:lamtemp} we plot $|\lambda|$ as a function of $T_{\rm eff}$ for systems taken from the RM data base of John Southworth's TEPCat \citep{southworth2011}\footnote{\url{http://www.astro.keele.ac.uk/jkt/tepcat/rossiter.html}, accessed at 17:00\,p.m. on 2016 August 16.}. At time of writing there were $102$ systems listed in this data base. To provide a relevant comparison sample for our systems, we select out the hot Jupiters in the data base by applying cuts on semi-major axis ($a<0.1\,{\rm au}$) and planet mass ($M_{\rm p}>0.3\,M_{\rm Jup}$). We then omit a small number of additional systems for a variety of reasons: WASP-2, WASP-23, and WASP-40 owing to indeterminate or non-significant measurements; \textit{CoRoT}-19, for which only the first half of the effect was observed; \textit{CoRoT}-1, \textit{CoRoT}-3, and XO-2 owing to poor sampling \footnote{We also omit the previous results for WASP-71 and WASP-79.}. Following this down-selection, we are left with a comparison sample of $88$ systems. To these systems we add our new results, plotting the DT-derived value of $\lambda$ as a function of spectroscopic $T_{\rm eff}$. Part of our intention in selecting the sample studied herein was to examine the uncertain boundary between `hot' and `cool' stars; unfortunately our results provide little new information to add to the definition of this transition.

\begin{figure}
	\centering
	\includegraphics[width=0.48\textwidth]{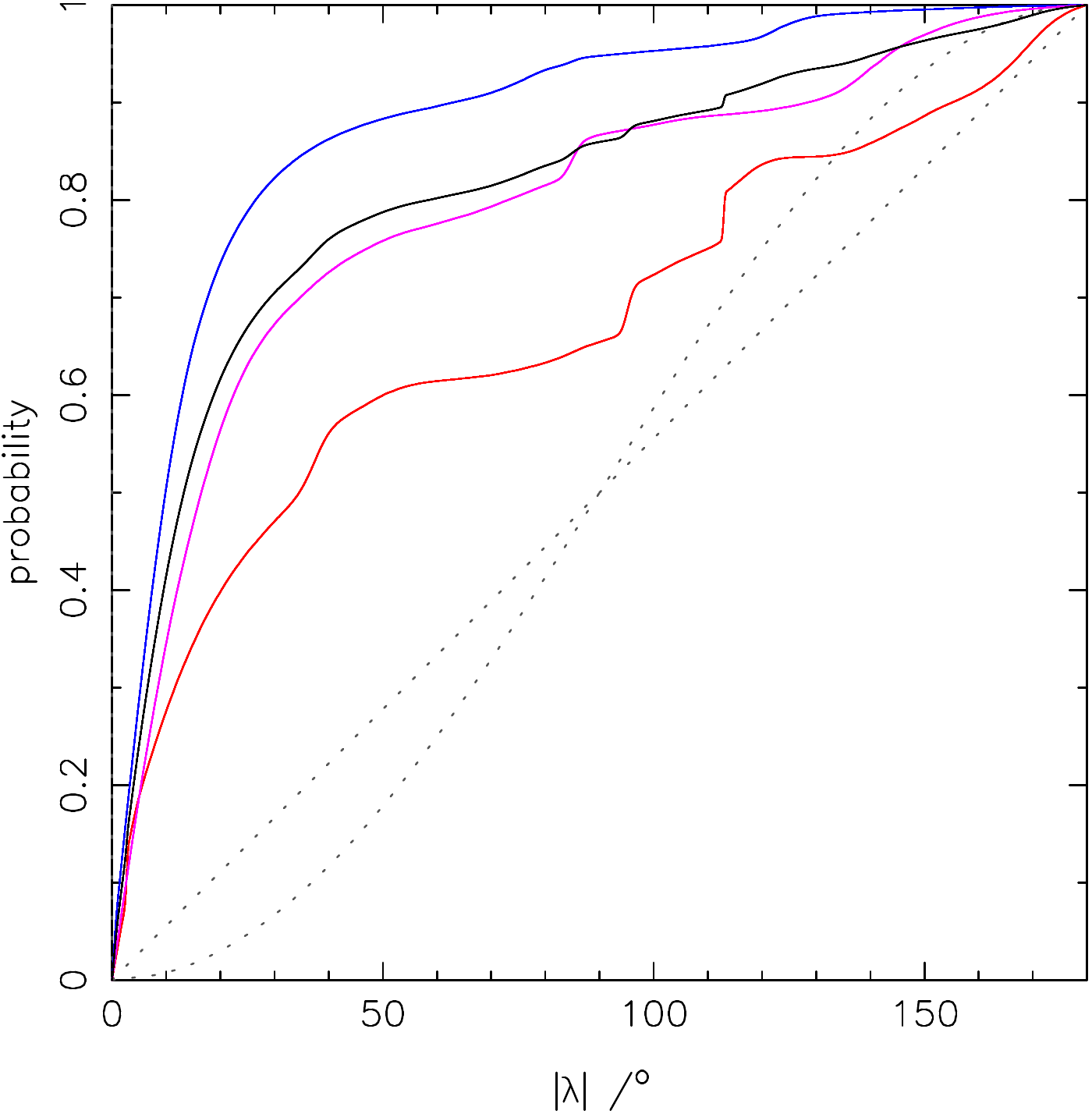}
	\caption{Cumulative probability distributions for $|\lambda|$ for three populations of systems from Fig.\,\ref{fig:lamtemp}: systems with $T_{\rm eff}<6150$\,K in blue, systems with $T_{\rm eff}>6350$\,K in red, and systems with $6150\leq T_{\rm eff}\leq6350$\,K in magenta. The black distribution represents the entire population shown in Fig.\,\ref{fig:lamtemp}. The vertical dotted line marks $30^\circ$, the angle above which systems are considered `misaligned'. The diagonal dotted line shows a uniform distribution in $\lambda$, while the curved dotted grey line denotes a uniform distribution in $\cos(\lambda)$. The distributions for the `hot' and `cold' populations are significantly different, while the `intermediate' population produces a distribution that is more similar to that of the complete population.}
	\label{fig:lamdist}
\end{figure}

We also examine the distribution of angles in a different fashion. For each system we create a skewed, two-dimensional Gaussian distribution using the reported $\lambda$ and $T_{\rm eff}$ as the mean coordinate, and using the upper and lower uncertainties as the standard deviations (for systems with only an upper or lower limit on $\lambda$, we assume a uniform distribution between the given value and either a lower bound of $0^\circ$, or an upper bound of $180^\circ$). We then sample this distribution $10,000$ times. The resulting values of $\lambda$ are then used to create the cumulative probability distributions for four different populations, classified using the sampled $T_{\rm eff}$ values (Fig.\,\ref{fig:lamdist}): `hot' systems (red); `cool' systems (blue); `transition' systems (magenta), and the full ensemble (black). We find that the distributions for the `hot' and `cool' populations are significantly different, as expected. The `hot' population has a more uniform probability distribution, albeit still biased somewhat towards aligned angles, while the `cool' population is strongly biased towards aligned systems. The `transition' population falls somewhere in between, and in fact agrees well with the distribution for the overall population. We carry out a two-dimensional Kolmogorov--Smirnov (KS) test to compare the distributions, finding that the hot, cold, and intermediate distributions are drawn from significantly different parent distributions with probability $>90\%$. Finally, we compare the number of misaligned and aligned system in each temperature bin, finding that the ratio of ${\rm aligned}:{\rm misaligned}$ is $4.8$ for `cool' systems, $1.8$ for `intermediate' systems, and $1.0$ for `hot' systems; as expected from previous work on the correlation between $T_{\rm eff}$ and $\lambda$, `cool' hot Jupiter systems show a strong bias towards alignment, whilst `hot' systems exhibit a uniform distribution of alignment angles.

\subsection{Alignment angle as a function of age}
\label{sec:lamage}
\citet{2011AA...534L...6T} noticed that, for stars with $M_{\rm s} \geq1.2{\rm M}_{\sun}$, all systems older than $\sim2.5$\,Gyr are mostly well aligned. This implies that the distribution of $\lambda$ changes with time, be it that misaligned planets realign or get destroyed.

The systems in our sample exhibit a range of different stellar masses, but by design have similar stellar temperatures and spectral types. Their ages should therefore be quite different, and it should be possible to use them to shed further light on the postulated trend of $\lambda$ decreasing with stellar age. We consider two sets of stellar models: the Padova models of \citet{2008AA...482..883M}, and the Yonsei-Yale models of \citet{2004ApJS..155..667D}. 

\begin{figure}
	\centering
	\includegraphics[width=0.48\textwidth]{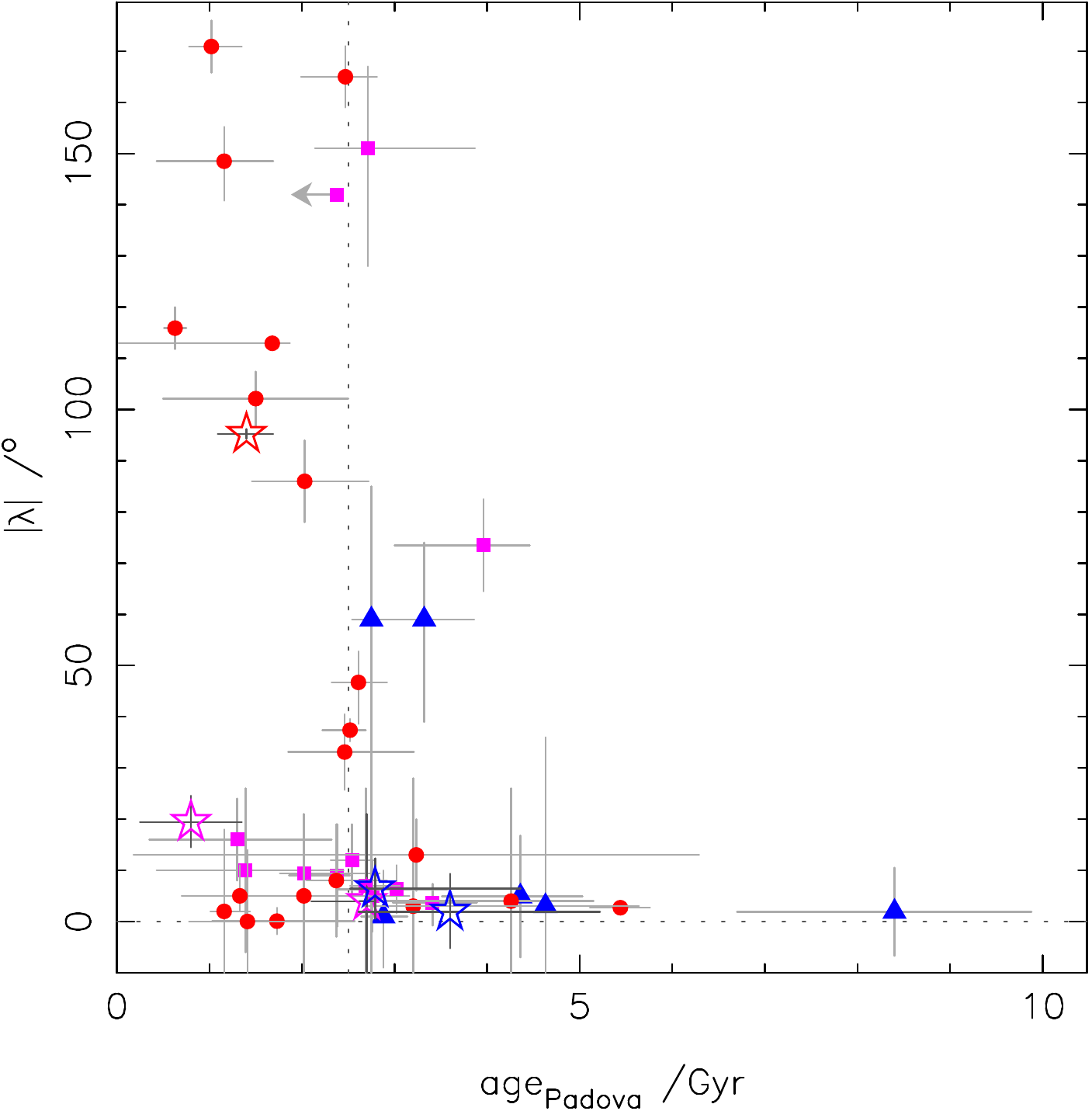}
	\caption{Absolute value of the alignment angle, $|\lambda|$, as a function of system age as determined using the Padova stellar evolution models for the systems shown in Fig.\,\ref{fig:lamtemp}. Key as in Fig.\,\ref{fig:lamtemp}. The vertical dashed line denotes the age of $2.5$\,Gyr by which it is posited that systems tend to realign. We find that the trend for misaligned systems to disappear at approximately $2.5$\,Gyr holds, but that the transition age is highly uncertain and depends strongly on stellar model choice.}
	\label{fig:lamage}
\end{figure}

We use the Padova models to recreate fig.\,2 of \citet{2011AA...534L...6T}. We calculate the ages for the six systems currently under discussion using the method described in appendix\,A of \citet{2014MNRAS.442.1844B}. We also calculate the ages for all systems in Fig.\,\ref{fig:lamtemp} with $M_\star\geq1.2$\,${\rm M}_{\sun}$ using the same method in order to provide a uniform sample. The results are shown in Fig.\,\ref{fig:lamage}. We find that the general trend for misaligned systems is weaker than found in \citet{2011AA...534L...6T}. Considering ages calculated using the Yonsei--Yale isochrones the pattern is weaker still, whereas it strengthens when employing the Geneva models of (\citealt{2012AA...541A..41M}, only on objects below the H-shell burning region). A full investigation of this effect is beyond the scope of this paper; it seems that the general trend for misaligned systems to be rare in older systems is weak and has some dependence on the choice of stellar models.

\subsection{Comparing techniques}
\label{sec:methodcompare}
A unique aspect of our study is that it allows us to compare the performance of three different models. Previous studies \citep[e.g.][]{2012MNRAS.423.1503B, 2013ApJ...771...11A} have compared the efficacy of RM modelling via RV fitting to DT, but the Bou{\'e} model is relatively new. Although it has been applied to a few systems \citep[e.g.][]{2013ApJ...774L...9A,2014AJ....148...29J}, to our knowledge no comparison has yet been made with more established models using real data. We note that \citet{2013ApJ...771...11A} measured the spin-orbit alignment of two systems using both in-transit RV fitting, and a method similar to our DT approach. However, their work had a different goal to ours; while they use the two methods to derive independent estimates of $\lambda$, examining separate wavelength regimes with the different methods, our goal was to examine the differences produced by the three methods when applied to the same data. More recently, \citet{2015arXiv151106402V} studied the RM effect in the XO-4 system and compared the results obtained using the Hirano model to those found by the model of \citet{2005ApJ...622.1118O, 2009ApJ...690....1O}. They found that the two models gave different values for $v\sin I_{\rm s}$ and $\lambda$, even when a Gaussian prior on stellar rotation was applied, leading to different interpretations of the system geometry. Furthermore, they suggest that the \citeauthor{2009ApJ...690....1O} model might lead to biased evaluations of system alignment.

\begin{figure}
	\centering
	\includegraphics{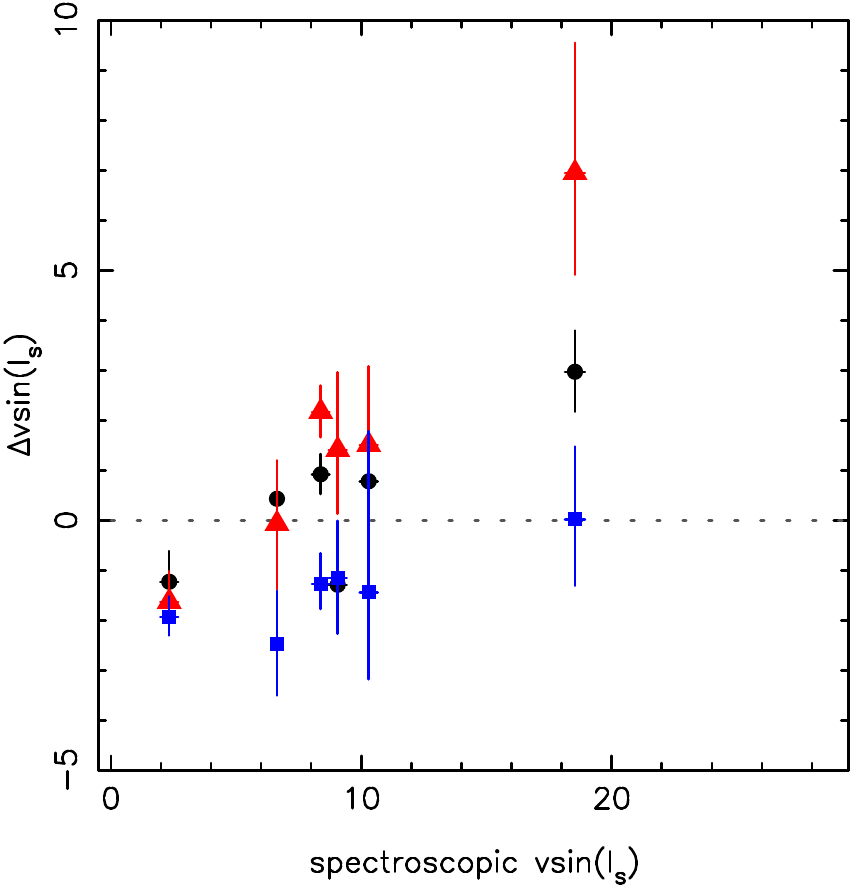}
	\caption{$\Delta v\sin I_{\rm s}$, the difference between the value returned by our model and the spectral analysis, as a function of the spectral line derived $v\sin I_{\rm s}$. Black circles represent values from DT, blue squares the values from the Bou{\'e} model, and red triangles the values from the Hirano model. Horizontal error bars are plotted, but are similar in size to the symbols and therefore not visible. The Bou{\'e} model clearly tends to underestimate the value of $v\sin I_{\rm s}$ compared to the spectral analysis, and compared to the other models.There are also trends in both the Hirano and DT results for the difference to the spectroscopic value to increase with the rotation rate of the star. This trend is more severe in the Hirano results than those from DT. }
	\label{fig:modelcomp}
\end{figure}

In Fig.\,\ref{fig:modelcomp}, we plot $\Delta v\sin I_{\rm s}$, the difference between the value returned by our model and the spectral analysis, as a function of the spectral line derived $v\sin I_{\rm s}$. We confirm that the Bou{\'e} model \textit{consistently undervalues $v\sin I_{\rm s}$ compared to the Hirano model} , which was their goal. This can also be seen in the various posterior probability distributions presented above. For four of the systems studied herein this leads to disagreement between the models, in the case of WASP-62 by $5\sigma$. Note though that the effect does not seem to be systematic, even if there is a tendency for the discrepancy to increase as the rotation rate of the star increases. The Bou{\'e} model also, to a lesser extent, tends to underestimate $v\sin I_{\rm s}$ compared to DT. These comparisons are less likely, however, to lead to disagreement than the comparison to the Hirano model. This arises as a result of the relatively large upper uncertainties on the Bou{\'e} values, coupled to the smaller numerical discrepancy between median values.

Compared to the spectroscopic values, all three models underestimate $v\sin I_{\rm s}$ for WASP-76, even when a prior on the impact parameter is applied (see Section\,\ref{sec:W76})\footnote{This not entirely surprising as its rotational broadening is close to the instrumental broadening. The spectroscopic value is likely overestimated in this case.}. But as the rotation rate of the star increases, the picture changes. The Bou{\'e} model continues to underestimate $v\sin I_{\rm s}$ up to the most rapidly rotating star, WASP-79, where the values are in agreement. In contrast, both the Hirano model and DT start to overestimate the rotation at approximately $v\sin I_{\rm s, spec}\approx8--10$\,km\,s$^{-1}$, with the discrepancy increasing with increasing $v\sin I_{\rm s}$. Using single value decomposition we carry out separate linear fits to the three data sets in Fig.\,\ref{fig:modelcomp}. We find slopes of $1.56$, $1.11$, and $1.25$ for the Hirano, Bou{\'e}, and DT models, respectively, indicating that the offset of the Hirano result is a stronger function of stellar rotation rate than for either of the other two models. The Bou{\'e} models produce the most consistent results, albeit biased towards lower velocity than spectral analysis.

As noted in Section\,\ref{sec:RMmodels}, the Bou{\'e} model relies on two parameters, $\beta_0$ and $\sigma_0$, to compute the line profile that is used to fit the RV CCFs. We investigated the effect of independently varying these parameters on the value of $v\sin I_{\rm s}$ returned by our best-fitting Bou{\'e} model. Using the same constraints as our reported results, we find that varying $\sigma_0$ has little effect on the stellar rotation velocity (or on any of the other reported parameters). However, we find that increasing $\beta_0$ leads to a more rapidly rotating star, and that increasing $\beta_0$ by a factor of $\approx3$ brings the reported $v\sin I_{\rm s}$ for WASP-61, WASP-62, and WASP-71 in line with spectroscopically assessed values.

Quality of fit is also an interesting point to compare between the Hirano and Bou{\'e} formulations. The form of the RM anomaly curves produced by the two models can be substantially different, as can be seen from, for example, Fig.\,\ref{fig:W61_RV}. This is reflected in the $\chi^2_{\rm red}$ values that we obtained. For WASP-71, -76, -78, and -79, there is little to choose between the different models. But for WASP-61 and -62 the story is quite different. In the former case the two models do still agree, but that agreement is weaker than for the other systems: the Bou{\'e} model gives $\chi^2_{\rm red}=1.4\pm0.2$, while the Hirano model gives $\chi^2_{\rm red}=1.2\pm0.2$\footnote{Compare this to WASP-78, for example, with $\chi^2_{\rm red; Bou\acute{e}}=\chi^2_{\rm red; Hirano}=0.8\pm0.2$.}. For WASP-62 though the models disagree at the $2-3\sigma$ level: $\chi^2_{\rm red; Bou\acute{e}}=1.9\pm0.3$, while $\chi^2_{\rm red; Hirano}=1.2\pm0.2$. It seems that for cases with disagreement over the quality of fit obtainable, that the Bou{\'e} model is worse off than the Hirano approach, at least for HARPS data. This is the opposite of what was expected given the basis of the Bou{\'e} model, which was developed to provide a more appropriate model for instruments without an iodine cell, such as HARPS.

\subsubsection{Response of RM models to $v\sin I_{\rm s}$ priors}
\begin{figure*}
	\centering
	\subfloat{
		\centering
		\includegraphics[width=0.48\textwidth]{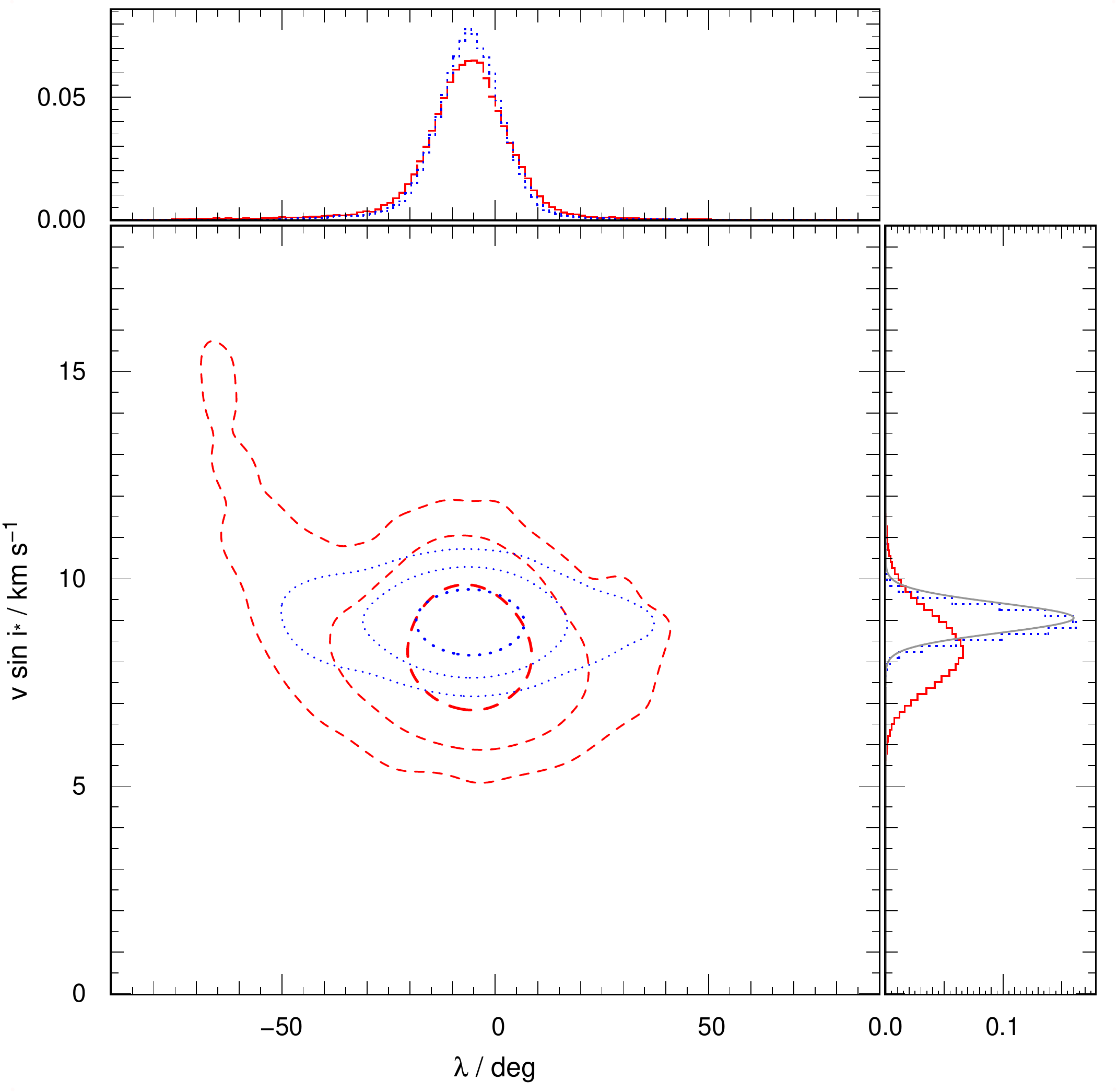}}
	\subfloat{
		\centering
		\includegraphics[width=0.48\textwidth]{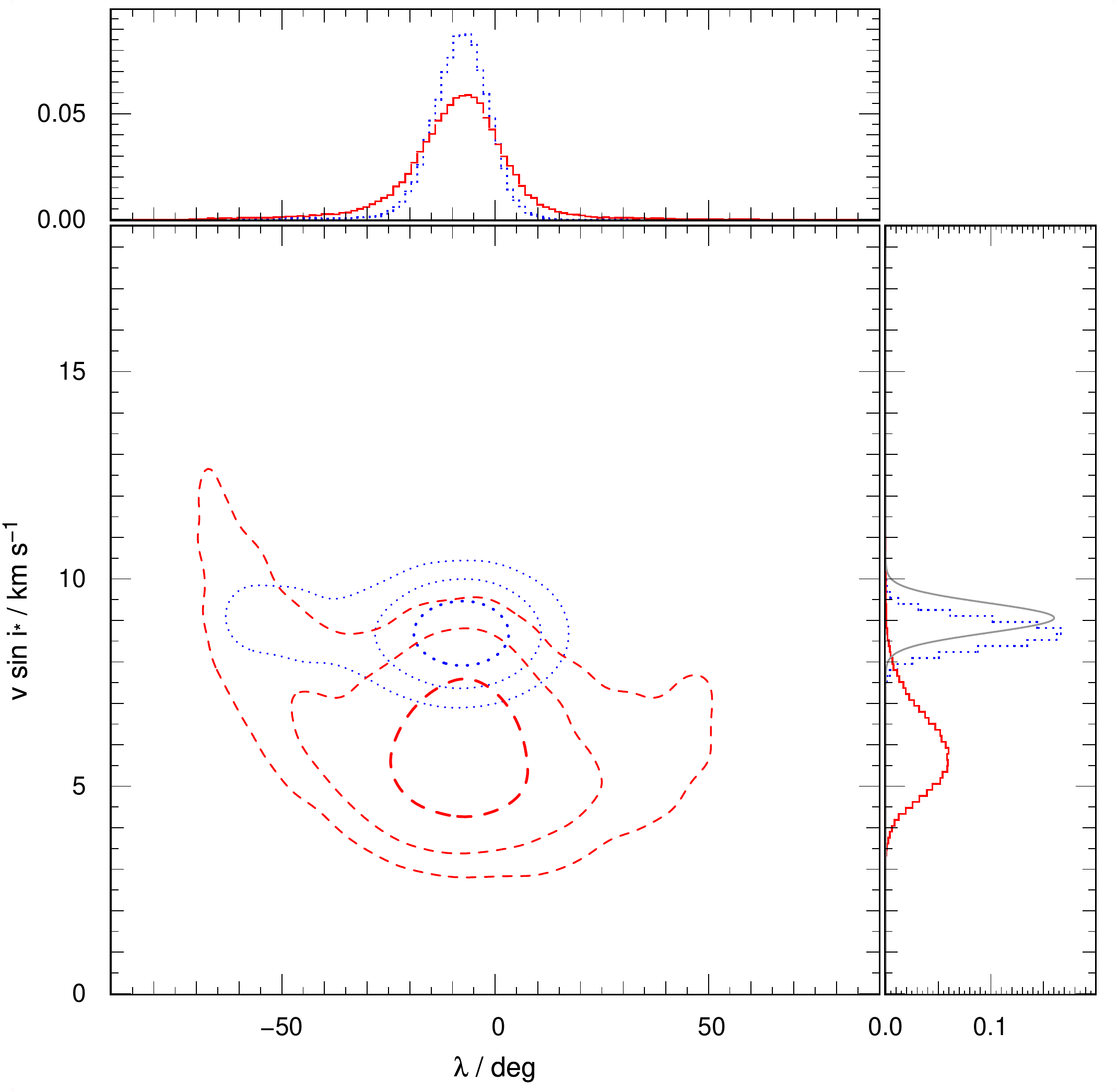}} \\
	\subfloat{
		\centering
		\includegraphics[width=0.48\textwidth]{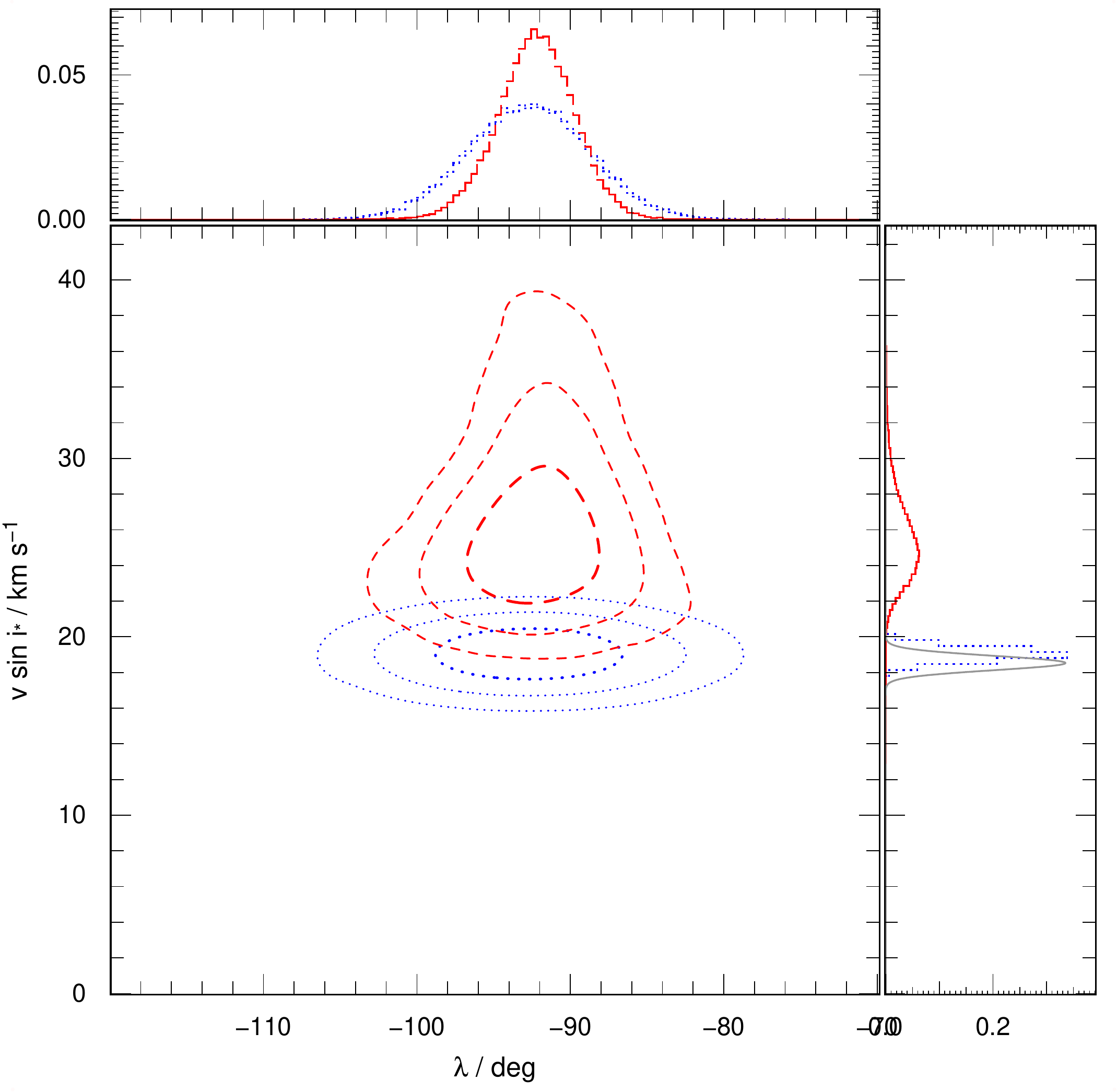}}
	\subfloat{
		\centering
		\includegraphics[width=0.48\textwidth]{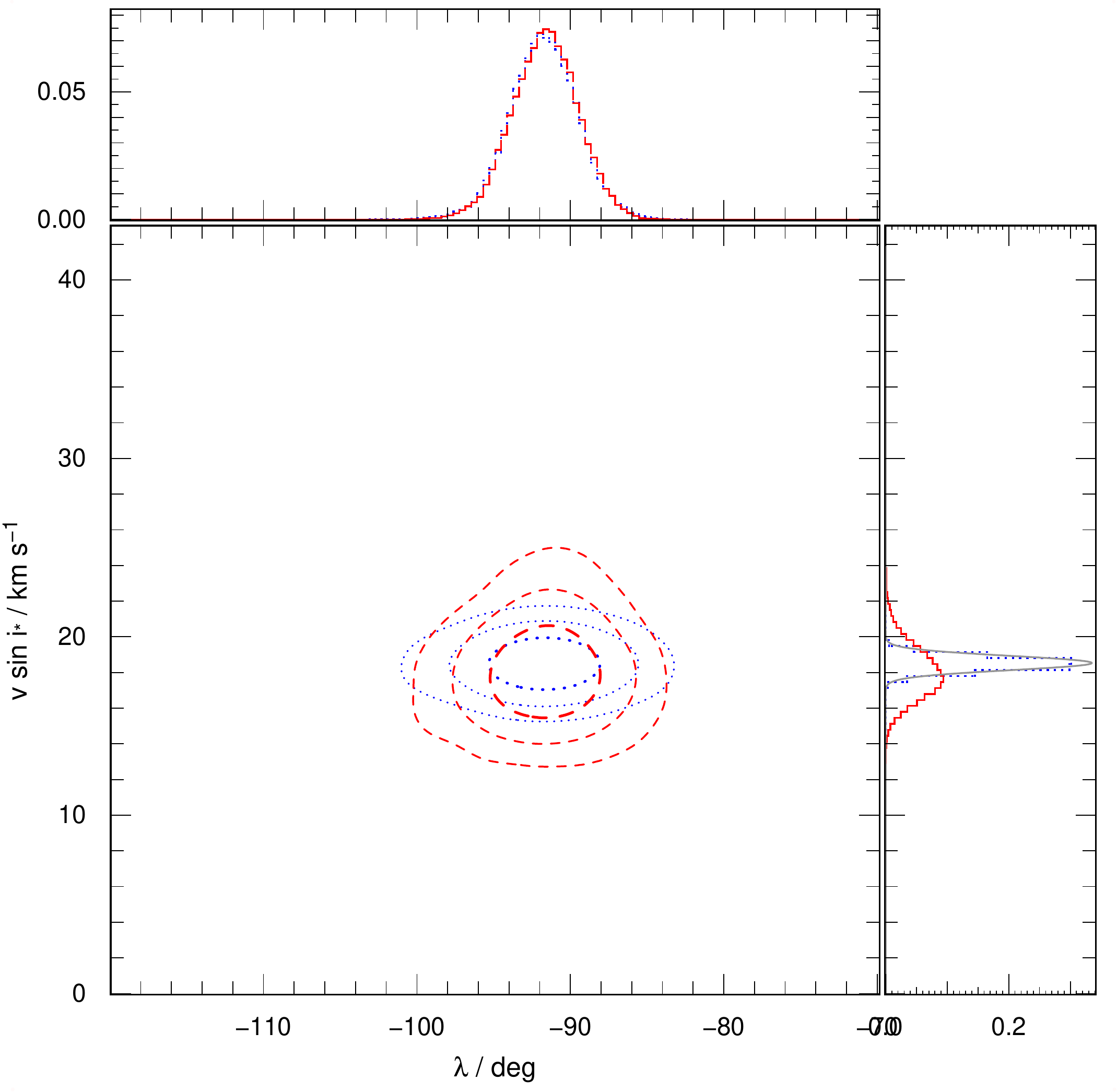}}
	\caption{A comparison of the posterior probability distributions produced by the Hirano and Bou{\'e} models both under the application of a Gaussian prior on $v\sin I_{\rm s}$ (blue, dotted contours), and when no prior is applied (red, dashed contours). Also displayed are the marginalized, one-dimensional distributions for the two parameters, with the additional, solid grey distribution in $v\sin I_{\rm s}$ representing the Bayesian prior. Left-hand column: data from the Hirano model. Right-hand column: data from the Bou{\'e} model. Top row: data for WASP-71. The distributions in the absence of a prior have a similar form, although the distribution from the Bou{\'e} model is centred at lower $v\sin I_{\rm s}$. Application of a prior on stellar rotation to this system has different effects on the two models. Bottom row: data for WASP-79. The absence of a prior leads to distributions of different form for the two models, with the distribution for the Hirano model being more extended, and centred at higher $v\sin I_{\rm s}$. Application of the prior limits the two models to similar ranges of $v\sin I_{\rm s}$, but retains the more extended distribution in $\lambda$ produced by the Hirano model.}
	\label{fig:prior_effect}
\end{figure*}

In general, the three models that we have considered produce probability distributions that appear similar when no prior is applied on $v\sin I_{\rm s}$ (see, for example, Fig.\,\ref{fig:W71_vsinI-lambda}). However, this is not always the case, as can be seen in Fig.\,\ref{fig:W62_vsinI-lambda}. Moreover, inspection of the distributions produced in the case of the application of a prior shows that these too can vary between models of the same system (see, for example, Fig.\,\ref{fig:prior_effect}). It seems, therefore, that the different RM models are affected in slightly different ways by the application of a stellar rotation prior.

We use WASP-71 and WASP-79 as examples to illustrate this. In Fig.\,\ref{fig:prior_effect} we show the change in distribution shape arising from the application of a prior on $v\sin I_{\rm s}$ when using the Bou{é} and Hirano models. For WASP-71, we can see that the distributions in the absence of a prior have a similar form, with a tail of negative $\lambda$ solutions that reaches towards more rapidly rotating stars, although the distribution from the Bou{\'e} model is centred around a lower value of $v\sin I_{\rm s}$. Application of a prior on stellar rotation to this system has different effects on the two models, with the Bou{\'e} model being restricted to a narrower range of $\lambda$, albeit with a tail towards large, negative values that is hard to pick out in the marginalized distribution. For WASP-79, we find that the absence of a prior leads to distributions of different form for the two models, with the distribution for the Hirano model being more extended, and centred at higher $v\sin I_{\rm s}$. Application of the prior limits the two models to similar ranges of $v\sin I_{\rm s}$, but retains the more extended distribution in $\lambda$ produced by the Hirano model.

This raises an interesting point. In what circumstances is it acceptable to enforce a prior on the stellar rotation? We assume that the estimates of $v\sin I_{\rm s}$ from RM modelling and spectral analysis \textit{should} give the same answer, though the precision will be different, and if this is the case then using the prior knowledge provided by the spectral analysis measurement (with an honest assessment of its uncertainty) is a valid approach. In the case of poor quality data (low SNR or missing data at ingress / egress) then this might in fact be the preferred approach, as application of the prior will guide the MCMC towards a physically motivated solution that is consistent with spectral broadening. On the other hand, if the data have sufficiently high SNR then the RM fit may give more precise constraints on $v\sin I_{\rm s}$ than the spectral analysis, implying that the fit is not strongly influenced by the prior. Note, however, that we are concerned here with the \textit{precision} of the measurement; the relative accuracy of the two estimates is still cause for concern, as shown by this work. Therefore, we would still urge readers to be cautious when applying a prior on $v\sin I_{\rm s}$ when modelling the RM effect. Although applying a prior can reduce correlations between parameters, it is clear from Section\,\ref{sec:methodcompare} that the `preferred' value of $v\sin I_{\rm s}$ is not necessarily that derived from spectral analysis.

\subsubsection{Comments on DT}
\label{sec:DTcomments}
One of the advantages that is commonly cited for DT over RM modelling is the ability of the former to break the degeneracy between $v\sin I_{\rm s}$ and $\lambda$ that can arise in low impact parameter systems. This is most commonly observed via the shape of the posterior probability distribution in $[v\sin I-\lambda]$ parameter space. In low impact systems the distribution produced by RM modelling tends to have a crescent shape, but in the DT method these parameters are determined in a fashion which renders them naturally uncorrelated, producing a smooth, elliptical distribution. However, our results for this set of systems demonstrate that this degeneracy breaking does not always take place.

Table\,\ref{tab:results} shows that both WASP-61 and WASP-76 are low impact parameter systems; recall here that we are considering the case for WASP-76 in which we apply a prior on $b$ but \textit{not} a prior on $v\sin I_{\rm s}$(see Table\,\ref{tab:W76models}), as the latter biases the posterior probability distribution. The distribution produced using the Bou{\'e} model has the expected crescent shape, as does the Hirano model. It is the tomography distribution though that is particularly interesting -- the crescent shape seen in the Hirano and Bou{\'e} distributions is still present , albeit with much shorter `arms' owing to the better constraint placed on $v\sin I_{\rm s}$. In fact, it is this improvement which seems to be the big advantage of tomography across our set of results. We note that using tomography does not necessarily help to restrict the range of $\lambda$ values investigated by the MCMC, nor does it necessarily remove the degeneracy in low impact systems. However, it does help to improve the constraints on $v\sin I_{\rm s}$, and with the right combination of priors help to distinguish between positive and negative minima in $\lambda$ parameter space.

\subsection{Other factors affecting fitting}
\label{sec:factors}
\subsubsection{Differential rotation}
\label{sec:diffrot}
The differences seen between the spectroscopic $v\sin I_{\rm s}$ and the results from our three analyses could arise in part from the effect of differential surface rotation. The spectroscopic measurements are derived from spectral lines, which arise from whole-disc observations and therefore give an indication of the dominant rotation rate of the star. The RM models that we have used here, however, all make use of the subplanet region in their calculations, i.e. the area of the stellar surface that lies directly beneath the planet's shadow. If this is at high latitudes (i.e. large impact parameter and/or a misaligned orbit), then the localized rotation rate that is measured would be slower than the whole disc measurement, which is dominated by the rotation rate at the equatorial regions. If not properly accounted for then this could lead to lower $v\sin I_{\rm s}$ values being returned by the RM modelling than by spectral analysis.

Differential rotation might therefore explain the discrepancy between our Bou{\'e} method results for WASPs-71 and -78, where we note that the impact parameter is of the order of $0.5$. It might also provide an explanation for the large discrepancy that we find in the WASP-76 system, where fitting an RM model drives the value of $v\sin I_{\rm s}$ downwards to values that are inconsistent with the spectral analysis, even in the presence of a prior on the impact parameter. It is unlikely to be the sole explanation, however.

\begin{figure}
	\centering
	\includegraphics{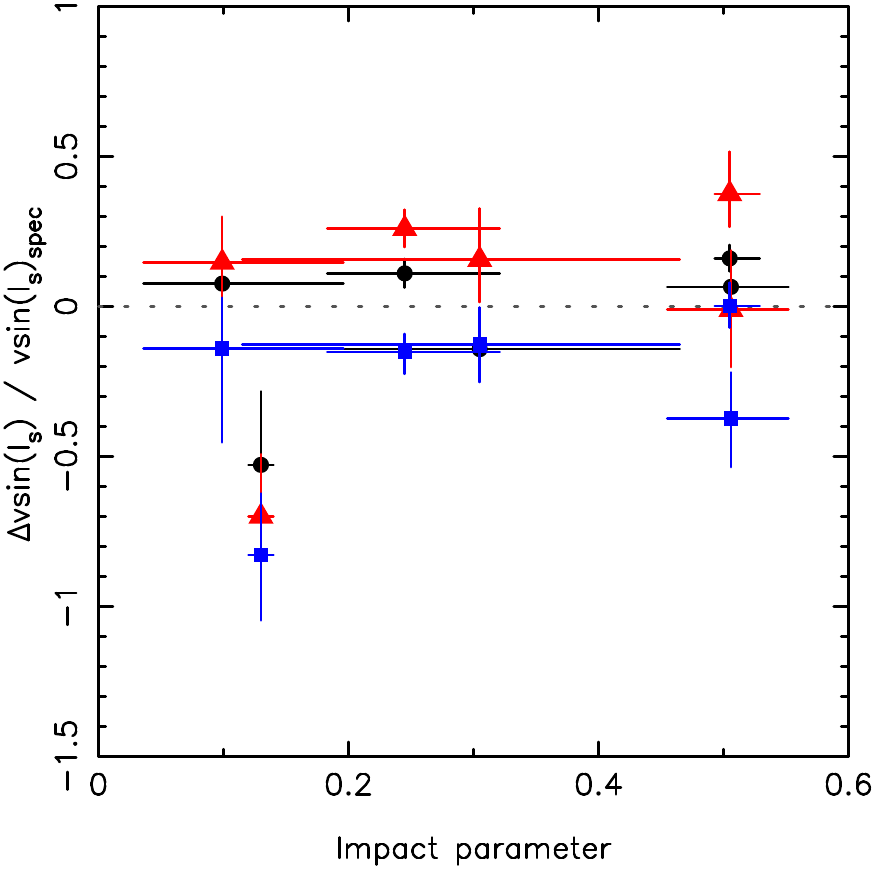}
	\caption{The fractional difference in stellar rotation, $\Delta v\sin I_{\rm s} / v\sin I_{\rm s, spec}$, as a function of the impact parameter, $b$. Black circles represent values from DT, blue squares the values from the Bou{\'e} model, and red triangles the values from the Hirano model. We find no trend, suggesting that differential rotation is not the source of the discrepancies that we observe, though we note that these are of the order of $10$\,percent or less in the majority of cases, approximately the expected magnitude of the differential rotation effect. The data for WASP-76 clearly show a much larger discrepancy than the other systems, further highlighting the unusual nature of this system's geometry.}
	\label{fig:diffrot}
\end{figure}

To test the influence of differential rotation on our results, we calculate the fractional difference in $v\sin I_{\rm s}=\Delta v\sin I_{\rm s}/v\sin I_{\rm s, spec}$, and plot it as a function of impact parameter in Fig.\,\ref{fig:diffrot}. We find no trend, suggesting that differential rotation is not the solution, though we note that the fractional difference in the majority of cases is of the order of (or less than) $10$\,percent. Fig.\,\ref{fig:diffrot} also further highlights the large discrepancy between our model-derived estimates of $v\sin I_{\rm s}$ and that given by spectral analysis for the case of WASP-76.

\subsubsection{Convective blueshift}
\label{sec:convblue}
The effect of the stellar surface convective blueshift might also be affecting the fit of our models to some of the stars in our sample. This effect originates in the movement of gas within the convective granules that make up the stellar surface, and was studied in relation to the RM effect by \citet{2011ApJ...733...30S}. The in-transit RV curve induced by convective blueshift is symmetrical, and for Solar-type host stars has a maximum amplitude of the order of $1$\,m\,s$^{-1}$. This effect is explicitly ignored by \citet{2011ApJ...742...69H} in the development of their model, while \citet{2013AA...550A..53B} make no mention of it at all.

Many studies neglect the contribution of convective blueshift to the RM effect, as the amplitude is much smaller than the uncertainties in their RV measurements. However, for slowly rotating stars it can be a significant effect; in the study of the HAT-P-17 system, for example, \citet{2013ApJ...772...80F} find that the amplitude of the RM anomaly is similar in magnitude to that of the convective blueshift effect. \citeauthor{2013ApJ...772...80F} further found that including the convective blueshift in their RM model led to changes in $\lambda$ of $\sim1\sigma$.

We calculate the mean uncertainties in our RV data and estimate the RM amplitude for our six systems. We then calculate the expected maximum amplitude of the convective blueshift effect, following equations 2 and 3 in \citet{2011ApJ...733...30S} and using the limb-darkening coefficients from our final solutions in concert with the four parameter limb-darkening law from \citet{2000AA...363.1081C}. We adopt $V_{\rm CB}=-750$\,m\,s$^{-1}$ for WASP-79, which is spectral type F5, and $V_{\rm CB}=-500$\,m\,s$^{-1}$ for the other systems, which are all of spectral type F7 or F8. The results are listed in Table\,\ref{tab:cbeffect}, and suggest that the convective blueshift effect might make a significant contribution to the modelling of WASP-62 and WASP-71. Our data also imply that the convective blueshift is significant for WASP-76, and may in fact be a dominant effect given the relative amplitudes of the fitted RM and predicted blueshift.

\begin{table*}
	\caption{The results of our brief convective blueshift analysis for the six systems studied herein.}
	\label{tab:cbeffect}
	\begin{tabular}{llllll}
		\hline \\ [2pt]
		System	& $V_{\rm CB}$/m\,s$^{-1}$	& $f$	 (R-band)	& $A_{\rm CB}$/m\,s$^{-1}$	& $K_{\rm RM}$/m\,s$^{-1}$	& $\bar{\sigma_{\rm RV}}$/m\,s$^{-1}$	\\ [2pt]
		\hline \\ [2pt]
		WASP-61	& $-500$					& $0.70$		& $2.8$					& $70$-$90$					& $29$							\\ [2pt]
		WASP-62	& $-500$					& $0.70$		& $4.0$					& $80$-$85$					& $7$							\\ [2pt]
		WASP-71	& $-500$					& $0.70$		& $1.4$					& $30$-$40$					& $9$							\\ [2pt]
		WASP-76	& $-500$					& $0.70$		& $3.8$					& $4$					& $4$							\\ [2pt]
		WASP-78	& $-500$					& $0.70$		& $2.9$					& $30$-$40$					& $20$							\\ [2pt]
		WASP-79	& $-750$					& $0.70$		& $5.9$					& $200$-$230$				& $24$							\\ [2pt]
		\hline \\
	\end{tabular}
\end{table*}

\citeauthor{2011ApJ...733...30S} admit that their model for the convective blueshift effect is first order only, and likely overly simple. For our purpose it serves well, but while preparing this manuscript, a more rigorous approach to modelling the convective blueshift was published by \citet{2016arXiv160102054C}. They conclude that for slow rotators ($v\sin I_{\rm s}\leq2$\,km\,s$^{-1}$) modelling the convective blueshift is more important than modelling the line profile correctly, but for faster rotators ($3\leq v\sin I_{\rm s}\leq10$\,km\,s$^{-1}$) the situation is reversed. They also find that neglecting to account for centre-to-limb variation when modelling the convective blueshift led to uncertainties in $\lambda$ of $10--20^\circ$ for aligned, central transits.

WASP-76 falls into the `slow rotators' category of \citeauthor{2016arXiv160102054C}, and thus the convective blueshift is indeed an important effect for this system. However, given the poor signal to noise of the system, we stand by our conclusion of a null detection. WASP-61, WASP-62, WASP-71, and WASP-76 all fall into their `faster rotators' category, such that the convective blueshift is the less important effect. However, all three of our models rely on the use of Gaussians for fitting either the CCF or the line profiles, such that our estimates of $\lambda$ may have underestimated uncertainties, or be inaccurate. Quantifying this effect, however, requires greater modelling capability than we possess, and estimating the effect based on the work of \citeauthor{2016arXiv160102054C} is problematic given the limited number of examples that they explore.

\section{Summary and conclusions}
\label{sec:summary}
We have presented new measurements of $\lambda$ for six WASP hot Jupiters: WASP-61, -62, -71, -76, -78, and -79. Using three different model (the Hirano and Bou{\'e} formulations for the RM effect, and DT) we investigated the possible alignment angles of the systems. We find that WASP-61 ($\lambda=4^\circ.0^{+17.1}_{-18.4}$), WASP-71 ($\lambda=-1^\circ.9^{+7.1}_{-7.5}$), and WASP-78 ($\lambda=-6^\circ.4\pm5.9$) are aligned, while WASP-79 ($\lambda=-95^\circ.2^{+0.9}_{-1.0}$) is strongly misaligned and on a retrograde orbit. WASP-62 ($\lambda=19.4^{\circ\,+5.1}_{\,\,\,-4.9}$) is slightly misaligned. We explore a range of possibilities for the orbit of WASP-76\,b, but are unable to constrain the alignment of the orbit beyond the general statement that it is likely strongly misaligned in the positive $\lambda$ direction.

Our result for WASP-71 disagrees with the larger angle published by \citet{2013AA...552A.120S}. Despite the use of HARPS data rather than the CORALIE data available to \citeauthor{2013AA...552A.120S}, we find little improvement in precision over their result, but for WASP-79 we find an improvement of more than a factor of 10 over the previous assessment by \citet{2013ApJ...774L...9A}, with which our new measurement agrees. WASP-76b's spin--orbit angle measurement is uncertain. However, because the host rotates slower than expected for its age and spectral type, as per \citet{2010ApJ...719..602S} we have an indication that the star is more likely pole-on than equator on. This is supported by the tests that were carried out in presence of priors on both $v\sin I_{\rm s}$ and $b$.

We show that the previously identified benefit of DT over RM modelling, namely the ability to break the degeneracy between $v\sin I_{\rm s}$ and $\lambda$ in low impact parameter cases, is not always applicable. However, tomography does consistently help to improve the precision of stellar rotation velocity measurements compared to RM modelling.

Using three different models has allowed us to compare and contrast their performance. We find that all three models give consistent values for $\lambda$ under the application of the same Bayesian priors, and when starting from the same initial conditions. We find that the Bou{\'e} model consistently underestimates the value of $v\sin I_{\rm s}$ compared to the Hirano model, and that it also tends to underestimate $v\sin I_{\rm s}$ compared to tomographic analysis. Moreover, we find that the estimates of $v\sin I_{\rm s}$ found from analysis of the RM effect often diverges from spectral analysis values, possibly a function of stellar rotation. We would therefore suggest caution when applying priors on $v\sin I_{\rm s}$.

\section*{Acknowledgements}
\label{sec:acknowledge}
The WASP Consortium consists of representatives from the Universities of Warwick, Keele, and St Andrews, along with the Isaac Newton Group (La Palma), the Instituto de Astrofisica de Canarias (Tenerife), and Observatoire de Haute Provence. DJAB would like to thank the UK Space Agency and the University of Warwick for their financial support, and acknowledges funding from the European Union Seventh Framework programme (FP7/2007- 2013) under grant agreement no. 313014 (ETAEARTH). AHMJT is a Swiss national science foundation fellow under grant PBGEP2-145594. This research has made use of NASA's Astrophysics Data System Bibliographic Services, the ArXiv preprint service hosted by Cornell University, and the TEPCat RM data base of John Southworth (\url{http://www.astro.keele.ac.uk/jkt/tepcat/rossiter.html}).

\bibliographystyle{mn}

\onecolumn
\begin{landscape}
\begingroup

\appendix
\section{Appendix}

\begin{table*}
\scriptsize
	\caption{A summary of the results for the six systems studied herein, as returned by MCMC modelling of the complete sets of photometric and spectroscopic data for each system, including DT modelling of the RM effect. For WASP-76, a prior on the impact parameter has been applied.}
	\label{tab:allresults}
	\begin{tabular}{lllllllll}
		\hline \\ [2pt]
		Parameter				& Symbol				& Units				& WASP-61				& WASP-62				& WASP-71				& WASP-76				& WASP-78				& WASP-79				\\ [2pt]
		\hline \\ [2pt]
		\multicolumn{8}{l}{\textit{Stellar parameters from spectral analysis}} \\
		Effective temperature	& $T_{\rm eff}$			&					& $6250\pm150$			& $6230\pm80$			& $6050\pm100$			& $6250\pm100$			& $6100\pm150$			& $6600\pm100$			\\
		Metallicity 				& [Fe/H]				& dex				& $-0.10\pm0.12$			& $0.04\pm0.08$			& $0.15\pm0.07$			& $0.19\pm0.10$			& $-0.35\pm014$			& $0.03\pm0.10$			\\
		Projected rotation velocity & $v\sin I_{\rm s}$		& km\,s$^{-1}$ 			& $10.29\pm0.36$			& $8.38\pm0.35$			& $9.06\pm0.36$			& $2.33\pm0.36$			& $6.63\pm0.16$			& $18.53\pm0.40$			\\
		Macroturbulent velocity	& $v_{\rm mac}$		& km\,s$^{-1}$ 			& $5.04$					& $4.66$					& $4.28$					& $4.84$					& $4.85$					& $6.96$ (extrapolated)		\\ [2pt] \\
		\multicolumn{8}{l}{\textit{Model parameters}} \\
		Orbital period 			& $P$				& days				& $3.855898\pm0.000003$	& $4.411940\pm0.000004$	& $2.903676\pm0.000008$	& $1.809886\pm0.000001$	& $2.175173\pm0.000003$	& $3.662392\pm0.000004$	\\
		Epoch of mid-transit 		& $t_0$ 				& ${\rm BJD}_{\rm TDB}$	& $6075.4585\pm0.0005$		& $5767.1533\pm0.0005$		& $5724.3320\pm0.0005$		& $5876.1895\pm0.0005$		& $6139.0303\pm0.0005$		& $6215.4556\pm0.0005$		\\
		Transit duration 		& $T_{\rm dur}$		& days				& $0.0427\pm0.0001$		& $0.0361\pm0.0002$		& $0.0712\pm0.0005$		& $0.0839\pm0.0004$		& $0.0914\pm0.0005$		& $0.0423\pm0.0002$		\\
		Planet:star area ratio		& $R_{\rm p}^2/R_\star^2$ & 					& $0.00800\pm0.00008$		& $0.0113\pm0.0002$		& $0.0041\pm0.0001$		& $0.0108\pm0.0001$		& $0.0082\pm0.0002$		& $0.0109\pm0.0001$		\\
		RV semi-amplitude		& $K$				& 					& $0.227\pm0.016$			& $0.061\pm0.003$			& $0.229\pm0.004$			& $0.112\pm0.001$			& $0.107\pm0.008$			& $0.090\pm0.008$			\\
		Systemic velocity		& $\gamma$			&					& $18.90^{+0.34}_{-0.43}$	& $14.86^{+0.17}_{-0.13}$	& $8.34^{+0.18}_{-0.34}$		& $-1.05\pm0.06$			& $0.48^{+0.39}_{-0.09}$		& $4.99^{+0.46}_{-0.06}$		\\
		Systemic velocity drift	& $\dot{\gamma}$		&					& $0$ (adopted)			& $0$ (adopted)			& $0$ (adopted)			& $0$ (adopted)			& $0$ (adopted)			& $0$ (adopted)			\\
		Impact parameter		& $b$				& 					& $0.10^{+0.10}_{-0.06}$		& $0.25^{+0.08}_{-0.06}$		& $0.30^{+0.16}_{-0.19}$		& $0.13^{+0.01}_{-0.01}$		& $0.51\pm0.05$			& $0.50\pm0.02$			\\
		Projected rotation velocity & $v\sin I_{\rm s}$		& km\,s$^{-1}$ 			& $11.1\pm0.7$				& $9.3\pm0.2$				& $7.8\pm0.3$				& $1.1^{+0.5}_{-0.4}$		& $7.1\pm0.5$				& $21.5\pm0.7$			\\
		Projected alignment angle	 & $\lambda$			& $^\circ$				& $4.0^{+17.1}_{-18.4}$		& $19.4^{+5.1}_{-4.9}$		& $-1.9^{+7.1}_{-7.5}$		& $64.6^{+10.0}_{-23.6}$		& $-6.4\pm5.9$				& $-95.2^{+0.9}_{-1.0}$		\\ [2pt] \\
		\multicolumn{8}{l}{\textit{Derived parameters}} \\
		Ingress / egress duration	& $T_{12}=T_{34}$ 		& days			& $0.0136	^{+0.0004}_{-0.0002}$ &	$0.0162^{+0.0012}_{-0.0008}$	& $0.0139^{+0.0021}_{-0.0014}$ & $0.0143\pm0.0002$		& $0.0216\pm0.002$			& $0.0190\pm0.0005$		\\
		Orbital inclination		& $i_{\rm p}$			& $^\circ$				& $89.3^{+0.4}_{-0.7}$		& $88.5^{+0.4}_{-0.7}$		& $85.8^{+2.4}_{-2.1}$		& $88.2\pm0.2$			& $81.3\pm1.1$			& $86.1\pm0.2$			\\
		Orbital eccentricity		& $e$				& 					& $0$ (adopted)			& $0$ (adopted)			& $0$ (adopted)			& $0$ (adopted)			& $0$ (adopted)			& $0$ (adopted)			\\
		Stellar mass			& $M_{\rm s}$			& ${\rm M}_{\sun}$			& $1.27\pm0.06$			& $1.28\pm0.04$			& $1.53^{+0.07}_{-0.06}$		& $1.46\pm0.06$			& $1.39^{+0.09}_{-0.08}$		& $1.39\pm0.06$			\\
		Stellar radius			& $R_{\rm s}$			& ${\rm R}_{\sun}$			& $1.39\pm0.03$			& $1.29^{+0.04}_{-0.03}$		& $2.17^{+0.18}_{-0.10}$		& $1.70\pm0.03$			& $2.35^{+0.10}_{-0.09}$		& $1.51^{+0.04}_{-0.03}$		\\
		Stellar surface gravity	& $\log g_{\rm s}$		& (cgs)				& $4.256^{+0.009}_{-0.013}$	& $4.321^{+0.013}_{-0.026}$	& $3.944^{+0.036}_{-0.050}$	& $4.144\pm0.008$			& $3.840\pm0.024$			& $4.226\pm0.010$			\\
		Stellar density			& $\rho_{\rm s}$		& $\rho_{\sun}$			& $0.476^{+0.007}_{-0.020}$	& $0.593^{+0.028}_{-0.055}$	& $0.147^{+0.022}_{-0.025}$	& $0.295\pm0.004$			& $0.108\pm0.010$			& $0.408\pm0.012$			\\
		Planet mass			& $M_{\rm p}$			& $M_{\rm Jup}$		& $2.06\pm0.16$			& $0.58\pm0.03$			& $2.14\pm0.08$			& $0.87\pm0.03$			& $0.86\pm0.08$			& $0.85\pm0.08$			\\
		Planet radius			& $R_{\rm p}$			& $R_{\rm Jup}$		& $1.21\pm0.03$			& $1.34^{+0.05}_{-0.03}$		& $1.35^{+0.13}_{-0.07}$		& $1.73\pm0.03$			& $2.06\pm0.10$			& $1.53\pm0.04$			\\
		Planet surface gravity	& $\log g_{\rm p}$ 		& (cgs)				& $3.509^{+0.031}_{-0.036}$	& $2.863^{+0.033}_{-0.038}$	& $3.424^{+0.046}_{-0.062}$	& $2.831\pm0.012$			& $2.671^{+0.042}_{-0.047}$	& $2.924^{+0.037}_{-0.040}$	\\
		Planet density			& $\rho_{\rm p}$ 		& $\rho_{\rm Jup}$		& $1.17^{+0.09}_{-0.10}$		& $0.24\pm0.03$			& $0.85^{+0.16}_{-0.17}$		& $0.173\pm0.007$			& $0.099\pm0.014$			& $0.240\pm0.023$			\\
		Scaled stellar radius		& $R_\star/a$			& 				& $0.1236^{+0.0018}_{-0.0006}$ & $0.1052^{+0.0025}_{-0.0015}$ &	 $0.219^{+0.016}_{-0.008}$	& $0.240\pm0.001$			& $0.297\pm0.009$			& $0.135\pm0.002$			\\
		Semi-major axis		& $a$				& au					& $0.0521\pm0.0008$		& $0.0571\pm0.0005$		& $0.0460\pm0.0006$		& $0.0330\pm0.0005$		& $0.0367\pm0.0008$		& $0.0519\pm0.0008$		\\
		Planet equilibrium temp. & $T_{{\rm P, A}=0}$ & K					& $1624.4^{+21.2}_{-14.6}$	& $1475.3^{+25.1}_{-20.0}$	& $2016.1^{+67.0}_{-52.5}$	& $2153.6^{+32.8}_{-30.7}$	& $2469.7^{+53.8}_{-55.9}$	& $1716.2^{+25.8}_{-24.4}$	\\
		System age			& ${\rm age}_{\rm Padova}$ & Gyr				& $2.7^{+0.1}_{-0.6}$		& $0.8\pm0.6$				& $3.6^{+1.6}_{-1.0}$		& $2.7^{+0.1}_{-0.6}$		& $2.8^{+1.6}_{-0.3}$		& $1.4\pm0.3$				\\
		\hline \\
	\end{tabular}
\end{table*}
\endgroup
\end{landscape}
\twocolumn

\begin{table}
	\caption{New in-transit radial velocities for WASP-61, obtained using the HARPS instrument on the night of 2012 December 22.}
	\label{tab:wasp61RVs}
	\begin{tabular}{lll}
		\hline \\
		Time					& RV		& Uncertainty \\ [2pt]
		${\rm BJD}_{\rm TDB}$	& km\,s$^{-1}$		& km\,s$^{-1}$ \\ [2pt]
		\hline \\
		$6283.577232$	& $19.03925$	& $0.03070$ \\
		$6283.586097$	& $18.95711$	& $0.02718$ \\
		$6283.594813$	& $18.97097$	& $0.02955$ \\
		$6283.604291$	& $18.99679$	& $0.02997$ \\
		$6283.612995$	& $19.03621$	& $0.02760$ \\ 
		$6283.622393$	& $18.99698$	& $0.03042$ \\
		$6283.630842$	& $19.05516$	& $0.03375$ \\
		$6283.640332$	& $19.03883$	& $0.02892$ \\
		$6283.649383$	& $18.99969$	& $0.03097$ \\
		$6283.658364$	& $18.99178$	& $0.02892$ \\
		$6283.667496$	& $19.07023$	& $0.02760$ \\
		$6283.676454$	& $19.00178$	& $0.02772$ \\
		$6283.685424$	& $18.88981$	& $0.03001$ \\
		$6283.694556$	& $18.86215$	& $0.03042$ \\
		$6283.703699$	& $18.84422$	& $0.03131$ \\
		$6283.712645$	& $18.83797$	& $0.02893$ \\
		$6283.721696$	& $18.88145$	& $0.03063$ \\
		$6283.730828$	& $18.90679$	& $0.03095$ \\
		$6283.739878$	& $18.85461$	& $0.02974$ \\
		$6283.748756$	& $18.92283$	& $0.03135$ \\
		$6283.758073$	& $18.99629$	& $0.02856$ \\
		$6283.766869$	& $18.91593$	& $0.02863$ \\
		$6283.776000$	& $18.91219$	& $0.02762$ \\
		$6283.784958$	& $18.89471$	& $0.02873$ \\
		$6283.794183$	& $18.91173$	& $0.02891$ \\
		$6283.803233$	& $18.96711$	& $0.02782$ \\
		$6283.812284$	& $18.92203$	& $0.02643$ \\
		\hline \\
	\end{tabular}
\end{table}

\begin{table}
	\caption{New in-transit radial velocities for WASP-62, obtained using the HARPS instrument on the night of 2012 October 12.}
	\label{tab:wasp62RVs}
	\begin{tabular}{lll}
		\hline \\
		Time					& RV		& Uncertainty \\ [2pt]
		${\rm BJD}_{\rm TDB}$	& km\,s$^{-1}$		& km\,s$^{-1}$ \\ [2pt]
		\hline \\
		$6212.667066$	& $14.98142$	& $0.00846$ \\
		$6212.675260$	& $14.96786$	& $0.00782$ \\
		$6212.683640$	& $14.98130$	& $0.00769$ \\
		$6212.690897$	& $15.02134$	& $0.00739$ \\
		$6212.698131$	& $15.04109$	& $0.00806$ \\
		$6212.705515$	& $15.03729$	& $0.00814$ \\
		$6212.712772$	& $15.04227$	& $0.00877$ \\
		$6212.720214$	& $15.04100$	& $0.00854$ \\
		$6212.727540$	& $15.03245$	& $0.00842$ \\
		$6212.734786$	& $15.01562$	& $0.00789$ \\
		$6212.742020$	& $15.01479$	& $0.00822$ \\
		$6212.749334$	& $14.97516$	& $0.00920$ \\
		$6212.757402$	& $14.96079$	& $0.00774$ \\
		$6212.764566$	& $14.92448$	& $0.00717$ \\
		$6212.771962$	& $14.91039$	& $0.00698$ \\
		$6212.779335$	& $14.89104$	& $0.00703$ \\
		$6212.786591$	& $14.87321$	& $0.00664$ \\
		$6212.793906$	& $14.85347$	& $0.00673$ \\
		$6212.801210$	& $14.86674$	& $0.00712$ \\
		$6212.808328$	& $14.87280$	& $0.00750$ \\
		$6212.815851$	& $14.87799$	& $0.00724$ \\
		$6212.823027$	& $14.90221$	& $0.00732$ \\
		$6212.830353$	& $14.93345$	& $0.00731$ \\
		$6212.838004$	& $14.95739$	& $0.00695$ \\
		$6212.844983$	& $14.95177$	& $0.00709$ \\
		$6212.852437$	& $14.96977$	& $0.00674$ \\
		$6212.859879$	& $14.95651$	& $0.00671$ \\
		$6212.866997$	& $14.94872$	& $0.00663$ \\
		$6212.874370$	& $14.96280$	& $0.00625$ \\
		$6212.881765$	& $14.95505$	& $0.00639$ \\
		\hline \\
	\end{tabular}
\end{table}

\begin{table}
	\caption{New in-transit radial velocities for WASP-71, obtained using the HARPS instrument on the night of 2012 October 26.}
	\label{tab:wasp71RVs}
	\begin{tabular}{lll}
		\hline \\
		Time					& RV		& Uncertainty \\ [2pt]
		${\rm BJD}_{\rm TDB}$	& km\,s$^{-1}$		& km\,s$^{-1}$ \\ [2pt]
		\hline \\
		$6226.542863$	& $7.84218$	& $0.01560$ \\
		$6226.552029$	& $7.83309$	& $0.01043$ \\
		$6226.563672$	& $7.84306$	& $0.00952$ \\
		$6226.574251$	& $7.86100$	& $0.01070$ \\
		$6226.584841$	& $7.87346$	& $0.01033$ \\
		$6226.595929$	& $7.83857$	& $0.01261$ \\
		$6226.606924$	& $7.85509$	& $0.00851$ \\
		$6226.617306$	& $7.86364$	& $0.00895$ \\
		$6226.628301$	& $7.85190$	& $0.00927$ \\
		$6226.639077$	& $7.82695$	& $0.00869$ \\
		$6226.649875$	& $7.81235$	& $0.00858$ \\
		$6226.660546$	& $7.81457$	& $0.00798$ \\
		$6226.671762$	& $7.80844$	& $0.00666$ \\
		$6226.682016$	& $7.78029$	& $0.00639$ \\
		$6226.693116$	& $7.76759$	& $0.00638$ \\
		$6226.703590$	& $7.76617$	& $0.00599$ \\
		$6226.714261$	& $7.75415$	& $0.00842$ \\
		$6226.725361$	& $7.75168$	& $0.00744$ \\
		$6226.736032$	& $7.75129$	& $0.00710$ \\
		$6226.746610$	& $7.73425$	& $0.00852$ \\
		$6226.757606$	& $7.73557$	& $0.00779$ \\
		$6226.768508$	& $7.74321$	& $0.00814$ \\
		$6226.779087$	& $7.73863$	& $0.00844$ \\
		$6226.789770$	& $7.74986$	& $0.00877$ \\
		$6226.801077$	& $7.72767$	& $0.01005$ \\
		$6226.811540$	& $7.74158$	& $0.00902$ \\
		$6226.822431$	& $7.74230$	& $0.01002$ \\
		\hline \\
	\end{tabular}
\end{table}

\begin{table}
	\scriptsize
	\caption{New in-transit radial velocities for WASP-76, obtained using the HARPS instrument on the night of 2012 November 11. Data marked with $\dagger$ were obtained during twilight, and were excluded from our analysis. }
	\label{tab:wasp76RVs}
	\begin{tabular}{lll}
		\hline \\
		Time					& RV		& Uncertainty \\ [2pt]
		${\rm BJD}_{\rm TDB}$	& km\,s$^{-1}$		& km\,s$^{-1}$ \\ [2pt]
		\hline \\
		$\dagger6243.496186$	& $-1.09877$	& $0.00429$ \\
		$\dagger6243.501185$	& $-1.07251$	& $0.00411$ \\
		$6243.505560$	& $-1.06883$	& $0.00434$ \\
		$6243.509403$	& $-1.07022$	& $0.00427$ \\
		$6243.513245$	& $-1.06395$	& $0.00426$ \\
		$6243.517053$	& $-1.06692$	& $0.00399$ \\
		$6243.520953$	& $-1.06354$	& $0.00374$ \\
		$6243.524726$	& $-1.07515$	& $0.00372$ \\
		$6243.528638$	& $-1.06760$	& $0.00382$ \\
		$6243.532411$	& $-1.06302$	& $0.00389$ \\
		$6243.536253$	& $-1.07220$	& $0.00401$ \\
		$6243.540107$	& $-1.06351$	& $0.00377$ \\
		$6243.543984$	& $-1.06557$	& $0.00394$ \\
		$6243.547757$	& $-1.06993$	& $0.00389$ \\
		$6243.551716$	& $-1.07947$	& $0.00363$ \\
		$6243.555523$	& $-1.08211$	& $0.00360$ \\
		$6243.559331$	& $-1.08265$	& $0.00363$ \\
		$6243.563139$	& $-1.08680$	& $0.00357$ \\
		$6243.567004$	& $-1.09083$	& $0.00368$ \\
		$6243.570777$	& $-1.09064$	& $0.00415$ \\
		$6243.574724$	& $-1.07989$	& $0.00475$ \\
		$6243.578613$	& $-1.09278$	& $0.00426$ \\
		$6243.582443$	& $-1.09255$	& $0.00374$ \\
		$6243.586216$	& $-1.09101$	& $0.00393$ \\
		$6243.590047$	& $-1.10042$	& $0.00402$ \\
		$6243.593994$	& $-1.11003$	& $0.00450$ \\
		$6243.597697$	& $-1.10157$	& $0.00414$ \\
		$6243.601482$	& $-1.10190$	& $0.00473$ \\
		$6243.605359$	& $-1.11008$	& $0.00458$ \\
		$6243.609317$	& $-1.10474$	& $0.00508$ \\
		$6243.613229$	& $-1.11579$	& $0.00506$ \\
		$6243.616933$	& $-1.11362$	& $0.00455$ \\
		$6243.620810$	& $-1.11277$	& $0.00423$ \\
		$6243.624618$	& $-1.11116$	& $0.00393$ \\
		$6243.628495$	& $-1.11463$	& $0.00366$ \\
		$6243.632337$	& $-1.11943$	& $0.00336$ \\
		$6243.636180$	& $-1.12098$	& $0.00324$ \\
		$6243.640022$	& $-1.12073$	& $0.00342$ \\
		$6243.643888$	& $-1.12748$	& $0.00340$ \\
		$6243.647672$	& $-1.12518$	& $0.00333$ \\
		$6243.651445$	& $-1.12467$	& $0.00358$ \\
		$6243.655380$	& $-1.11811$	& $0.00404$ \\
		$6243.659257$	& $-1.11684$	& $0.00391$ \\
		$6243.663100$	& $-1.11818$	& $0.00379$ \\
		$6243.666977$	& $-1.12417$	& $0.00353$ \\
		$6243.670785$	& $-1.11357$	& $0.00367$ \\
		$6243.674523$	& $-1.12522$	& $0.00331$ \\
		$6243.678389$	& $-1.12550$	& $0.00354$ \\
		$6243.682266$	& $-1.12776$	& $0.00358$ \\
		$6243.685958$	& $-1.12669$	& $0.00415$ \\
		$6243.689997$	& $-1.13571$	& $0.00423$ \\
		$6243.693805$	& $-1.13646$	& $0.00431$ \\
		$6243.697589$	& $-1.14239$	& $0.00450$ \\
		$6243.701640$	& $-1.14299$	& $0.00434$ \\
		$6243.705343$	& $-1.13806$	& $0.00382$ \\
		$6243.709221$	& $-1.14263$	& $0.00385$ \\
		$6243.713005$	& $-1.15742$	& $0.00359$ \\
		$6243.716917$	& $-1.14773$	& $0.00337$ \\
		$6243.720725$	& $-1.14841$	& $0.00317$ \\
		$6243.724567$	& $-1.15342$	& $0.00334$ \\
		$6243.728479$	& $-1.15186$	& $0.00361$ \\
		$6243.732865$	& $-1.15442$	& $0.00348$ \\
		$6243.736743$	& $-1.15420$	& $0.00356$ \\
		$6243.740481$	& $-1.15511$	& $0.00332$ \\
\hline \\
	\end{tabular}
\end{table}

\begin{table}
	\caption{New in-transit radial velocities for WASP-78, obtained using the HARPS instrument on the night of 2012 November 2.}
	\label{tab:wasp78RVs}
	\begin{tabular}{lll}
		\hline \\
		Time					& RV		& Uncertainty \\ [2pt]
		${\rm BJD}_{\rm TDB}$	& km\,s$^{-1}$		& km\,s$^{-1}$ \\ [2pt]
		\hline \\
		$6234.615044$	& $0.51919$	& $0.02351$ \\
		$6234.625588$	& $0.53285$	& $0.02095$ \\
		$6234.636583$	& $0.54676$	& $0.02041$ \\
		$6234.647463$	& $0.50238$	& $0.01932$ \\
		$6234.658054$	& $0.52525$	& $0.02047$ \\
		$6234.669049$	& $0.56150$	& $0.02096$ \\
		$6234.679721$	& $0.52599$	& $0.02182$ \\
		$6234.690624$	& $0.52157$	& $0.02240$ \\
		$6234.701411$	& $0.54176$	& $0.02561$ \\
		$6234.712326$	& $0.51550$	& $0.02363$ \\
		$6234.723020$	& $0.54373$	& $0.02355$ \\
		$6234.733819$	& $0.52400$	& $0.02220$ \\
		$6234.744606$	& $0.47403$	& $0.02093$ \\
		$6234.755613$	& $0.42754$	& $0.02057$ \\
		$6234.766181$	& $0.46334$	& $0.01833$ \\
		$6234.776563$	& $0.45630$	& $0.01975$ \\
		$6234.787547$	& $0.43274$	& $0.02144$ \\
		$6234.798554$	& $0.45586$	& $0.02073$ \\
		$6234.809538$	& $0.41622$	& $0.01953$ \\
		$6234.820337$	& $0.42134$	& $0.01862$ \\
		$6234.830916$	& $0.46127$	& $0.01650$ \\
		$6234.841622$	& $0.43577$	& $0.01648$ \\
		$6234.852421$	& $0.43693$	& $0.01464$ \\
		$6234.863405$	& $0.47287$	& $0.01599$ \\
		$6234.874203$	& $0.46658$	& $0.01625$ \\
	\hline \\
	\end{tabular}
\end{table}

\begin{table}
	\caption{New in-transit radial velocities for WASP-79, obtained using the HARPS instrument on the night of 2012 November 13.}
	\label{tab:wasp78RVs}
	\begin{tabular}{lll}
		\hline \\
		Time					& RV		& Uncertainty \\ [2pt]
		${\rm BJD}_{\rm TDB}$	& km\,s$^{-1}$		& km\,s$^{-1}$ \\ [2pt]
		\hline \\
		$6244.650501$	& $5.02953$	& $0.01838$ \\
		$6244.664587$	& $5.02735$	& $0.01808$ \\
		$6244.676010$	& $4.99396$	& $0.02304$ \\
		$6244.684135$	& $5.08237$	& $0.02248$ \\
		$6244.691878$	& $5.06599$	& $0.02969$ \\
		$6244.699830$	& $5.15211$	& $0.02258$ \\
		$6244.706045$	& $5.16701$	& $0.02396$ \\
		$6244.714807$	& $5.16033$	& $0.03097$ \\
		$6244.721369$	& $5.18134$	& $0.02347$ \\
		$6244.728615$	& $5.22263$	& $0.02053$ \\
		$6244.735860$	& $5.18547$	& $0.02309$ \\
		$6244.743453$	& $5.20207$	& $0.02014$ \\
		$6244.750501$	& $5.20501$	& $0.01865$ \\
		$6244.757400$	& $5.23074$	& $0.02281$ \\
		$6244.765629$	& $5.23317$	& $0.02347$ \\
		$6244.772191$	& $5.19395$	& $0.02390$ \\
		$6244.779992$	& $5.18341$	& $0.02337$ \\
		$6244.787249$	& $5.16418$	& $0.02480$ \\
		$6244.794634$	& $5.23456$	& $0.02440$ \\
		$6244.802076$	& $5.15278$	& $0.02212$ \\
		$6244.808974$	& $5.13525$	& $0.02376$ \\
		$6244.816705$	& $5.10280$	& $0.02294$ \\
		$6244.823731$	& $4.99968$	& $0.02512$ \\
		$6244.830722$	& $5.00030$	& $0.02408$ \\
		$6244.839263$	& $4.97480$	& $0.03132$ \\
		$6244.846231$	& $4.91483$	& $0.03050$ \\
		$6244.854194$	& $4.97432$	& $0.02788$ \\
		$6244.861243$	& $4.95226$	& $0.02404$ \\
		$6244.868141$	& $4.91467$	& $0.02497$ \\
	\hline \\
	\end{tabular}
\end{table}

\begin{table}
	\caption{New photometric observations for WASP-61, obtained in white light using EulerCam on the night of 2012 December 22. We present here the first ten data; the full table is available online as supplementary data.}
	\label{tab:wasp79RVs}
	\begin{tabular}{lll}
		\hline \\
		Time					& RV		& Uncertainty \\ [2pt]
		${\rm BJD}_{\rm TDB}$	& km\,s$^{-1}$		& km\,s$^{-1}$ \\ [2pt]
		\hline \\
		$6244.650501$	& $5.02953$	& $0.01838$ \\
		$6244.664587$	& $5.02735$	& $0.01808$ \\
		$6244.676010$	& $4.99396$	& $0.02304$ \\
		$6244.684135$	& $5.08237$	& $0.02248$ \\
		$6244.691878$	& $5.06599$	& $0.02969$ \\
		$6244.699830$	& $5.15211$	& $0.02258$ \\
		$6244.706045$	& $5.16701$	& $0.02396$ \\
		$6244.714807$	& $5.16033$	& $0.03097$ \\
		$6244.721369$	& $5.18134$	& $0.02347$ \\
		$6244.728615$	& $5.22263$	& $0.02053$ \\
		\hline \\
	\end{tabular}
\end{table}

\begin{table}
	\caption{New photometric observations for WASP-78, obtained using EulerCam on the nights of 2012 November 2 (I-band) and 2012 November 26 (R-band). We present here the first ten data from each set of observations; the full table is available online as supplementary data.}
	\label{tab:wasp79RVs}
	\begin{tabular}{lll}
		\hline \\
		Time					& RV		& Uncertainty \\ [2pt]
		${\rm BJD}_{\rm TDB}$	& km\,s$^{-1}$		& km\,s$^{-1}$ \\ [2pt]
		\hline \\
		\multicolumn{3}{l}{$I$ band} \\
		$6234.604847$	& $0.000644$	& $0.000813$ \\
		$6234.606256$	& $0.000162$	& $0.000803$ \\
		$6234.607646$	& $0.000035$	& $0.000798$ \\
		$6234.609009$	& $0.000748$	& $0.000793$ \\
		$6234.610402$	& $0.001041$	& $0.000784$ \\
		$6234.611760$	& $0.001415$	& $0.000780$ \\
		$6234.613124$	& $-0.000817$	& $0.000777$ \\
		$6234.614500$	& $0.002074$	& $0.000771$ \\
		$6234.615886$	& $0.001722$	& $0.000762$ \\
		$6234.617286$	& $0.001553$	& $0.000758$ \\
		\multicolumn{3}{l}{$R$ band} \\
		$6258.536656$	& $0.000103$	& $0.000813$ \\
		$6258.538001$	& $0.000892$	& $0.000801$ \\
		$6258.539247$	& $0.000002$	& $0.000891$ \\
		$6258.540377$	& $-0.000612$	& $0.000885$ \\
		$6258.541491$	& $0.000998$	& $0.000873$ \\
		$6258.542653$	& $0.000649$	& $0.000867$ \\
		$6258.543765$	& $0.000564$	& $0.000861$ \\
		$6258.544914$	& $0.001276$	& $0.000857$ \\
		$6258.546089$	& $0.002385$	& $0.000850$ \\
		$6258.547215$	& $0.000933$	& $0.000853$ \\
		\hline \\
	\end{tabular}
\end{table}

\begin{table}
	\caption{New photometric observations for WASP-79, obtained in the $R$ band using EulerCam on the nights of 2012 November 11 and December 4. We present here the first 10 data from each set of observations; the full table can be viewed online.}
	\label{tab:wasp79RVs}
	\begin{tabular}{lll}
		\hline \\
		Time					& RV		& Uncertainty \\ [2pt]
		${\rm BJD}_{\rm TDB}$	& km\,s$^{-1}$		& km\,s$^{-1}$ \\ [2pt]
		\hline \\
		\multicolumn{3}{l}{2012 November 11} \\
		$6244.656675$	& $-0.000354$	& $0.000533$ \\
		$6244.657558$	& $0.000972$	& $0.000532$ \\
		$6244.658458$	& $-0.003319$	& $0.000534$ \\
		$6244.659337$	& $-0.002535$	& $0.000533$ \\
		$6244.660232$	& $-0.001641$	& $0.000527$ \\
		$6244.661104$	& $-0.000249$	& $0.000527$ \\
		$6244.662007$	& $-0.002509$	& $0.000528$ \\
		$6244.662901$	& $-0.003302$	& $0.000529$ \\
		$6244.663807$	& $-0.005031$	& $0.000525$ \\
		$6244.664692$	& $-0.001716$	& $0.000523$ \\
		\multicolumn{3}{l}{2012 December 4} \\
		$6266.629315$	& $0.001236$	& $0.000564$ \\
		$6266.629960$	& $0.001505$	& $0.000564$ \\
		$6266.630613$	& $0.001733$	& $0.000564$ \\
		$6266.631321$	& $0.000342$	& $0.000505$ \\
		$6266.632120$	& $0.001124$	& $0.000505$ \\
		$6266.632936$	& $-0.000455$	& $0.000505$ \\
		$6266.633727$	& $-0.001233$	& $0.000505$ \\
		$6266.634437$	& $-0.001463$	& $0.000559$ \\
		$6266.635094$	& $-0.000475$	& $0.000558$ \\
		$6266.635777$	& $-0.000373$	& $0.000558$ \\
	\hline \\
	\end{tabular}
\end{table}

\end{document}